\documentclass[12pt]{article}

\usepackage{graphicx}
\usepackage{cite}

\def\beq{\begin{eqnarray}}
\def\eeq{\end{eqnarray}}
\def\bea{\begin{eqnarray*}}
\def\eea{\end{eqnarray*}}
\def\wtilde{\widetilde}

\newcommand{\Kahler}{K\"ah\-ler~}

\makeatletter
\def\@cite#1#2{{[{#1}]\if@tempswa\typeout
{IJCGA warning: optional citation argument
ignored: `#2'} \fi}}
\newcount\@tempcntc
\def\@citex[#1]#2{\if@filesw\immediate\write\@auxout{\string\citation{#2}}\fi
  \@tempcnta\z@\@tempcntb\m@ne\def\@citea{}\@cite{\@for\@citeb:=#2\do
    {\@ifundefined
       {b@\@citeb}{\@citeo\@tempcntb\m@ne\@citea\def\@citea{,}{\bf ?}\@warning
       {Citation `\@citeb' on page \thepage \space undefined}}%
    {\setbox\z@\hbox{\global\@tempcntc0\csname b@\@citeb\endcsname\relax}%
     \ifnum\@tempcntc=\z@ \@citeo\@tempcntb\m@ne
       \@citea\def\@citea{,}\hbox{\csname b@\@citeb\endcsname}%
     \else
      \advance\@tempcntb\@ne
      \ifnum\@tempcntb=\@tempcntc
      \else\advance\@tempcntb\m@ne\@citeo
      \@tempcnta\@tempcntc\@tempcntb\@tempcntc\fi\fi}}\@citeo}{#1}}
\def\@citeo{\ifnum\@tempcnta>\@tempcntb\else\@citea\def\@citea{,}%
  \ifnum\@tempcnta=\@tempcntb\the\@tempcnta\else
   {\advance\@tempcnta\@ne\ifnum\@tempcnta=\@tempcntb \else \def\@citea{--}\fi
    \advance\@tempcnta\m@ne\the\@tempcnta\@citea\the\@tempcntb}\fi\fi}
\makeatother

\newcommand{\drawsquare}[2]{\hbox{%
\rule{#2pt}{#1pt}\hskip-#2pt%  left vertical
\rule{#1pt}{#2pt}\hskip-#1pt%  lower horizontal
\rule[#1pt]{#1pt}{#2pt}}\rule[#1pt]{#2pt}{#2pt}\hskip-#2pt%  upper horizontal
\rule{#2pt}{#1pt}}% right vertical

\def\vbr{\vphantom{\sqrt{F_e^i}}}% vertical brace for tables

\newcommand{\Yfund}{\drawsquare{7}{0.6}}%  fundamental
\newcommand{\Ysymm}{\drawsquare{7}{0.6}\hskip-0.6pt%
	\drawsquare{7}{0.6}}%  symmetric second rank tensor
\newcommand{\Yasymm}{\drawsquare{7}{0.6}\hskip-7.6pt%
	\raisebox{7pt}{\drawsquare{7}{0.6}}}%  antisymmetric second rank

\newcommand{\fund}{\drawsquare{6.5}{0.4}}%  fundamental

\newcommand{\asymm}{\raisebox{-3pt}{\drawsquare{6.5}{0.4}\hskip-6.9pt%
        \raisebox{6.5pt}{\drawsquare{6.5}{0.4}}}}%  antisymmetric second rank

\newcommand{\Ythrees}{\raisebox{-.5pt}{\drawsquare{6.5}{0.4}}\hskip-0.4pt%
          \raisebox{-.5pt}{\drawsquare{6.5}{0.4}}\hskip-0.4pt% 
          \raisebox{-.5pt}{\drawsquare{6.5}{0.4}}}%  symmetric third rank
\newcommand{\Yfours}{\raisebox{-.5pt}{\drawsquare{6.5}{0.4}}\hskip-0.4pt%
          \raisebox{-.5pt}{\drawsquare{6.5}{0.4}}\hskip-0.4pt% 
          \raisebox{-.5pt}{\drawsquare{6.5}{0.4}}\hskip-0.4pt% 
          \raisebox{-.5pt}{\drawsquare{6.5}{0.4}}}%  symmetric fourth rank
\newcommand{\Ythreea}{\raisebox{-3.5pt}{\drawsquare{6.5}{0.4}}\hskip-6.9pt%
        \raisebox{3pt}{\drawsquare{6.5}{0.4}}\hskip-6.9pt
        \raisebox{9.5pt}{\drawsquare{6.5}{0.4}}}

\newcommand{\Yadjoint}{\raisebox{-3.5pt}{\drawsquare{6.5}{0.4}}\hskip-6.9pt%
        \raisebox{3pt}{\drawsquare{6.5}{0.4}}\hskip-0.4pt
        \raisebox{3pt}{\drawsquare{6.5}{0.4}}}%  SU(3) adjoint
\newcommand{\Ysquare}{\raisebox{-3.5pt}{\drawsquare{6.5}{0.4}}\hskip-0.4pt%
        \raisebox{-3.5pt}{\drawsquare{6.5}{0.4}}\hskip-13.4pt%
        \raisebox{3pt}{\drawsquare{6.5}{0.4}}\hskip-0.4pt%
        \raisebox{3pt}{\drawsquare{6.5}{0.4}}}%  4 boxes in a square
\newcommand{\Yoneoone}{\raisebox{-3.5pt}{\drawsquare{6.5}{0.4}}\hskip-6.9pt%
        \raisebox{3pt}{\drawsquare{6.5}{0.4}}\hskip-6.9pt%
        \raisebox{9.5pt}{\drawsquare{6.5}{0.4}}\hskip-0.4pt%
        \raisebox{9.5pt}{\drawsquare{6.5}{0.4}}}%

\def\beq{\begin{eqnarray}}
\def\eeq{\end{eqnarray}}
\def\bea{\begin{eqnarray*}}
\def\eea{\end{eqnarray*}}

\begin{document}

\setcounter{section}{0}
\setcounter{equation}{0}
\setcounter{footnote}{0}

%%%%%%%%%%%%%%%%%%%%%%%%%%%%%%%%%%%%%%%%%%%%%%%%%%%%%%%%%%%%%%
%%%%%%%%%%%%%%%%%%%%%%%%%%%%%%%%%%%%%%%%%%%%%%%%%%%%%%%%%%%%%%
%%%%%%%%%%%%%%%%%%%%%%%%%%%%%%%%%%%%%%%%%%%%%%%%%%%%%%%%%%%%%%

%Create the title page

\title{TASI-2002 Lectures:\\
Non-perturbative Supersymmetry}

\author{John Terning\footnote{email address:  {\tt terning@lanl.gov}} \\
Theory Division T-8\\
 Los Alamos National Laboratory\\
 Los Alamos, NM  87545 }

\maketitle

\vspace{2.5cm}

\begin{abstract}
These lectures contain a pedagogical review of non-perturbative results from holomorphy and Seiberg duality with applications
to dynamical SUSY breaking. Background material on anomalies, instantons, unitarity bounds from superconformal symmetry, and gauge mediation are also included.
 \end{abstract}
 
 \thispagestyle{empty}

\newpage
\tableofcontents
\newpage

%%%%%%%%%%%%%%%%%%%%%%%%%%%%%%%%%%%
%%%%%%%%%%%%%%%%%%%%%%%%%%%%%%%%%%%%
\section{Holomorphy}
\setcounter{equation}{0}
Two of the early exciting observations about ${\mathcal N}=1$ SUSY 
\cite{susy}
gauge theories were the absence of quadratic loop corrections
to scalar masses and the absence of renormalization  for many
superpotentials \cite{Amati:ft,Shifman:1986zi,Shifman:1991dz,Grisaru:1979wc}. 
While the former property was an immediately obvious
consequence of Bose-Fermi symmetry, the latter
 was only much later realized to follow in a very simple way from
the holomorphy \cite{Seiberg:1993vc,Seiberg:1994bp} of the superpotential, 
that is from the fact that
superpotentials are functions of chiral superfields but not of the
complex conjugates of these fields. Holomorphy arguments laid the
groundwork for the later development of Seiberg duality 
\cite{Seiberg:1995ac,IntriligatorSeiberg}.

%%%%%%%%%%%%%%%%%%%%%%%%%%%
\subsection{Non-renormalization Theorems}

Couplings in the superpotential can always be regarded as background
fields, and so superpotentials are also holomorphic functions of
coupling constants.  The fact that superpotentials are holomorphic 
can be used to prove powerful non-renormalization theorems
\cite{Seiberg:1993vc}. If we integrate out physics above a scale $\mu$ (i.e.
calculate the Wilsonian effective action\footnote{As opposed to the 1PI effective
action, see ref. \cite{Kaplunovsky:1994fg,Burgess:1995aa} for a related discussion.})
then the 
effective superpotential must also be a holomorphic function of the
couplings. Consider a theory renormalized at some scale $\Lambda$ with
a superpotential:
\beq
W_{\rm tree} = {{m}\over{2}} \phi^2 + {\lambda\over{3}}\phi^3 . 
\eeq
Here $\phi$ is a chiral superfield. I will also refer to the scalar component
as $\phi$ and the fermion component by $\psi$.

In general for ${\mathcal N}=1$ SUSY theories there is at most
one independent symmetry
generator\footnote{For ${\mathcal N}=2$ there is an $SU(2)_R$ $R$
symmetry and for  ${\mathcal N}=4$ there is an $SU(4)_R \sim SO(6)_R$ $R$
symmetry.}, referred to as the $R$ charge\footnote{The existence of
a discrete symmetry often referred to a $R$-parity is completely
independent of the continuous $R$ symmetry discussed here 
\cite{Martin:1997ns}.},
which  doesn't commute with the SUSY generators: 
\beq
[R,Q_\alpha]=-Q_\alpha,
\eeq
so we have $R[\psi] = R[\phi] -1$, $R[\theta] = 1$. Since
the Lagrangian in our toy model has Yukawa couplings,
\beq
{\mathcal L} \supset \frac{\lambda}{3} \phi \psi \psi~,
\eeq
which must have zero $R$ charge, we have
\beq
3R[\phi] -2=0~,
\eeq
and therefore $R[W]=2$. More generally we could get the same result by noting: 
\beq
{\mathcal L}_{\rm int} = \int d^2 \theta W~.
\eeq
By convention superfields are labeled by the $R$ charge of 
the lowest component, which in our example of a chiral
superfield is the scalar component. By convention the gaugino assigned
$R$ charge 1.

In our toy model we can make the following charge
assignments:
\begin{equation}
\begin{array}{ccc}
\quad & U(1) & U(1)_R \\
\phi & 1 & 1 \\
m & - 2 & 0 \\
 \lambda & -3 & -1 \\
\end{array}
\end{equation}
where we are treating the mass and coupling as background spurion
fields in order to keep track of all the symmetry information.
  Note that non-zero
values for $m$ and $\lambda$ explicitly break both $U(1)$ symmetries,
but these symmetries still lead to selection rules.

If we consider the effective superpotential generated by integrating out
modes from $\Lambda$ down to some scale $\mu$, then the symmetries and 
holomorphy of the effective superpotential restrict
it to be of the form
\beq
 W_{\rm eff}
&=& m \phi^2 \, h\left( \frac{\lambda \phi}{m} \right) \nonumber \\
&=& \sum_n a_n \lambda^n m^{1-n} \phi^{n+2}~,
\eeq
since $m \phi^2$ has $U(1)$ charge 0 and $R$-charge 2, and  
$\lambda \phi / m$ has  $U(1)$ charge 0 and $R$-charge 0.
The weak coupling limit $\lambda \rightarrow 0$ restricts $n \ge 0$, and
the massless limit $m  \rightarrow 0$ restricts $n \le 1$ so

\beq
W_{\rm eff}= {{m}\over{2}} \phi^2 +  \frac{\lambda}{3} \phi^3 = W_{\rm tree}~.
\eeq
Thus we have shown that the superpotential is not renormalized.

%%%%%%%%%%%%%%%%%%%%%%%%%%%%%%%%%%
\subsection{Wavefunction Renormalization}
Next consider the kinetic terms:
\beq
{\mathcal L}_{\rm kin.} = Z \partial_\mu \phi^* \partial^\mu \phi +i Z \psi^\dagger
\overline{\sigma}^\mu \partial_\mu \psi~,
\eeq
where the wavefunction renormalization factor is a non-holomorphic function
\beq
Z=Z(m,\lambda,m^\dagger,\lambda^\dagger,\mu,\Lambda)~.
\eeq
If we integrate out modes down to $\mu > m$ we have (at one-loop order)
\beq
Z=  1 + c \lambda \lambda^\dagger \ln\left({{\Lambda^2}\over{\mu^2}}\right)~,
\eeq
where $c$ is a constant determined by the  perturbative calculation.
If we integrate out modes down to scales below $m$ we have
\beq
Z=  1+ 
 + c \lambda \lambda^\dagger \ln\left({{\Lambda^2}\over{m m^\dagger}}\right)
\eeq
So there is wavefunction renormalization, and the couplings of canonically
normalized fields run.  In our example the running mass and running
coupling are given by
\beq
{{m}\over{Z}}, {{\lambda}\over{Z^{{3}\over{2}}}}~.
\eeq

%%%%%%%%%%%%%%%%%%%%%%%%%%%%%%%%%%%%%%%%
\subsection{Integrating Out}
Consider a model with two different chiral superfields:
\beq
W = {{1}\over{2}} M \phi_H^2 +{{\lambda}\over{2}} \phi_H \phi^2
\label{holo:3:superpot}
\eeq
This model has three global $U(1)$ symmetries:
\beq
\begin{array}{cccc}
 & U(1)_A & U(1)_B & U(1)_R \\
 \phi_H & 1 & 0 & 1 \\
 \phi & 0 & 1 & {{1}\over{2}} \\
 M & -2 & 0 & 0 \\
 \lambda & -1 & -2 & 0 \\
 \end{array}
\eeq
where  $U(1)_A$ and $U(1)_B$ are spurious symmetries for $M$, $\lambda \ne 0$.
If we want to integrate out modes down to $\mu < M$, we must 
integrate out $\phi_H$.
An arbitrary term in the effective superpotential has the form 
\beq
\phi^j M^k \lambda^p ~.
\eeq
To preserve the symmetries we must have $j=4$, $p=2$, and $k=-1$.
By comparing with tree-level perturbation theory we can determine
the coefficient:
\beq
W_{\rm eff} = -{{\lambda^2 \phi^4}\over{8 M}}~.
\label{holo:3:supereff}
\eeq
We could also derive this exact result using the algebraic equation of motion
\beq
{{\partial W}\over{\phi_H}}=M \phi_H +{{\lambda}\over{2}}  \phi^2= 0~.
\eeq
We can simply solve this equation for $\phi_H$ and plug the result back into
the superpotential (\ref{holo:3:superpot}), which yields the perturbative
effective superpotential
(\ref{holo:3:supereff}).

Another interesting example is 
\beq
W = {{1}\over{2}} M \phi_H^2 +{{\lambda}\over{2}} \phi_H \phi^2 +{{y}\over{6}} \phi_H^3~.
\eeq
The $\phi_H$  equation of motion yields
\beq
\phi_H=-\frac{M}{y}\left(1 \pm \sqrt{1-\frac{ \lambda y \phi^2}{ M^2} }\right)~.
\eeq
Note that as $y \rightarrow 0$, the two vacua approach $\phi_H=- \lambda \phi /(2 M)$ (as in
the previous example) and  $\phi_H=\infty$.
 Integrating out  $\phi_H$ yields
\beq
W_{\rm eff}= {{M^3}\over{3 y^2}}
\left(1-\frac{ 3 \lambda y \phi^2}{2 M^2}
\pm \left(1-\frac{ \lambda y \phi^2}{ M^2}\right) 
\sqrt{1-\frac{ \lambda y \phi^2}{ M^2}}\right)~.
\eeq
The singularities in $W_{\rm eff}$ indicate points in the parameter space
and the space of $\phi$ VEVs where $\phi_H$ becomes
massless and we shouldn't have integrated it out.  The mass of $\phi_H$ can be found by calculating
\beq
\frac{\partial^2 W}{\partial \phi_H^2}= M+y \phi_H~,
\eeq
and substituting in the solution of the equation of motion:
\beq
\frac{\partial^2 W}{\partial \phi_H^2}= \mp M  \sqrt{1-\frac{ \lambda y \phi^2}{ M^2} }~.
\eeq
We could also apply the holomorphy analysis to this problem. First, we can maintain the
symmetries of the previous example by  assigning charges (-3,0,-1)
under $U(1)_A \times U(1)_B \times U(1)_R$  to the coupling $y$.
Then requiring that the effective superpotential has no dependence on $\phi_H$ and maintains
the three symmetries we find that it must have the form
\beq
W_{\rm eff}=\frac{M^3}{y^2}  f\left(\frac{  \lambda y \phi^2}{ M^2}\right)~,
\eeq
for some function $f$, just as we found from integrating out $\phi_H$.

%%%%%%%%%%%%%%%%%%%%%%%%%%%%%%%%%%%%%%
\subsection{The Holomorphic Gauge Coupling}

Using the superspace notation\footnote{For an excellent review see ref. \cite{Martin:1997ns}.}
$y^\mu \equiv x^\mu - i \theta \sigma^\mu  \theta$
(where $\theta$ is an anti-commuting Grassmannian variable)
we can write a chiral supermultiplet consisting of a scalar $\phi$, a fermion $\psi$ and
an auxiliary field $F$ as a chiral superfield\beq
\Phi=\phi(y)+ \sqrt{2} \theta \, \psi(y) + \theta^2 F(y)~.
\eeq
We can also use this notation to represent
an $SU(N)$ gauge super-multiplet as a chiral superfield
\beq
W^a_\alpha = -i\lambda^a_\alpha(y) +\theta_\alpha D^a(y) 
-(\sigma^{\mu \nu} \theta)_\alpha F^a_{\mu\nu}(y)
-(\theta \theta)\sigma^\mu D_\mu \lambda^{a\dagger}(y)~,
\eeq
where the index $a$ labels an element of the adjoint representation,
running from $1$ to $N^2-1$, $\lambda^a$ is the gaugino field,
$F^a_{\mu\nu}$ is the usual gauge
field strength,and $D^a$ is the
auxiliary field. We have used the notation that the $\sigma^i$ are the usual Pauli matrices and
\beq
\sigma^\mu&=&(I,\sigma^i)\\
 {\overline \sigma}^\mu&=& (I,-\sigma^i)\\
\sigma^{\mu\nu} &=& {{i}\over{4}}(\sigma^\mu {\overline \sigma}^\nu - 
\sigma^\nu {\overline \sigma}^\mu)~.
\eeq
The field strength can be written as
\beq
F^a_{\mu\nu} =\partial_\mu A^a_\nu -\partial_\nu A^a_\mu +g f^{abc} 
A^b_\mu A^c_\nu ~,
\eeq
where $f^{abc}$ is the structure constant of the gauge group
determined through the adjoint generators $T^a$ by
\beq
[T^a,T^b]=i f^{abc} ~.
\eeq
Finally, using the notation
\beq
\tau \equiv {{\theta_{\rm YM}}\over{2 \pi}} +{{4 \pi i}\over{g^2}}~,
\eeq
we can write the SUSY Yang-Mills action as a superpotential term
\beq
&&\frac{1}{16 \pi i} \int d^4 x \int d^2 \theta \, \tau \, W^a_\alpha
W^a_\alpha + h.c. \\
&&=\int d^4 x\left[ -\frac{1}{4 g^2} F^{a \mu \nu}F^a_{\mu\nu} 
-{{\theta_{\rm YM}}\over{32\pi^2}}
F^{a\mu\nu}{\wtilde F}^a_{\mu\nu} +{{i}\over{g^2}} \lambda^{a\dagger} 
\overline{\sigma}^\mu D_\mu \lambda^a +{{1}\over{2 g^2}}D^a D^a \right]~,\nonumber
\label{holo:SYMaction}
\eeq
where
\beq
{\wtilde F}^a_{\mu\nu}={{1}\over{2}} \epsilon^{\mu\nu\alpha\beta}
F^a_{\alpha\beta}~.
\eeq
Note that the gauge coupling $g$ only appears only in $\tau$ which is 
a holomorphic parameter, but the gauge fields are
not canonically normalized. 
To go to a canonically normalized basis\footnote{Due to subtleties with
the measure in the path integral there is a non-trivial relation
between the running of the holomorphic gauge coupling and the running of 
the canonical gauge coupling 
\cite{Shifman:1986zi,Shifman:1991dz,Dine:1994su,Arkani-Hamed:1997mj}.} 
we would rescale the fields by
\beq
(A^a_\mu, \lambda^a_\alpha, D^a) \rightarrow g 
(A^a_\mu, \lambda^a_\alpha, D^a)
\eeq
Recall that the one-loop running of the gauge coupling $g$ is
given by the renormalization group equation:
\beq
\mu \frac{d g}{d \mu} = - \frac{b}{16 \pi^2} g^3 ~,
\eeq
where for an $SU(N)$ gauge theory with $F$ flavors and ${\mathcal N}=1$ 
supersymmetry, 
\beq
b=3N-F~.
\eeq
The solution for the running coupling is
\beq
\frac{1}{g^2(\mu)} = - \frac{b}{8 \pi^2} \ln \left( \frac{|\Lambda|}{\mu}
\right)~,
\eeq
where $|\Lambda|$ is the intrinsic scale of the non-Abelian gauge theory
that enters through dimensional transmutation.
We can then write the one-loop running version of
our holomorphic parameter $\tau$ as
\beq
\tau_{1-{\rm loop}} &=& {{\theta_{\rm YM}}\over{2 \pi}} +{{4 \pi i}\over{g^2(\mu)}}\\
&=& \frac{1}{2 \pi i} \ln \left[ \left(\frac{|\Lambda|}{\mu}\right)^b
 e^{i \theta_{\rm YM}} \right]~.
\eeq
We can then define a holomorphic intrinsic scale
\beq
\Lambda &\equiv& |\Lambda| e^{i \theta_{\rm YM}/b}\\
&=& \mu e^{2 \pi i \tau/b}~,
\label{holo:holomorphicscale}
\eeq
or equivalently
\beq
\tau_{1-{\rm loop}} = \frac{b}{2 \pi i} \ln \left(\frac{\Lambda}{\mu}\right)~.
\label{holo:taupert}
\eeq

In order to take account of non-perturbative effects we need to understand
the term  in the action proportional to $\theta_{\rm YM}$.
This $F {\wtilde F}$ term violates a 
discrete symmetry: CP 
(the product of charge conjugation and parity).  The CP violating term
can be rewritten as
\beq
 F^{a\mu\nu}{\wtilde F}^a_{\mu\nu} = 4 \epsilon^{\mu \nu \rho \sigma}
\partial_\mu {\rm Tr} \left( A_\nu \partial_\rho A_\sigma + \frac{2}{3}
A_\nu A_\rho A_\sigma \right) ~.
\eeq
Thus the CP violating term is a total derivative and can have no effect
in perturbation theory since it integrates to terms at the boundary
of space-time.
Nevertheless it is well known that it can have non-perturbative
effects.  To see this consider the following semi-classical instanton configuration
of the gauge field:
\beq
A^a_\mu(x) =  \frac{ h^a_{\mu \nu}(x-x_0)^\nu}{(x-x_0)^2 + \rho^2}~,
\label{holo:instantonconfig}
\eeq
where $h^a_{\mu \nu}$ describes how the instanton is oriented in the
gauge space and spacetime.
Equation (\ref{holo:instantonconfig}) 
represents an instanton configuration of size $\rho$
centered about the point $x_0^\nu$. Such instantons have a non-trivial,
topological
winding number, $n$, which takes integer values.
The CP violating term measures the winding number:
\beq
\frac{\theta_{\rm YM}}{32\pi^2} \int d^4x \, 
F^{a\mu\nu}{\wtilde F}^a_{\mu\nu} = n\, \theta_{\rm YM}~.
\eeq

Since the path integral has the form
\beq
\int {\mathcal D} A^a {\mathcal D} \lambda^a {\mathcal D} D^a \, e^{i S}~,
\eeq
and the action $S$ depends on  $\theta_{\rm YM}$ 
only through a term which is an integer times $\theta_{\rm YM}$
it follows that 
\beq
\theta_{\rm YM} \rightarrow \theta_{\rm YM} + 2 \pi~,
\eeq
is a symmetry of the theory since it has no effect on the path integral.

The Euclidean action of a instanton configuration can be bounded,
since
\beq
0 &\le& \int d^4x Tr\left( F_{\mu \nu} \pm {\wtilde F}_{\mu\nu}\right)^2
= \int d^4x Tr\left(2  F^2 \pm 2 F {\wtilde F}\right)~,
\eeq
so we have
\beq
 \int d^4x Tr  F^2 \ge |\int d^4x Tr F {\wtilde F} | = 32 \pi^2 |n|~.
\eeq

  Thus one instanton effects are
suppressed by
\beq
e^{-S_{\rm int}} = e^{\frac{-8 \pi^2}{g^2(\mu)} + i \theta_{\rm YM} } = 
\left(\frac{\Lambda}{\mu}\right)^b~.
\label{one-instanton}
\eeq

If we integrate down to the scale $\mu$ we have the effective superpotential
\beq
W_{\rm eff}={{\tau(\Lambda;\mu)}\over{16\pi i}} W^a_\alpha W^a_\alpha~.
\eeq
Since the physics must be periodic in $\theta_{\rm YM}$,
\beq
\Lambda \rightarrow e^{2 \pi i/b} \Lambda~,
\eeq
is a symmetry.
To allow for non-perturbative corrections we can write the
most general form of $\tau$ as:
\beq
\tau(\Lambda;\mu)
 = {{b}\over{2 \pi i}}\ln\left( {{\Lambda}\over{\mu}}\right) +
f( \Lambda;\mu)~,
\eeq
where $f$ is a holomorphic function of $\Lambda$.  Since $\Lambda
\rightarrow 0$ corresponds to weak coupling where we must recover the
perturbative result (\ref{holo:taupert}), $f$ must have Taylor series
representation in positive powers of $\Lambda$.  Since plugging
\beq
\Lambda \rightarrow e^{2 \pi i/b} \Lambda
\eeq
into the perturbative term already shifts $\theta_{\rm YM}$ by $2 \pi$,
$f$ must be invariant under this transformation, so the Taylor series
must be in positive powers of $\Lambda^b$.  Thus in general we can
write:
\beq
\tau(\Lambda;\mu) = 
{{b}\over{2 \pi i}}\ln\left( {{\Lambda}\over{\mu}}\right) 
+ \sum_{n=1}^\infty a_n
\left( {{\Lambda}\over{\mu}}\right)^{b n} 
\label{holo:taunonpert}
\eeq
So the holomorphic gauge coupling only receives one-loop corrections and
non-perturbative $n$-instanton corrections, or in other words in
perturbation theory there is no additional running beyond one-loop.

%%%%%%%%%%%%%%%%%%%%%%%%%%%%%%%%
%%%%%%%%%%%%%%%%%%%%%%%%%%%%%%%%%

\section{Review of Anomalies and Instantons}
\setcounter{equation}{0}

The reader should feel free to skip over this section if they already have a basic familiarity
with anomalies and instantons.

%%%%%%%%%%%%%%%%%%%%%%%%%%%%%%%%%%%%%
\subsection{Anomalies in the Path Integral}
\label{holo:sec:fujikawa}
\begin{center}
\begin{figure}
\begin{center}
\includegraphics[width=0.4\hsize]{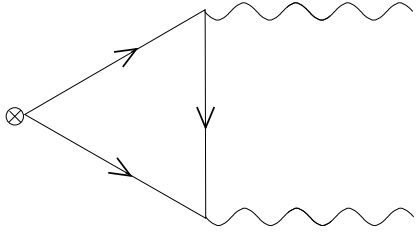}
\end{center}
\caption{The fermion triangle which contributes to the anomaly. One also
adds the crossed graph where the gauge bosons are interchanged.}
\label{holo:Fermiontriangle}
\end{figure}
\end{center}

Consider some fermions $\psi$ coupled to a gauge field $A^a$:
\beq
S_{\rm fermion}=\int d^4 x i \psi^\dagger {\overline \sigma}^\mu (\partial_\mu 
+i A_\mu^a T^a) \psi
\eeq
Under a position dependent chiral rotation
\beq
\psi &\rightarrow& e^{i \alpha(x)} \psi \\
S_{\rm fermion} &\rightarrow& S_{\rm fermion}
- \int d^4 x \alpha(x) \partial_\mu
(\psi^\dagger {\overline \sigma}^\mu \psi)
\eeq
Thus, since $\alpha= $ constant is a symmetry of the action,
classically we find
\beq
\partial_\mu (\psi^\dagger {\overline \sigma}^\mu \psi) = \partial_\mu j^\mu_A =0
\eeq
Quantum mechanically this is not true: the global current has an anomaly.  
One way to see this is to calculate the diagram with a fermion triangle and
the global current and and two gauge currents at the three vertices, as in 
Figure \ref{holo:Fermiontriangle}.

One finds  directly from this calculation
that the global current is not conserved:
\beq
\partial_\mu j^\mu_A \propto F^{a\mu\nu}{\wtilde F}^a_{\mu\nu}~.
\eeq

Another way to see this is from
Fujikawa's path integral derivation \cite{Fujikawa:1979ay} of the anomaly.
First define
\beq
\not\!\!D \equiv i {\overline \sigma_E}^\mu(\partial_\mu + i A_\mu) \\
\overline{\not\!\!D} \equiv i  \sigma_E^\mu(\partial_\mu - i A_\mu) 
\eeq
where $\sigma^i$ are the usual Pauli matrices and
\beq
\sigma_E^\mu=(iI,\sigma^i)~,\\
 {\overline \sigma_E}^\mu= (iI,-\sigma^i)~,\\
\sigma_E^{\mu\nu} = {{1}\over{4}}(\sigma_E^\mu {\overline \sigma_E}^\nu - 
\sigma_E^\nu {\overline \sigma_E}^\mu)~.
\eeq
Then
\beq
\int d^4 x  \psi^\dagger {\not\!\!D} \psi =
\int d^4 x  \psi {\overline {\not\!\!D}} \psi^\dagger
\eeq
and we can define orthonormal eigenfunctions $f_n$ and $g_n$ 
\beq
D^2 f_n={\overline {\not\!\!D}}{\not\!\!D} f_n & =& -\lambda_n^2 f_n ~,\\
{\overline D}^2 g_n = {\not\!\!D}{\overline {\not\!\!D}} g_n &=&-\lambda_n^2 g_n~,\\
{\not\!\!D} f_n &=& \lambda_n g_n ~,\\
{\overline {\not\!\!D}} g_n &=&- \lambda_n f_n ~,
\eeq
with the usual completeness properties
\beq
\sum_n f_n^*(x) f_n(y) &=& \delta(x-y)~,\\
\sum_n g_n^*(x) g_n(y) &=& \delta(x-y)~,\\
Tr \int d^4 x f_n^*(x) f_m(x) &=& \delta_{n m} ~,\\
Tr \int d^4 x g_n^*(x) g_m(x) &=& \delta_{n m} ~.
\eeq
In general $D^2$ and 
${\overline D}^2$ have different numbers of  zero eigenvalues.
We can expand the fermion fields in this basis:
\beq
\psi(x) = \sum_n a_f f_n(x)~,\\
\psi^\dagger(x)= \sum_n b_n g_n(x)~.
\eeq
The partition function in the background gauge field $A_\mu$ is given by a path integral over the fermion fields which in this basis reduces to
\beq
Z[A_\mu]=\int {\mathcal D} \psi {\mathcal D} \psi^\dagger  \, e^{-S}
= \int \Pi_{n m} d a_n d b_m   \, e^{-S}~.
\eeq
Under a chiral rotation
\beq
a_n \rightarrow a_n^\prime &=&  C_{n m} a_m ~, \\
b_n \rightarrow b_n^\prime &=& {\overline C}_{n m} b_m ~,
\eeq
where the transformation matrices are defined by
\beq 
C_{n m} &=& {\rm Tr} \int d^4 x e^{i \alpha(x)} f_n^*(x) f_m(x)~,\\
{\overline C}_{n m} &=& {\rm Tr} \int d^4 x e^{-i \alpha(x)} g_m^*(x) g_n(x)~.
\eeq
So the path integral measure changes by
\beq
\Pi_{n m} d a_n d b_m \rightarrow ({\rm det} C {\rm det} {\overline C})^{-1}
\Pi_{n m} d a_n d b_m~.
\eeq
where
\beq
({\rm det} C {\overline C})^{-1} =\exp\left( -i \int d^4 x\alpha(x) A(x) \right)~,\\
A(x) = {\rm Tr} \sum_n \left( f_n^*(x)f_n(x) - g^*_n(x)g_n(x) \right) ~.
\eeq
Since the partition function, $Z$, is independent of the chiral
rotation parameter $\alpha$, 
if we take a functional derivative with respect to
it we find:
\beq
0 = {{\delta Z}\over{\delta \alpha}}|_{\alpha=0} =
i\langle \partial_\mu j^\mu(x) -A(x) \rangle ~.
\eeq
To evaluate $A$, we use a smooth regulator function $R(z)$  to suppress the effects
of the highest eigenmodes.  $R(z)$ is chosen such that $R(0)=1$, $R(\infty)=0$, $R'(\infty)=0$, $R''(\infty)=0$, $\ldots$ For example, $e^{-z}$ 
satisfies these conditions. Inserting the regulator we have:
\beq
A(x)&& = \begin{array}{c} {\rm lim} \\M \rightarrow \infty \end{array}
{\rm Tr} \sum_n R(\lambda_n^2/M^2)\left( f_n^*(x)f_n(x) - g^*_n(x)g_n(x) \right)~,
\label{regulatedanomaly}\\
&&= \begin{array}{c} {\rm lim} \\M \rightarrow \infty \end{array}
{\rm Tr} \sum_n \left( f_n^*(x)R(-D^2/M^2)f_n(x) - 
g^*_n(x)R(-{\overline D}^2/M^2)g_n(x) \right)~,\nonumber\\
&&= \begin{array}{c} {\rm lim}\\y\rightarrow x\end{array}
 \begin{array}{c} {\rm lim} \\M \rightarrow \infty \end{array}
{\rm Tr} \left(R(-D^2/M^2) - 
R(-{\overline D}^2/M^2) \right) \delta(y-x)~.
\eeq
To simplify this first note that 
\beq
D^2 = \partial^2 -A^2 -\sigma^{\mu\nu}(F_{\mu\nu} - 2A_{[\mu}\partial_{\nu]})\equiv \partial^2 -A^2-F^+ +{\mathcal O}(\partial . A),\nonumber \\
{\overline D}^2 = \partial^2 -A^2 +
{\overline \sigma}^{\mu\nu}(F_{\mu\nu} - 2A_{[\mu}\partial_{\nu]})\equiv \partial^2 -A^2+ F^-  +{\mathcal O}(\partial . A).
\eeq
  For simplicity we will chose a gauge where $\partial . A=0$.
Using the Fourier transform of $\delta(y-x)$, one finds by Taylor expanding $R(z)$ around 
$(p^2+A^2)/M^2$ that
\beq
A(x) 
&=&
\begin{array}{c} {\rm lim}\\ M \rightarrow \infty \end{array}
{\rm Tr}\int {{d^4p}\over{(2\pi)^4}} \sum_{n=0}^\infty{{1}\over{n !}}\nonumber\\
&&\,\,\,\,\,\,
 \left[ \left(\frac{F^+}{M^2}\right)^n- 
 \left(-\frac{F^-}{M^2}\right)^n
 \right] R^{(n)}\left({{p^2+A^2}\over{M^2}}\right)~.
\eeq
 The $n=0$ in the series vanishes trivially, the Dirac trace of the $n=1$ term vanishes, and terms with $n > 2$ vanish in the large $M$ limit. Defining $z=p^2/M^2$ we are left with
\beq
A(x) 
 &=&{{1}\over{32 \pi^2}} \int_0^\infty z dz R^{(2)}(z)  \epsilon^{\mu\nu\alpha\beta}
 {\rm Tr} F_{\mu\nu} F_{\alpha\beta}~,\nonumber\\
 &=&{{1}\over{16 \pi^2}}  {\rm Tr} F_{\mu\nu} {\wtilde F}_{\mu\nu}~.
\eeq
So the anomaly is given by
\beq
\partial_\mu j^\mu(x_E) = {{1}\over{16 \pi^2}} F^{a \mu\nu} {\wtilde F}^a_{\mu\nu}~.
\eeq
Now consider integrating this equation over spacetime. The left-hand
side is given by integrating Eq. (\ref{regulatedanomaly})
which yields
\beq 
\int d^4x A & =&  \begin{array}{c} {\rm lim} \\M \rightarrow \infty \end{array}\sum_n R(\lambda_n^2/M^2)
{\rm Tr}  \int d^4x \left( f_n^*(x)f_n(x) - g^*_n(x)g_n(x) \right)~,\nonumber \\
& =& n_\psi - n_{\psi^\dagger}~,
\eeq
 which is just the number of  
fermion zero-modes minus the number of  
anti-fermion zero-modes, since all other modes occur in pairs and cancel in the difference. So
we have 
\beq
n_\psi - n_{\psi^\dagger}&=&
{{1}\over{32 \pi^2}}\int d^4x \, F^a_{\mu\nu} {\wtilde F}^a_{\mu\nu}\nonumber
 \\
&=&n~,
\eeq
where $n$ is the winding number (Pontryagin number) of the gauge field configuration.
This result is known as the Atiyah-Singer index theorem.

%%%%%%%%%%%%%%%%%%%%%%%%%%%%%%%%%%%%%%%%%%%
\subsection{Gauge Anomalies}
There are also potentially anomalies for three-point 
functions of gauge bosons which are
proportional to 
\beq
A^{abc} \equiv {\rm Tr} [T^a \{T^b,T^c\} ]~.
\eeq
These anomalies are potentially non-vanishing for example in 
the case with one $U(1)$ gauge boson and and two
$SU(N)$ gauge bosons, or with three $SU(N)$ gauge bosons for $N \ge 3$.

The anomaly for two fermions in two different representations ${\bf R_1}$ and ${\bf R_2}$
simply add
\beq
A^{abc}({\bf R_1} \oplus {\bf R_2}) = A^{abc}({\bf R_1}) + A^{abc}({\bf R_2})~,
\eeq
while for a fermion that is in two different representations ${\bf R_1}$ and ${\bf R_2}$
of two different groups one finds:
\beq
A^{abc}({\bf R_1} \otimes {\bf R_2}) = 
{\rm dim}({\bf R_1}) A^{abc}{\bf (R_2})+{\rm dim}({\bf R_2}) A^{abc}({\bf R_1})~.
\eeq

For the fundamental (defining) representation we define
\beq
d^{abc} \equiv {\rm Tr} [T_F^a \{T_F^b,T_F^c\}]~.
\eeq
It is then convenient to define an anomaly factor $A({\bf R})$, which
measures the anomaly
relative to the to the fundamental representation:
\beq
A^{abc}({\bf R}) = A({\bf R}) d^{abc}~.
\eeq
So the gauge anomaly for a theory vanishes if
\beq
\sum_i A({\bf R_i}) =0
\eeq

The dimension, index,  and anomaly coefficient 
of the smallest $SU(N)$ representations are listed below.
\beq \begin{array}{|c|c|c|c|} \hline
{\rm Irrep} & {\rm dim({\bf r})} & 2T({\bf r}) & A({\bf r}) \\ \hline
\Yfund & N & 1 & 1 \\
{\bf  Ad} & N^2-1 & 2N & 0 \\
\Yasymm & \frac{N(N-1)}{2} & N-2 & N-4 \\
\Ysymm & \frac{N(N+1)}{2} & N+2 & N+4 \\
\Ythreea & \frac{N(N-1)(N-2)}{6} & \frac{(N-3)(N-2)}{2} & 
\frac{(N-3)(N-6)}{2} \\
\Ythrees & \frac{N(N+1)(N+2)}{6} & \frac{(N+2)(N+3)}{2} & 
\frac{(N+3)(N+6)}{2} \\
\Yadjoint & \frac{N(N-1)(N+1)}{3} & N^2-3 & N^2-9 \\
\Ysquare & \frac{N^2(N+1)(N-1)}{12} & \frac{N(N-2)(N+2)}{3} &
           \frac{N(N-4)(N+4)}{3} \\
\Yfours & \frac{N(N+1)(N+2)(N+3)}{24} & \frac{(N+2)(N+3)(N+4)}{6} &
          \frac{(N+3)(N+4)(N+8)}{6} \\
\Yoneoone & \frac{N(N+1)(N-1)(N-2)}{8} & \frac{(N-2)(N^2-N-4)}{2} & 
            \frac{(N-4)(N^2-N-8)}{2} \\  \hline \end{array} 
\eeq

If the gauge anomaly does not vanish, then the theory can only
make sense as a spontaneously broken theory, since two triangle
graphs back to back generate a mass for the gauge bosons.
Alternatively if we start with an anomaly free gauge theory
and give masses to some subset of the fields so that the
anomaly no longer cancels, then the low energy effective theory
has an anomaly, but we can only produce such masses if the 
gauge symmetry is spontaneously broken.

%%%%%%%%%%%%%%%%%%%%%%%%%%%%%%%%%%%%%%%%%%
\subsection{Review of Instantons}
\label{holo:sec:instantons}
Recall that instantons\footnote{For an excellent review see ref. \cite{ColemanInstantons}.} 
are Euclidean solutions of 
\beq
D^\mu F^a_{\mu\nu}=0~,
\eeq
characterized by a size $\rho$, which,  as $|x| \rightarrow \infty$,
approach pure gauge:
\beq
A_\mu(x)\rightarrow i U(x) \partial_\mu U(x)^\dagger~.
\eeq
In general, 
instantons break one linear combination of axial $U(1)$ symmetries.
Consider the axial symmetry that has charge +1 for all (left-handed) fermions.
We have
\beq
\partial_\mu j^\mu_A \propto F^{a\mu\nu}{\wtilde F}^a_{\mu\nu}
\eeq
Integrating this current in the instanton background with winding number
$n$, one finds:
\beq
\int d^4x \partial_\mu j^\mu_A =n\left[\sum_r n_r \, 2 T({\bf r})\right]~,
\eeq
where $n_r$ is the number of fermions in the representation $r$.
Thus instantons can create or annihilate fermions. Also an axial
rotation of the fermions
\beq
\psi \rightarrow e^{i \alpha} \psi~,
\eeq
is equivalent to a shift of $\theta_{\rm YM}$
\beq
\theta_{\rm YM}\rightarrow \theta_{\rm YM} - \alpha \sum_r n_r \, 2 T({\bf r})~.
\eeq
In the instanton background the gauge covariant derivative can be diagonalized:
\beq
{\overline \sigma}^\mu D_\mu \psi_i = \lambda_i \psi_i~.
\eeq
For a fermion in representation ${\bf r}$ one finds $2 T({\bf r})$ zero eigenvalues
corresponding to $2 T({\bf r})$ ``zero-modes''.
(In the anti-instanton background $\psi^\dagger$ has $2 T({\bf r})$ zero 
eigenvalues.)

Consider a fundamental of $SU(N)$ with $T(\fund) = {{1}\over{2}}$. We can write
$\psi_i$ in terms of a Grassmann variable  and a complex eigenfunction
$\psi_i = \xi_i f_i(x)$.  We then have
\beq
\psi = \xi_0 f_0 + \sum_i \xi_i f_i ~,\\
\psi^\dagger =   \sum_i \xi_i^\dagger f_i^*~,
\eeq
where $f_0$ corresponds to the zero eigenvalue.
The path integration over this particular fermion is then
\beq
\int {\mathcal D} \psi {\mathcal D}\psi^\dagger = \int d \xi_0 \int \prod_{ij} 
d\xi_i d\xi^\dagger_j~.
\eeq
So
\beq
\int {\mathcal D} \psi {\mathcal D}\psi^\dagger \exp\left(-\int \psi^\dagger 
{\overline \sigma}^\mu D_\mu \psi\right) 
&&=\int d \xi_0 \int \prod_{ij} 
d\xi_i d\xi^\dagger_j\exp\left( -\sum_n \lambda_n \xi^\dagger_n \xi_n\right)\nonumber
\\
&&=
\int d \xi_0 \int \prod_{ij} 
d\xi_i d\xi^\dagger_j\prod_n\left(1- \lambda_n \xi^\dagger_n \xi_n\right)
\nonumber \\
&&=
\int d \xi_0 \prod_n\ \lambda_n \nonumber \\
&&=0~,
\eeq
since integrating a constant over a Grassmann variable vanishes:
\beq
\int d \xi_0=0~.
\eeq
However if we insert the fermion field into the path integral we find
\beq
\int {\mathcal D} \psi {\mathcal D}\psi^\dagger \exp\left(-\int \psi^\dagger 
{\overline \sigma}^\mu D_\mu \psi\right) \psi(x) 
=f_0(x) \,\prod_n\ \lambda_n ~.
\eeq
Thus for an instanton amplitude to be non-vanishing it must be accompanied
by one external fermion leg for each zero-mode.

At distances much larger than the instanton size `t Hooft \cite{'tHooft:fv}
showed that instantons
produce effective interactions for a quark $Q_{n j}$ with color index
$n$ and flavor index $j$:
\beq
{\mathcal L}_{\rm inst} = c \det \overline{Q}^{i n}Q_{n j} + h.c.
\eeq
This interaction respects the non-Abelian $SU(F) \times SU(F)$ flavor symmetry
but breaks the $U(1)_A$ symmetry.

%%%%%%%%%%%%%%%%%%%%%%%%%%%%%%%%%%%%%
\subsection{Instantons in Broken Gauge Theories}
In a theory with scalars that carry gauge quantum numbers, VEVs of the
scalars prevent us from finding solutions of the classical 
Euclidean equation of motion.
However we can find approximate solutions \cite{Csaki:1998vv} when we drop the scalar contribution
to the gauge current:
\beq
D^\mu F^a_{\mu\nu}= 0\\
D^\mu D_\mu \phi^j +{{\partial V(\phi)}\over{\partial \phi^*_j}} = 0~,
\eeq
As $|x| \rightarrow \infty$ we want the gauge field to go to pure gauge
and the scalar field to go to its VEV (up to a gauge transformation):
\beq
A_\mu(x)\rightarrow i U(x) \partial_\mu U(x)^\dagger\\
\phi^j \rightarrow U(x) \langle\phi^j\rangle~,
\eeq
where $\langle\phi^j\rangle$ is a vacuum solution.  For small
($\rho < 1/(gv)$) instantons with a completely broken gauge symmetry
we find Euclidean actions:
\beq
S_{\rm inst} = {{8\pi}\over{g^2}}\\
S_{\phi}=8 \pi^2 \rho^2 v^2~.
\eeq
Integrating over instanton locations and sizes we find
\beq
&&\int d^4 x_0 \int {{d\rho}\over{\rho^5}} e^{-S_{\rm inst}-S_{\phi}} \nonumber \\
&&=\int d^4 x_0 \int {{d\rho}\over{\rho^5}} 
\left(\rho \Lambda\right)^b e^{-8\pi^2 \rho^2v^2}~,
\eeq
which is dominated at 
\beq
\rho^2 ={{b}\over{16\pi^2 v^2}}~.
\eeq
Thus the integration is convergent: 
breaking the gauge symmetry provides an exponential infrared cutoff.

Note that since $A_\mu$ is related to an element of the gauge group at 
$|x|\rightarrow\infty$,
the topological character of the instanton relies on
\beq
U: S^3\rightarrow SU(2) \subset G
\eeq
If the scalar fields break the gauge group $G$ down to $H$, then there 
will still be pure instanton in the $H$ gauge theory if $SU(2)\subset H$.
If the instantons in $G/H$ can be gauge rotated into $SU(2)\subset H$, then
all $G$ instanton effects can be accounted for by the effective theory
through $H$ instantons.  If not, we must add new interactions in the
effective theory in order to match the physics properly.
Examples of when this is necessary \cite{Csaki:1998vv}  include
\beq
&&SU(N) \,\,{\rm breaks}\,\, {\rm completely} \nonumber\\
&&SU(N) \rightarrow U(1) \nonumber\\
&&SU(N) \times SU(N) \rightarrow SU(N)_{\rm diag} \nonumber\\
&&SU(N) \rightarrow SO(N)\nonumber
\eeq
This obviously happens whenever there are a different number of zero modes for $G$
and $H$ instantons. In general new interactions must be included in the
effective theory when $\pi_3(G/H)$ is non-trivial.

%%%%%%%%%%%%%%%%%%%%%%%%%%%%%%%%%%%%%%%%%
%%%%%%%%%%%%%%%%%%%%%%%%%%%%%%%%%%%%%%%%%
\section{Gaugino Condensation}
\setcounter{equation}{0}

\label{holo:sec:gaugino}

We now turn to the chiral symmetry transformations and condensation of
gauginos. Note that in pure $SU(N)$ SUSY Yang-Mills 
(that is with $F=0$ flavors, so the only fermion is a gaugino)
the $U(1)_R$ symmetry is broken by instantons. We can see this by calculating
the mixed triangle anomaly between one $U(1)_R$ current and two gluons
(see Figure \ref{holo:Fermiontriangle}).

The anomaly is proportional to a group theoretical factor (the anomaly
index) which arises by summing (tracing)
over all possible fermions in the loop
taking in to account the Abelian and non-Abelian charges.
In this example the anomaly index is given by
the $R$-charge of the gaugino (which is 1)
times the index of the adjoint representation ($T(Ad)= N$), so  
the anomaly  is non-vanishing.
 
Because of the anomaly, the
chiral rotation
\beq
\lambda^a \rightarrow e^{i\alpha} \lambda^a~,
\label{holo:u1R}
\eeq
is equivalent to a shift in the coefficient of the CP
violating term in the the action, 
$F^{a\mu\nu}{\wtilde F}^a_{\mu\nu}$,
\beq
\theta_{\rm YM}\rightarrow \theta_{\rm YM} - 
2N \alpha ~.
\eeq
The factor $2N$ arises because
the gaugino $\lambda^a$ is in the adjoint representation of
the gauge group and thus has $2N$ zero-modes in a one-instanton background.
Thus the chiral rotation is only a symmetry
when
\beq
\alpha={{k\pi}\over{N}}~,
\eeq
where $k$ is an integer, 
so the $U(1)_R$ symmetry is explicitly broken down to a 
discrete $Z_{2N}$ subgroup.
Treating the holomorphic gauge coupling $\tau$ as a background (spurion)
chiral superfield we can define a spurious
symmetry given by
\beq
\lambda^a \rightarrow e^{i\alpha} \lambda^a
~, \nonumber \\
\tau \rightarrow \tau + \frac{N \alpha}{\pi}~,
\label{holo:spuriousR}
\eeq
which leaves the path integral  invariant.

Assuming that SUSY Yang-Mills has no massless particles, just 
massive color-singlet composites, then holomorphy and symmetries
determine the effective superpotential to be:
\beq
W_{\rm eff}&=& a \mu^3 e^{{2\pi i \tau}\over{N}}\nonumber \\
&=& a \Lambda^3 ~.
\label{holo:Weffgaugino}
\eeq
This is the unique form because under the spurious $U(1)_R$ rotation
(\ref{holo:spuriousR}) the superpotential 
(which has $R$-charge 2) transforms as
\beq
W_{\rm eff} \rightarrow e^{2 i\alpha} W_{\rm eff}~.
\eeq
Since the 
holomorphic gauge coupling parameter $\tau$ transforms linearly
under the spurious $U(1)_R$ rotation, it must appear an exponential
with the coefficient given in (\ref{holo:Weffgaugino}).

Again treating $\tau$ as a background chiral superfield, then
 in the SUSY Yang Mills action (\ref{holo:SYMaction}) the $F$ component
of $\tau$ (which we will refer to as $F_\tau$) acts as a source
for $\lambda^a \lambda^a$.  With our assumption that there are no
massless degrees of freedom, the  effective action  at low energies 
is given by just the effective superpotential (\ref{holo:Weffgaugino}).
Thus the gaugino condensate is given by
\beq
\langle \lambda^a \lambda^a \rangle &=& 16 \pi i {{\partial}\over{\partial F_\tau}}
\ln Z\nonumber\\
&=& 16 \pi i {{\partial}\over{\partial F_\tau}} \int d^2 \theta W_{\rm eff}\nonumber\\
&=& 16 \pi i {{\partial}\over{\partial \tau}}  W_{\rm eff}\nonumber\\
&=& 16 \pi i {{2\pi i}\over{N}} a \mu^3 e^{{2\pi i \tau}\over{N}}~.
\eeq
Dropping the non-perturbative corrections\footnote{Which only
contribute a phase.} to the running of 
$\tau$  and plugging  $b=3N$ into (\ref{holo:taunonpert})
we find
\beq
\langle \lambda^a \lambda^a \rangle 
&=& -{{32 \pi^2}\over{N}}   a \Lambda^3 ~,
\label{holo:gauginocondensate}
\eeq
thus there is a non-zero gaugino condensate\footnote{For related work using instantons, see ref. 
\cite{Davies:1999uw,Novikov:ic,Shifman:ie}.}.
The presence of this condensate means that the vacuum does not respect
the discrete $Z_N$ symmetry since
\beq
\langle \lambda^a \lambda^a \rangle \rightarrow e^{2 i \alpha}
\langle \lambda^a \lambda^a \rangle~,
\eeq
is not invariant for all possible values of $\alpha = k\pi/N$.  In fact
it is only invariant for $k=0$ or $k=N$.
So we see that $Z_{2N}$ symmetry is spontaneously
broken to $Z_2$, and that there should be $N$
degenerate but distinct vacua.  
It is easy to check that the symmetry transformation
$\theta_{\rm YM}\rightarrow \theta_{\rm YM} +2 \pi$
sweeps out $N$ different values for $\langle \lambda^a \lambda^a \rangle$.
It should be kept in mind that at this point we have not justified
the assumption of no massless degrees of freedom which is crucial to
the calculation.

%%%%%%%%%%%%%%%%%%%%%%%%%%%%%%
\section{The Affleck-Dine-Seiberg Superpotential}
\setcounter{equation}{0}

%%%%%%%%%%%%%%%%%%%%%%%%%%%%%%
\subsection{Symmetry and Holomorphy}
Consider $SU(N)$ SUSY QCD with $F$ flavors (that is
there are $2NF$ chiral supermultiplets) where $F <N$.
We will denote the 
quarks and their superpartner squarks
that transform in the $SU(N)$ fundamental (defining) representation
by $Q$ and $\Phi$ respectively, and use $\overline{Q}$ and
$\overline{\Phi}$ for the quarks and squarks
 in the anti-fundamental representation.
The theory has an $SU(F) \times  SU(F) \times  U(1)\times   U(1)_R$
global symmetry.  The quantum numbers of the chiral supermultiplets
are summarized in the following table\footnote{As usual only the 
$R$-charge of the squark is given, and $R[Q]=R[\Phi]-1$.} where $\fund$ 
denotes 
the fundamental representation of the group.
\beq
\begin{tabular}{c|c|cccc}
& $SU(N)$ & $SU(F)$  & $SU(F)$ & $U(1)$ & $U(1)_R$ \\
\hline
$\Phi$, $Q$ & \fund & \fund & ${\bf 1}$ & 1 & ${{F-N}\over{F}}$ 
$\vphantom{\raisebox{3pt}{\asymm}}$\\
$\overline{\Phi}$, $\overline{Q}$ & $\overline{\fund}$   
& ${\bf 1}$  & $\overline{\fund}$
& -1 & ${{F-N}\over{F}}$ $\vphantom{\raisebox{3pt}{\asymm}}$\\
\end{tabular}
\eeq
The $SU(F) \times SU(F)$ global symmetry
is the analog of the $SU(3)_L \times SU(3)_R$ 
chiral symmetry of non-supersymmetric
QCD with 3 flavors, while the $U(1)$ is the 
analog\footnote{Up to a factor of $N$.} 
of baryon number since quarks (fermions in the fundamental representation
of the gauge group) and anti-quarks 
(fermions in the anti-fundamental representation
of the gauge group) have opposite charges. There is an additional
$U(1)_R$ relative to non-supersymmetric
QCD since in the supersymmetric theory there is also a gaugino.

Recall that the auxiliary $D^a$ fields for this theory are given by
\beq
D^a = g( \Phi^{*jn}(T^a)^m_n \Phi_{mj} - 
\overline{\Phi}^{jn}(T^a)^m_n \overline{\Phi}^*_{mj})~,
\eeq
where $j$ is a flavor index that runs from $1$ to $F$,
$m$ and $n$ are color indices that run from $1$ to $N$,
the index $a$ labels an element of the adjoint representation,
running from $1$ to $N^2-1$, and $T^a$ is a gauge group generator.
The $D$-term potential is given by:
\beq
V= \frac{1}{2g^2} D^a D^a ~.
\eeq
Classically there is a (D-flat) moduli space (space of VEVs 
where the potential
$V$ vanishes) which is given by
\beq
\langle \overline{\Phi}^{*}\rangle =
\langle \Phi \rangle = \left( \begin{array}{ccc}
v_1 & & \\
& \ddots &  \\
& & v_F \\
0 & \ldots & 0 \\
\vdots  & & \vdots \\
0 & \ldots & 0 \\
\end{array} \right)
\eeq
where $\langle \Phi \rangle$ is a matrix with $N$ rows and $F$ columns
and we have used global and gauge symmetries to rotate 
$\langle \Phi \rangle$ to a simple
form.
At a generic point in the moduli space the $SU(N)$
gauge symmetry is broken to $SU(N-F)$. There are 
\beq
N^2-1-((N-F)^2-1)=2NF - F^2
\eeq
broken generators, so of the original $2NF$ chiral supermultiplets
only $F^2$ singlets are left massless. This is because in the supersymmetric
Higgs mechanism  a massless vector supermultiplet ``eats'' an entire
chiral supermultiplet to form a massive vector supermultiplet.
We can describe the remaining $F^2$ 
light degrees of freedom in a gauge invariant
way by an $F \times F$ matrix field
\beq
M^j_i =  \overline{\Phi}^{j n}\Phi_{n i}~,
\eeq
where we sum over the color index n.
Since $M$ can be written as a  chiral superfield which is a product of
of chiral superfields\footnote{That is there is a fermionic
superpartner of $M$ (the ``mesino'') given by
$M_\psi=  \overline{Q}^{j n}\Phi_{n i}+ 
\overline{\Phi}^{j n} Q_{n i}$.}, 
then, because of holomorphy, 
the only renormalization of $M$ is the product
of wavefunction renormalizations for $\Phi$ and $\overline{\Phi}$.

The axial $U(1)_A$ symmetry which assigns each quark a charge 1 is 
explicitly broken
by instantons, while the $U(1)_R$ symmetry remains unbroken.
To check this we can calculate the corresponding mixed anomalies between
the global current and two gluons.
For $U(1)_R$ we multiply the $R$-charge by the index of the representation,
and sum over fermions.  The gaugino contributes $1 \cdot N$ while each of
the $2 F$ quarks contributes $(\frac{F-N}{F}-1)\frac{1}{2}$.  Adding these
together we find that the mixed anomaly for $U(1)_R$ vanishes:
\beq
A_{Rgg} = N+\left(\frac{F-N}{F}-1\right)\frac{1}{2}=0  ~.
\eeq
For the $U(1)_A$ anomaly, the gauginos do not contribute since they have
no $U(1)_A$ charge and we find the anomaly coefficient is non-vanishing:
\beq
A_{Agg} = 1 \cdot 2 F \cdot \frac{1}{2}~.
\eeq
  To keep track of selection rules arising from the
broken $U(1)_A$ we can define a spurious
symmetry in the usual way.  The transformations
\beq
Q\rightarrow e^{i\alpha}Q ~,\nonumber\\
\overline{Q}\rightarrow e^{i\alpha}\overline{Q}~,\nonumber\\
\theta_{\rm YM}\rightarrow\theta_{\rm YM}+2F\alpha~,
\eeq
leave the path integral  invariant.
Under this transformation the holomorphic intrinsic scale 
(\ref{holo:holomorphicscale}) transforms as
\beq
\Lambda^b \rightarrow e^{i2F \alpha}\Lambda^b~.
\eeq
To construct the effective superpotential we can
make use of the following chiral superfields: $W^a$, $\Lambda$,
and $M$. Their charges under the $U(1)_R$ and spurious $U(1)_A$ symmetries
are given in the following table.
\beq
\begin{array}{c|cc}
 &U(1)_A & U(1)_R \\ \hline
W^a W^a& 0 & 2 \\
\Lambda^b & 2F & 0 \\
{\rm det}M & 2F & 2(F-N) \\
\end{array}
\eeq

Note that ${\rm det}M$ is the only $SU(F)\times SU(F)$ invariant we can
make out of $M$. To be invariant, a general non-perturbative 
term in the Wilsonian superpotential must have the form
\beq
\Lambda^{bn} (W^aW^a)^m({\rm det}M)^p~.
\eeq
As usual to preserve the periodicity of $\theta_{YM}$ 
we can only
have powers of $\Lambda^{b}$ (for $m=1$ we can still have the perturbative
term $b \log \Lambda W^aW^a$ because the change in the path integral
measure due to the anomaly).
Since the superpotential is neutral under $U(1)_A$ and has
charge 2 under $U(1)_R$, the two symmetries require:
\beq
0 = n \,2F +p\,2F \\
2=2 m +p\,2(F-N)~.
\eeq
The solution of these equations is
\beq
n=-p= {{1-m}\over{N-F}}~.
\eeq
Since $b=3N-F>0$ we can only have a sensible weak-coupling limit
($\Lambda \rightarrow 0$) if $n\ge 0$,
which implies $p\le 0$ and (because $N > F$)
$m \le 1$.  Since $W^aW^a$ contains derivative terms,
locality requires $m \ge 0$ and that $m$ is integer valued. 
In other words, since we trying to find a Wilsonian effective action
(which corresponds to performing the path integral over 
field modes with momenta larger than a scale $\mu$) which is valid
at low-energies (momenta below $\mu$) it must have a sensible derivative
expansion.
So there are only two possible
terms in the effective superpotential:
$m=0$ and $m=1$.  The $m=1$ term is just the tree-level field
strength
term.  The coefficient of this term 
is restricted by the periodicity of $\theta_{\rm YM}$ to be 
proportional to $b \ln \Lambda$.  So we see that the gauge coupling
receives no non-perturbative renormalizations.
The other  term ($m=0$) is the Affleck-Dine-Seiberg superpotential
\footnote{First discussed by Davis, Dine, and Seiberg \cite{DavisDineSeiberg}
and explored in more detail by Affleck, Dine,  and Seiberg
in \cite{Affleck:1984mk}}:
\beq
W_{\rm ADS} = C_{N,F} \left( {{\Lambda^{3N-F}}\over{ {\rm det}M
}}\right)^{{1}\over{N-F}} ~,
\eeq
where $C_{N,F}$ is in general renormalization scheme dependent.

%%%%%%%%%%%%%%%%%%%%%%%%%%%%%%%%%%%%
\subsection{Consistency of $W_{\rm ADS}$: Moduli Space}
We can check whether the Affleck-Dine-Seiberg superpotential is consistent 
by constructing effective
theories with fewer colors or flavors by going out in the classical
moduli space or by
adding mass terms for some of the flavors.  
Consider giving a large vacuum expectation value
(VEV), $v$, to one
flavor. This breaks the gauge symmetry to $SU(N-1)$ and one
flavor is partially ``eaten'' by the Higgs mechanism (since there
are $2N-1$ broken generators)
so the effective theory has $F-1$ flavors. There are $2F-1$ additional
gauge singlet chiral supermultiplets left over as well
since 
\beq
2NF-(2N-1)-(2F-1)= 2(N-1)(F-1)~ .  
\eeq

We can write
an  
effective theory for the  $SU(N-1)$ gauge theory with 
$F-1$ flavors (since the gauge singlets only interact with the fields
in the effective gauge theory by the exchange of heavy gauge bosons they
must decouple from the gauge theory at low energies, i.e. they interact
only through irrelevant operators with dimension greater than 4). 
The running holomorphic gauge coupling, $g_L$, in the low-energy theory
is given by
\beq
{{8\pi^2}\over{g_L^2(\mu)}} = b_L \ln\left({{\mu}\over{\Lambda_L}}\right)~,
\eeq
where $b_L$ is the standard $\beta$-function coefficient of the
low-energy theory
\beq
b_L= 3(N-1)-(F-1)= 3 N -F - 2~,
\eeq
and $\Lambda_L$ is the holomorphic intrinsic scale of the low-energy
effective theory
\beq
\Lambda_L &\equiv& |\Lambda_L|e^{i \theta_{\rm YM}/b_L} \nonumber \\
&=& \mu e^{2 \pi i \tau_L/b_L}~,
\label{holo:holomorphicscale_L}
\eeq
This coupling 
should match onto the running coupling of the high-energy theory
\beq
{{8\pi^2}\over{g^2(\mu)}} = b \ln\left({{\mu}\over{\Lambda}}\right)~,
\eeq
at the scale of the heavy gauge boson threshold\footnote{
The fact that the heavy gauge boson mass is $v$ rather than $g v$ is
a result of having an unconventional normalization for the quark and
squark fields, this is necessary to maintain $g_L$ and $\Lambda_L$ 
as holomorphic parameters. For related discussions on this point
see \cite{Shifman:1986zi,Shifman:1991dz,Dine:1994su}.} $v$:
\beq
{{8\pi^2}\over{g^2(v)}} =
{{8\pi^2}\over{g_L^2(v)}}+c~,
\eeq
where $c$ represents scheme dependent
corrections, which vanish in the 
$\overline{\rm DR}$ renormalization 
scheme\footnote{$\overline{\rm DR}$ uses dimensional regularization through
dimensional reduction with modified minimal subtraction 
\cite{Siegel:1979wq,Finnell:1995dr}.}.
So we have
\beq
\left({{\Lambda}\over{v}}\right)^b = 
\left({{\Lambda_L}\over{v}}\right)^{b_L}~,
\nonumber\\
{{\Lambda^{3N-F}}\over{v^2}} = \Lambda^{3N-F-2}_{N-1,F-1}~,
\label{holo:matchingoneflavorvev}
\eeq
where we have started labeling (for later convenience)
the intrinsic scale of the low-energy effective theory
with a subscript showing the number of colors and
flavors in the gauge theory that it corresponds to:
$\Lambda_{N-1,F-1} = \Lambda_L$.
If we  represent the light $(F-1)^2$ degrees of freedom (corresponding
to the gauge invariant combinations of the chiral superfields
that are fundamentals under the remaining gauge symmetry) 
as an $(F-1) \times (F-1)$ matrix $\wtilde M$ then we have
\beq
{\rm det}M = v^2{\rm det}\wtilde M +\ldots~,
\eeq
where $\ldots$ represents terms involving the decoupled gauge singlet
fields.
Plugging these results into the  Affleck-Dine-Seiberg superpotential
for $N$ colors and $F$ flavors (which we denote by $W_{\rm ADS}(N,F)$)
and using 
\beq
\left({{\Lambda^{3N-F}}\over{v^2}}\right)^\frac{1}{N-F}
 = \left( \Lambda^{3N-F-2}_{N-1,F-1}\right)^\frac{1}{(N-1)-(F-1)}~,
\eeq
(which follows from equation (\ref{holo:matchingoneflavorvev})) we reproduce 
$W_{\rm ADS}(N-1,F-1)$ provided that 
$C_{N,F}$ is only a function of $N-F$.

Giving equal VEVs to all flavors we break the gauge symmetry from
$SU(N)$ down to $ SU(N-F)$, and all the flavors are ``eaten''.
Through the same method of matching running
couplings,
\beq
\left(\frac{\Lambda}{v}\right)^{3N-F} = 
\left(\frac{\Lambda_{N-F,0}}{v}\right)^{3(N-F)}~,
\eeq
we then have 
\beq
{{\Lambda^{3N-F}}\over{v^{2F}}} = \Lambda^{3(N-F)}_{N-F,0}~.
\eeq
So the effective superpotential is given by
\beq
W_{\rm eff}=C_{N,F} \Lambda^{3}_{N-F,0}~.
\eeq
Which agrees with the result (\ref{holo:Weffgaugino}) derived
from holomorphy arguments for gaugino condensation
in pure SUSY Yang-Mills, as  described
in section \ref{holo:sec:gaugino}.

%%%%%%%%%%%%%%%%%%%%%%%%%%%%%%%%%%%%%
\subsection{Consistency of $W_{\rm ADS}$: Mass Perturbations}
Now consider giving a mass, $m$, to one flavor.  Below this
mass threshold the low-energy effective theory is 
an $SU(N)$ gauge theory with $F-1$ flavors.  Matching the holomorphic
gauge coupling of the effective theory to that of the underlying theory
at the scale $m$ gives:
\beq
\left({{\Lambda}\over{m}}\right)^b = \left({{\Lambda_L}\over{m}}\right)^{b_L}
\nonumber\\
m\Lambda^{3N-F} = \Lambda^{3N-F+1}_{N,F-1}
\eeq
Using holomorphy the superpotential must have the form
\beq
W_{\rm exact} = \left( {{\Lambda^{3N-F}}\over{ {\rm det}M
}}\right)^{{1}\over{N-F}} f(t)~,
\eeq
where
\beq
t= m M^F_F\left( {{\Lambda^{3N-F}}\over{ {\rm det}M
}}\right)^{{-1}\over{N-F}}~,
\eeq
and $f(t)$ is an as yet undetermined function. Note that since
$m M^F_F$ is actually a mass term in the underlying superpotential, it
has $U(1)_A$ charge 0, and $R$-charge 2, so $t$ has $R$-charge 0.

Taking the weak coupling, small mass 
limit $\Lambda\rightarrow 0$, $m \rightarrow 0$, we must recover our previous
results with the addition of a small mass term, hence
\beq
f(t) = C_{N,F} +t~.
\eeq
However in this double limit $t$ is still arbitrary so this is the
exact form of $f(t)$.  Thus we find
\beq
W_{\rm exact} = C_{N,F}\left( {{\Lambda^{3N-F}}\over{ {\rm det}M
}}\right)^{{1}\over{N-F}} +m M^F_F~.
\label{holo:W_exact:massterm}
\eeq

The equations of motion for $M^F_F$ and $M^j_F$
\beq
\frac{\partial W_{\rm exact}}{\partial M^F_F}&=&
C_{N,F}\left( {{\Lambda^{3N-F}}\over{ {\rm det}M
}}\right)^{{1}\over{N-F}}\left(\frac{-1}{N-F}\right) +m =0 ~,\\
\frac{\partial W_{\rm exact}}{\partial M^j_F}&=&
C_{N,F}\left( {{\Lambda^{3N-F}}\over{ {\rm det}M
}}\right)^{{1}\over{N-F}}\left(\frac{-1}{N-F}\right)
\frac{{\rm cof}(M^j_F)}{{\rm det}M}=0~,
\eeq
(where ${\rm cof}(M^F_i)$ is the cofactor of the matrix element $M^F_i$)
  imply that 
\beq
\frac{C_{N,F}}{N-F}\left( {{\Lambda^{3N-F}}\over{ {\rm det}M
}}\right)^{{1}\over{N-F}}= m M^F_F~,
\label{holo:eq.motion_MFF}
\eeq
and  that the cofactor of $M^F_i$ is zero.
Thus $M$ has 
the block diagonal form
\beq
 M =
 \left( \begin{array}{cc}
{\wtilde M} &0 \\
0 &  M^F_F 
\end{array} \right)~,
\eeq
where ${\wtilde M}$ is an $(F-1)\times(F-1)$ matrix.
Plugging  the solution (\ref{holo:eq.motion_MFF}) into 
the exact superpotential (\ref{holo:W_exact:massterm}) we find
\beq
W_{\rm exact}(N,F-1) = \left(N-F+1\right)
\left({{C_{N,F}}\over{N-F}}\right)^{{N-F}\over{N-F+1}} 
\left( {{\Lambda^{3N-F+1}_{N,F-1}}\over{ {\rm det}{\wtilde M}
}}\right)^{{1}\over{N-F+1}}~,
\eeq
which is just $W_{ADS}$ up to an overall constant.
Thus,for consistency, we have a recursion relation:
\beq
C_{N,F-1}=
\left(N-F+1\right)
\left({{C_{N,F}}\over{N-F}}\right)^{{N-F}\over{N-F+1}}~.
\eeq
An instanton calculation is reliable 
for $F=N-1$ since the non-Abelian gauge group is completely broken, 
and such a calculation \cite{Novikov:ic,Finnell:1995dr}  
determines $C_{N,N-1}=1$ in the
$\overline{\rm DR}$ scheme, and hence
\beq
C_{N,F}=N-F~.
\eeq

We can also consider adding masses for all the flavors.  Holomorphy determines
the superpotential to be 
\beq
W_{\rm exact} = C_{N,F}\left( {{\Lambda^{3N-F}}\over{ {\rm det}M
}}\right)^{{1}\over{N-F}} +m^i_j M^j_i~,
\eeq
where $m^i_j$ is the quark (and squark) mass matrix.
The equation of motion for $M$ gives
\beq
M^j_i &=& (m^{-1})^j_i \left( {{\Lambda^{3N-F}}\over{ {\rm det}\, M
}}\right)^{{1}\over{N-F}} ~.
\label{holo:MVEV}
\eeq
Taking the determinant and plugging the result back in to 
(\ref{holo:MVEV}) gives
\beq
 {\bar \Phi}^j \Phi_i   =
M^j_i& =& (m^{-1})^j_i \left({\rm det}m\, \Lambda^{3N-F}\right)^{{1}\over{N}}~.
\eeq
Note that since the result involves the $N$-th root there are $N$ distinct
vacua that differ by the phase of $M$. This is in precise accord with the Witten
index argument \cite{Witten:df}.

Matching the holomorphic gauge coupling at the mass thresholds gives
\beq
 \Lambda^{3N-F}{\rm det}\,m = \Lambda_{N,0}^{3N}~.
\eeq
So
\beq
W_{eff}= N \Lambda_{N,0}^3~,
\eeq
which agrees with our holomorphy result\footnote{Which assumed a mass gap.}, 
equation (\ref{holo:Weffgaugino}),
for pure SUSY Yang-Mills
and determines the parameter $a=N$ up to a phase.
So the gaugino condensate (\ref{holo:gauginocondensate})
 is given by
\beq
\langle \lambda^a \lambda^a \rangle=  -32 \pi^2 e^{2 \pi i k/N}
\Lambda_{N,0}^3 ~,
\label{holo:gauginocondensate2}
\eeq
where $k=1...N$.
Thus starting with $F=N-1$ flavors
we can consistently derive the correct
Affleck-Dine-Seiberg
effective superpotential for $0\le F < N-1$, and gaugino 
condensation for $F=0$. This retroactively justifies the assumption
that there was a mass gap in SUSY Yang Mills.

%%%%%%%%%%%%%%%%%%%%%%%%%%%%%%%%
\subsection{Generating $W_{\rm ADS}$ from Instantons}
\begin{center}
\begin{figure}
\begin{center}
\includegraphics[width=0.8\hsize]{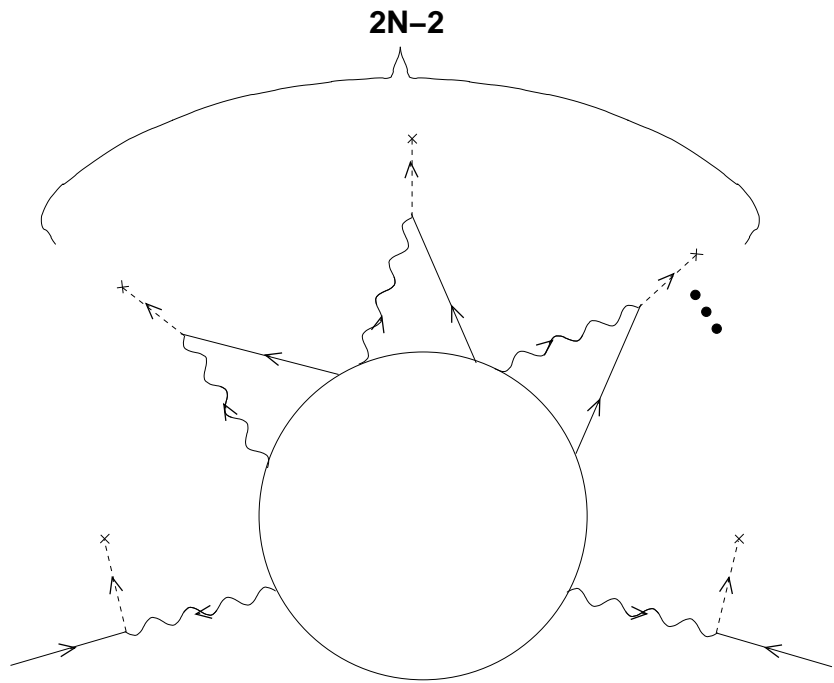}
\end{center}
\caption{Instanton with 2N-2  quark legs (solid, straight lines)
 and 2N gaugino legs (wavy lines), connected by 2N squark VEVs (dashed lines
with crosses).}
\label{holo:instanton}
\end{figure}
\end{center}

Recall that  the Affleck-Dine-Seiberg superpotential
\beq
W_{\rm ADS} \propto \Lambda^{{b}\over{N-F}}~,
\eeq
while instanton effects are suppressed by\footnote{See Eq. (\ref{one-instanton}).}
\beq
e^{-S_{\rm inst}} \propto \Lambda^b~.
\eeq
So for $F=N-1$ is is possible that instantons can generate
$W_{\rm ADS}$. Since $SU(N)$ can be completely broken in this case,
we can do a reliable instanton calculation.
When all VEVs are equal 
the ADS superpotential predicts quark masses of order
\beq
{{\partial^2 W_{ADS}}\over{\partial \Phi_i \partial{\overline \Phi}^j}} \sim
{{\Lambda^{2N+1}}\over{v^{2N}}}~,
\eeq
and a vacuum energy density of order
\beq
\left| {{\partial W_{ADS}}\over{\partial \Phi_i}} \right|^2 \sim
\left|{{\Lambda^{2N+1}}\over{v^{2N-1}}}  \right|^2~.
\eeq
Looking at a single instanton vertex  we find $2N$ gaugino legs 
(corresponding to $2N$ zero-modes) and $2F=2N-2$ quark legs,
as shown in Figure \ref{holo:instanton}. All
the quark legs can be connected to gaugino legs by the insertion of
a scalar VEV.  The remaining two gaugino legs can be converted to two
quark legs by the insertion of two more VEVs.  Thus a fermion mass
is generated.
From the instanton calculation we find the quark mass is given by
\beq
m &\sim& e^{- 8 \pi^2/g^2(\rho)} \frac{v^{2 N}}{\rho^{2N-1}} \nonumber \\
 &\sim& \left(\frac{\Lambda}{\rho}\right)^{b}
 \frac{v^{2 N}}{\rho^{2N-1}} \nonumber \\
&\sim& \left(\frac{\Lambda}{\rho}\right)^{2N+1}
 \frac{v^{2 N}}{\rho^{2N-1}} ~.
\eeq
 The dimensional analysis works because the only other
scale in the problem is the instanton size $\rho$, and the integration
over $\rho$ is dominated by the region around
\beq
\rho^2 ={{b}\over{16\pi^2 v^2}}~.
\eeq
Forcing the quark legs to end at the same space time point generates the
$F$ component of $M$, and hence a vacuum energy of the right size.
From our previous arguments we recall that we can derive the ADS
superpotential
 for smaller values of $F$ from the case $F=N-1$, so in particular we
can derive gaugino condensation for zero flavors  from this instanton
calculation with $N-1$ flavors.

%%%%%%%%%%%%%%%%%%%%%%%%%%%%%%%%%%%%%%
\subsection{Generating $W_{\rm ADS}$ from Gaugino Condensation}
For $F<N-1$ instantons cannot generate $W_{\rm ADS}$ since at a generic
point in the classical moduli space
with ${\rm det}M \ne 0$  the $SU(N)$ gauge group breaks to 
$SU(N-F) \supset SU(2)$.  Matching the gauge coupling in the effective
theory at a generic point in the classical moduli space gives
\beq
 \Lambda^{3N-F} = \Lambda_{N-F,0}^{3(N-F)}{\rm det}M~.
\label{holo:matching}
\eeq
In the far infrared the effective theory splits into and $SU(N-F)$ gauge
theory and $F^2$ gauge singlets described by $M$.  However these
sectors can be coupled by irrelevant operators.  Indeed they must be, since
by themselves the $SU(N-F)$ gauginos have an anomalous $R$ symmetry.  The
$R$ symmetry of the underlying theory was spontaneously broken by squark
VEVs
but it should not be anomalous.  An analogous situation occurs in QCD where
the $SU(2)_L \times SU(2)_R$ chiral symmetry is spontaneously broken
and the axial anomaly of the quarks is reproduced in the low-energy
theory by an irrelevant operator (the Wess-Zumino term \cite{Wess:yu}) 
which manifests itself in 
the anomalous decay $\pi^0 \rightarrow \gamma \gamma$.  

In SUSY QCD the  correct term is
in fact present since 
the effective holomorphic coupling in the low-energy effective theory,
\beq
\tau = 
\frac{3(N-F)}{2 \pi i}\ln\left( {{\Lambda_{N-F,0}}\over{\mu}}\right) ~,
\label{holo:tau_effective}
\eeq
depends on $\ln {\rm det}M$ through
the matching condition (\ref{holo:matching}).  The relevant term 
in the low-energy Lagrangian is (factoring out $3(N-F)/(32 \pi^2)$)
\beq
&&\int d^2 \theta \ln {\rm det}M W^a W^a +h.c.\nonumber \\
&&= \left[ Tr(F_M M^{-1}) \lambda^a \lambda^a + 
{\rm Arg}({\rm det}M)
F^{a\mu\nu} {\wtilde F}^a_{\mu\nu} + \ldots\right] +h.c.
\eeq
where $F_M$ is the auxiliary field (i.e. the coefficient of $\theta^2$ in superspace
notation) for $M$.
The second term can be seen to arise through triangle diagrams involving
the fermions in the massive gauge supermultiplets.  
Note ${\rm Arg}({\rm det}M)$ transforms under a chiral rotation
in the correct manner to be the
Goldstone boson of the spontaneously broken $R$ symmetry:
\beq
{\rm Arg}({\rm det}M)\rightarrow {\rm Arg}({\rm det}M)+2 F \alpha~.
\eeq
The equation of motion for $F_M$ gives
\beq
F_M &=& \frac{\partial W}{\partial M} \nonumber \\
&=&M^{-1} \langle \lambda^a \lambda^a \rangle \nonumber \\ 
&\propto& M^{-1} \Lambda^3_{N-F,0}\nonumber \\ 
&\propto&  M^{-1} 
\left({{\Lambda^{3N-F}}\over{ {\rm det}M}}\right)^{{1}\over{N-F}}~,
\eeq
This gives a vacuum energy density that agrees with  the  
Affleck-Dine-Seiberg
calculation.
This potential energy implies that a non-trivial superpotential was 
generated for $M$, and since the only superpotential consistent with
holomorphy and symmetry is $W_{\rm ADS}$ we can conclude that for 
$F<N-1$ flavors gaugino condensation generates $W_{\rm ADS}$.

%%%%%%%%%%%%%%%%%%%%%%%%%%%%%%%%%%%
\subsection{Vacuum Structure}
Now that we believe $W_{\rm ADS}$ is correct what does it tells us about
the
vacuum structure of the theory?  It is easy to see that 
\beq
V&=& \sum_i  |\frac{\partial W}{\partial Q_i}|^2+ |\frac{\partial W}{\partial {\overline Q_i}}|^2 \nonumber\\
&=&  \sum_i   |F_i|^2 + |{\overline F_i}|^2~,
\eeq
is
minimized as ${\rm det}M \rightarrow \infty$, so there is a ``run-away
vacuum'', or more strictly speaking no vacuum. A loop-hole in this argument
would seem to be that we have not included wavefunction renormalization
effects, which could produce wiggles or even local minima in the potential,
but it could not produce new vacua unless the renormalization factors were
singular.  It is usually assumed that this cannot happen unless there 
are particles that become massless at some point in the field space, which
would also produce a singularity in the superpotential.  This
is just what happens at ${\rm det}M = 0$, 
where the massive gauge supermultiplets become massless. 
So we do not yet understand the
theory without VEVs.

%%%%%%%%%%%%%%%%%%%%%%%%%%%%%%%%%%%%%%%%%
%%%%%%%%%%%%%%%%%%%%%%%%%%%%%%%%%%%%%%%%%%%
\section{`t Hooft's Anomaly Matching}
\setcounter{equation}{0}
`t Hooft made one of the  most important advances in understanding
strongly coupled theories with composite degrees of freedom
by pointing out that the anomalies of the constituents and the
composites must match \cite{tHooft}.

Consider an asymptotically free gauge theory, with a global
symmetry group $G$. We can easily compute the anomaly for three
global $G$ currents in the ultraviolet by looking
at triangle diagrams of the fermions. We will call the result $A^{UV}$.
Now imagine that we weakly gauge $G$ with a gauge coupling
$g \ll 1$.  If $A^{UV}\ne 0$,
then we can add some spectators that only have $G$ gauge couplings, they
can be chosen such that their $G$ anomaly is $A^S=- A^{UV}$, so the
total $G$ anomaly vanishes.  Now construct the effective theory
at a scale less than the strong interaction scale. If we compute
the $G$ anomaly at this scale we add up the triangle diagrams of
light fermions, which consist of the spectators and strongly interacting
or composite fermions.  If $G$ is  not spontaneously
broken by the strong interactions
its anomaly must still vanish\footnote{If $G$ is spontaneously broken
the anomaly will still be reproduced by the interactions of the
Goldstone bosons, this is the origin of the Wess-Zumino term 
\cite{Wess:yu}.}, so 
\beq
0= A^{IR} + A^S~.
\eeq
Thus we have
\beq
A^{IR}=A^{UV}~.
\eeq
Taking $g \rightarrow 0$ decouples the weakly coupled gauge bosons
but does not change the three point functions of currents.

%%%%%%%%%%%%%%%%%%%%%%%%%%%%%%%%%%%%%
%%%%%%%%%%%%%%%%%%%%%%%%%%%%%%%%%%%%%
\section{Duality for SUSY QCD}
\setcounter{equation}{0}

Theoretical effort
in the mid 1990s (mainly due to Seiberg
\cite{Seiberg:1995ac,Seiberg:1994pq}) led
to a dramatic break-through in the understanding of strongly
coupled ${\mathcal N}=1$ SUSY gauge theories\footnote{For other review of these
developments see \cite{IntriligatorSeiberg,Peskin:1997qi,Shifman:1995ua}.}.  After this work we now
have a detailed understanding of the infrared (IR) behavior of 
many strongly-coupled theories, including  the phase structure
such theories.
  
The phase of a gauge theory can be understood by considering
the potential $V(R)$ between two static test charges a distance
$R$ apart\footnote{Holding the charges fixed for a time $T$ corresponds
to a Wilson loop of area $T \cdot R$.}. Up to an additive constant
we expect the functional form of the potential will fall into one
of the following categories:
\beq
\begin{array}{ll}
{\rm Coulomb}: & V(R) \sim \frac{1}{R} \\
{\rm free} \,\, {\rm electric}: & V(R) \sim \frac{1}{R\ln(R \Lambda)}\\
{\rm free} \,\, {\rm magnetic}: & V(R) \sim \frac{\ln(R \Lambda)}{R} \\
{\rm Higgs}:  & V(R) \sim {\rm constant}\\
{\rm confining}: & V(R) \sim \sigma R~.
\end{array}
\eeq

The explanation of these functional forms is as follows.
In a gauge theory where the coupling doesn't run (e.g.. at an IR fixed point
or in QED at energies below the electron mass)
then we expect to simply have a Coulomb potential. In a gauge theory where
the coupling runs to zero in the IR
(e.g.. QED with massless electrons) there is an inverse logarithmic correction to
squared gauge coupling and hence to the potential. Since
electric and magnetic charges are inversely related by the Dirac quantization
condition, the squared charge 
of a static magnetic monopole grows logarithmically in the IR due to the
renormalization by loops of massless electrons.
Using electric-magnetic duality to exchange electrons with monopoles one
finds that the logarithmic correction to the potential for static electrons
renormalized by massless monopole loops appears in the numerator since
the coupling grows in the IR.
In a Higgs phase the gauge bosons are massive
so there are no long range forces. In a confining phase\footnote{More precisely
a confining phase with area law confinement. Note however that
with dynamical quarks in the fundamental representation of the gauge group
we can produce quark anti-quark pairs which screen the test charge, so there
is no long range force, that is the flux tube breaks.} we expect a tube of confined
gauge flux between the charges, which, at large distances, acts like
a string with constant mass per unit length, and thus gives rise to a linear
potential. 

The familiar electric-magnetic duality exchanges electrons and magnetic
monopoles and hence the free electric phase with the free magnetic phase.
Mandelstam and `t Hooft \cite{'tHooftMandelstam} 
conjectured that duality could also exchange the Higgs
and confining phases and that confinement can be thought of as a dual
Meissner effect arising from a monopole condensate\footnote{Seiberg and
Witten later showed that this is actually the case in certain ${\mathcal N}=2$
SUSY gauge theories \cite{Seiberg:1994rs}.}. 
Electric-magnetic duality also exchanges an Abelian Coulomb phase with
another Abelian Coulomb phase. Seiberg 
\cite{IntriligatorSeiberg,Seiberg:1994pq} conjectured that analogs 
of the first and last of these dualities actually
occur in the IR of non-Abelian ${\mathcal N}=1$ SUSY gauge theories
and demonstrated that his conjecture satisfies many non-trivial 
consistency checks.

%%%%%%%%%%%%%%%%%%%%%%%%%%%%%%%%%%%%
\subsection{The Classical Moduli Space for $F \ge N$}
\label{dual:classmodFgeN}

Consider $SU(N)$ SUSY QCD with $F$ flavors and $F \ge N$.
This theory has a global $SU(F) \times SU(F) \times U(1) \times U(1)_R$
symmetry.  The quantum numbers\footnote{As usual only the $R$-charge of the 
squark is given, and $R[Q]=R[\Phi]-1$.}  of
the squarks and quarks are summarized in table \ref{dual:table:SUN_F>N}.
\beq
\begin{tabular}{c|c|cccc}
& $SU(N)$ & $SU(F)$  & $SU(F)$ & $U(1)$ & $U(1)_R$ \\
\hline
$\Phi$, $Q$ & \fund & \fund & ${\bf 1}$ & 1 & ${{F-N}\over{F}}$ 
$\vphantom{\raisebox{3pt}{\asymm}}$\\
$\overline{\Phi}$, $\overline{Q}$ & $\overline{\fund}$   & ${\bf 1}$  
& $\overline{\fund}$
& -1 & ${{F-N}\over{F}}$ $\vphantom{\raisebox{3pt}{\asymm}}$\\
\label{dual:table:SUN_F>N}
\end{tabular}
\eeq
The $SU(F) \times SU(F)$ global symmetry
is the analog of the $SU(3)_L \times SU(3)_R$ 
chiral symmetry of non-supersymmetric
QCD with 3 flavors, while the $U(1)$ is the 
analog\footnote{Up to a factor of $N$.} 
of baryon number since quarks (fermions in the fundamental representation
of the gauge group) and anti-quarks 
(fermions in the anti-fundamental representation
of the gauge group) have opposite charges. There is an additional
$U(1)_R$ relative to non-supersymmetric
QCD since in the supersymmetric theory there is also a gaugino.

Recall that the $D$-terms for this theory are given in terms of the
squarks by
\beq
D^a = g( \Phi^{*jn}(T^a)^m_n \Phi_{mj} - 
\overline{\Phi}^{jn}(T^a)^m_n \overline{\Phi}^*_{mj})~,
\eeq
where $j$ is a flavor index that runs from $1$ to $F$,
$m$ and $n$ are color indices that run from $1$ to $N$,
the index $a$ labels an element of the adjoint representation,
running from $1$ to $N^2-1$, and $T^a$ is a gauge group generator.
The  $D$-term potential is:
\beq
V= {{1}\over{2g^2}} D^a D^a ~,
\eeq
where we sum over the index $a$.
Define
\beq
D^n_m &\equiv &\langle \Phi^{*jn}\Phi_{mj}\rangle ~,\\
\overline{D}^n_m &\equiv & \langle \overline{\Phi}^{jn}\overline{\Phi}^*_{mj}\rangle~.
\eeq
$D^n_m$ and $\overline{D}^n_m$ are $N \times N$ positive semi-definite
Hermitian matrices.
In a SUSY vacuum state the vacuum energy vanishes and we must have:
\beq
D^a \equiv g T^{a m}_n(D^n_m -\overline{D}^n_m)=0~.
\eeq
Since $T^a$ is a complete basis for traceless matrices, we must have that the second
matrix is proportional to the identity:
\beq
D^n_m -\overline{D}^n_m= \rho I~.
\label{dual:Dident}
\eeq
$D^n_m$ can be diagonalized by an $SU(N)$ gauge transformation
\beq
D^\prime=U^\dagger D U~,
\eeq
so we can take $D^n_m$ to have the form:
\beq
D = \left( \begin{array}{cccc}
|v_1|^2 & & & \\
& |v_2|^2 & &  \\
& & \ddots &  \\
& & & |v_N|^2 
\end{array} \right)~.
\eeq
In this basis, because of Eq. (\ref{dual:Dident}),
$\overline{D}^n_m$ must also be diagonal, with eigenvalues  
$|{\overline v}_i|^2$.
This tells us that
\beq
|v_i|^2=|{\overline v}_i|^2 +\rho~.
\eeq
Since $D^n_m$ and $\overline{D}^n_m$ are invariant under flavor
transformations,
we can use $SU(F) \times SU(F)$ flavor transformations to put 
$\langle \Phi \rangle$ and $\langle \overline{\Phi}\rangle$ in the form
\beq
\langle \Phi \rangle = \left( \begin{array}{cccccc}
v_1 & & & 0 & \ldots & 0\\
& \ddots & &\vdots & & \vdots  \\
& & v_N & 0   & \ldots & 0\\
\end{array} \right)~,
\eeq
\beq
\langle \overline{\Phi}\rangle =
\left( \begin{array}{ccc}
{\overline v}_1 & & \\
& \ddots &  \\
& & {\overline v}_N \\
0 & \ldots & 0 \\
\vdots  & & \vdots \\
0 & \ldots & 0 \\
\end{array} \right)~.
\eeq
Thus we have a space of degenerate vacua,
which is referred to as a moduli space of vacua.  The vacua are
physically distinct since,  for example, different values of the VEVs 
correspond to 
different masses for the gauge bosons.

With a VEV for a single flavor turned on we break the gauge
symmetry down to $SU(N-1)$.
At a generic point in the moduli space the $SU(N)$
gauge symmetry is broken completely and there are
$2NF -(N^2-1)$ massless chiral supermultiplets left over.
We can describe these light degrees of freedom in 
a gauge invariant way by scalar ``meson'' and ``baryon''
fields and their superpartners:
\beq
M^j_i &=&  \overline{\Phi}^{j n}\Phi_{n i}~,\\
B_{i_1,\ldots,i_N} &=&  \Phi_{n_1 i_1}\ldots\Phi_{n_N i_N}
\epsilon^{n_1,\ldots,n_N}~,\\
{\overline B}^{\,i_1,\ldots,i_N} &=&  {\overline \Phi}^{\,n_1 i_1}\ldots
{\overline\Phi}^{\,n_N i_N}
\epsilon_{n_1,\ldots,n_N}~.
\eeq
The fermion partners of these fields are 
the corresponding products of scalars and one fermion.
There are constraints relating the fields $M$ and $B$,  since the $M$ has $F^2$ components,  
$B$ and ${\overline B}$ each
have  $\left(\begin{array}{c}F\\N\end{array}\right)$
components, and all three are
constructed out of the same $2NF$ underlying squark fields.  For
example, at the classical level, there is a relationship between the
product of the $B$ and ${\overline B}$ eigenvalues and the product
of the non-zero eigenvalues of $M$:
\beq
 B_{i_1,\ldots,i_N} 
{\overline B}^{\,j_1,\ldots,j_N}= M^{j_1}_{[i_1}\ldots M^{j_N}_{i_N]}~,
\eeq
where $[\,]$ denotes antisymmetrization.

Up to flavor
transformations
the moduli can be written as:
\beq
\langle M\rangle &=&
\left( \begin{array}{cccccc}
 v_1{\overline v}_1 & & & & &\\
& \ddots &  & & &   \\
& & v_N {\overline v}_N &  & & \\
& & & 0 & &\\
& & & & \ddots & \\
& & & & & 0
\end{array} \right) ~,\\
\langle B_{1,\ldots,N} \rangle &=&  v_1\ldots v_N~,\\
\langle {\overline B}^{1,\ldots,N}\rangle &=&  {\overline v}_1 \ldots
{\overline v}_N~,
\eeq
with all other components set to zero.
We also see that the rank of $M$ is at most $N$.  
If it is less than $N$, then
$B$ or ${\overline B}$ (or both) vanish. If the rank of $M$ is $k$,
then $SU(N)$ is broken to $SU(N-k)$ with $F-k$ massless flavors.

%%%%%%%%%%%%%%%%%%%%%%%%%%%%%%%%
\subsection{The Quantum  Moduli Space for $F \ge N$}
Recall that the ADS superpotential made no sense for
$F \ge N$ however the vacuum solution
\beq
M^j_i 
 =(m^{-1})^j_i \left({\rm det}m \Lambda^{3N-F}\right)^{{1}\over{N}}~,
\label{dual:vevs}
\eeq
is still sensible.  Giving large masses, $m_H$, to flavors $N$ through $F$
and matching the gauge coupling at the mass thresholds gives
\beq
 \Lambda^{3N-F}{\rm det}m_H = \Lambda_{N,N-1}^{2N+1}~.
\eeq
The low-energy effective theory has $N-1$ flavors and an ADS
superpotential.
If we give small masses, $m_L$, to the light flavors we  have
\beq
M^j_i 
& =&(m_L^{-1})^j_i \left({\rm det}m_L
\Lambda_{N,N-1}^{2N+1}\right)^{{1}\over{N}} \nonumber\\
& =&(m_L^{-1})^j_i \left({\rm det}m_L {\rm det}m_H \Lambda^{3N-F}
\right)^{{1}\over{N}}~.
\eeq
Since the masses are holomorphic parameters of the theory, this
relationship can only break down at isolated singular
points, so equation (\ref{dual:vevs})
is true for generic masses and VEVs. For $F \ge N$ we can take 
$m^i_j\rightarrow 0$ with components of $M$ finite or zero. So
the vacuum degeneracy is not lifted and there is
a quantum moduli space \cite{Seiberg:1994bz}    for $F \ge N$, however the classical constraints
between $M$, $B$ and ${\overline B}$ may be modified.
  Thus we can parameterize the quantum moduli space\footnote{There
is a theorem which shows that in general a moduli space
is parameterized by the gauge invariant operators, see \cite{Luty:1995sd}.} 
by
$M$, $B$ and ${\overline B}$. When these fields have large values (with maximal
rank) then the squark VEVs are large (compared to $\Lambda$)
and the gauge theory is broken in
the perturbative regime. As the meson and baryon fields approach zero (the origin of moduli 
space) then the gauge couplings become stronger, and at the origin we
would naively 
expect a singularity since the gluons are becoming massless at this
point. We shall  see that this expectation is too naive.

%%%%%%%%%%%%%%%%%%%%%%%%%%%%%%%%%%%%%%
\subsection{Infrared Fixed Points}

For $F \ge 3N$ we lose asymptotic freedom, so the theory can
be understood as a weakly coupled low-energy effective theory.
For $F$ just below $3N$ we have an infrared fixed point.
This was pointed out by Banks and Zaks \cite{Banks:nn} as a
general property of gauge theories. By considering the large $N$
limit of $SU(N)$ with $F/N$ infinitesimally below the point where
asymptotic freedom is lost they showed that the $\beta$ function
has a perturbative fixed point.  Here we will apply their argument
to SUSY QCD.
Novikov, Shifman, Vainshtein and Zakharov
\cite{Novikov:uc} have shown that the exact $\beta$ function
for the running canonical\footnote{For 
discussions of how the holomorphic gauge coupling is related to
the canonical gauge coupling
see \cite{Shifman:1986zi,Shifman:1991dz,Dine:1994su,Arkani-Hamed:1997mj}.} 
gauge coupling is given by
\beq
\beta(g) &=& - {{g^3}\over{16 \pi^2}} 
{{\left(3 N -F(1-\gamma)\right)}\over{1-N{{g^2}\over{8\pi^2}} }}~.
\label{dual:exactbeta}
\eeq
where $\gamma$ is the anomalous dimension of the quark mass term.
In perturbation theory one finds
\beq
\gamma &=& - {{g^3}\over{8 \pi^2}} {{N^2-1}\over{N}} + {\mathcal O}(g^4)~.
\eeq
So
\beq
16 \pi^2\beta(g) &=& - g^3
\left(3 N -F\right)
- {{g^5}\over{8 \pi^2}}\left(3N^2 -FN -F {{N^2-1}\over{N}}\right)\nonumber\\
&& + {\mathcal O}(g^7)~.
\eeq
Now take the number of flavors to be infinitesimally close
to the point where asymptotic freedom is lost.  For $F=3N- \epsilon N$
we have
\beq
16 \pi^2\beta(g) &=& - g^3\epsilon N
- {{g^5}\over{8 \pi^2}}\left(3(N^2-1)  + {\mathcal O}(\epsilon)\right)\nonumber\\
&& + {\mathcal O}(g^7)~.
\eeq
So there is an approximate solution 
of the condition $\beta=0$ where there first two terms cancel.  This
solution
corresponds to a perturbative infrared (IR) fixed point at 
\beq
g_*^2 = {{8 \pi^2}\over{3}}  {{N}\over{N^2-1}}\, \epsilon~,
\eeq
and we can safely neglect the  ${\mathcal O}(g^7)$ terms since they are higher
order in $\epsilon$.

Without any masses this fixed point gauge theory
theory is scale invariant when the coupling is set to the fixed point value
$g=g_*$.
A general result of field theory 
is that a scale invariant theory of fields
with spin $\le 1$ is actually conformally invariant \cite{Callan:ze}.  In a
conformal SUSY theory the SUSY algebra is extended to a superconformal
algebra. A particular $R$-charge enters the superconformal
algebra in an important way, we will refer to this superconformal $R$-charge
as $R_{\rm sc}$.
Mack found \cite{Mack} that in the case of conformal symmetry there are lower bounds
on the dimensions of gauge invariant fields.
In the superconformal case\footnote{A brief review is given in the next section.}
it was shown for ${\cal N}=1$ by Flato and Fronsdal \cite{Flato:1983te} (and for general ${\cal N}$ by Dobrev and Petkova \cite{Dobrev:qv})
that the dimensions of the scalar component of gauge invariant chiral and anti-chiral superfields
are given by
\beq
d &=& {{3}\over{2}} R_{\rm sc}, \,\,{\rm for}\,{\rm chiral}\,
\,{\rm super}\,{\rm fields}~,\\
d &=& -{{3}\over{2}} R_{\rm sc}, \,\,{\rm for}\,{\rm anti}-{\rm chiral}\,
\,{\rm super}\,{\rm fields}~.
\eeq
Since the charge of a product of fields is the sum of the 
individual charges,
\beq
R_{\rm sc}[{\mathcal O}_1 {\mathcal O}_2] = R_{\rm sc}[{\mathcal O}_1] + 
R_{\rm sc}[{\mathcal O}_2]~,
\eeq
we have the result that for chiral superfields dimensions simply add:
\beq
D[{\mathcal O}_1 {\mathcal O}_2] = D[{\mathcal O}_1] + 
D[{\mathcal O}_2]~.
\eeq
This is a highly non-trivial statement since in general the dimension
of a product of fields is affected by renormalizations that are 
independent of the renormalizations of the individual fields.
In general the $R$ symmetry of a SUSY gauge theory
seems ambiguous\footnote{A general procedure for determining
the superconformal $R$ symmetry was given by Intriligator and Wecht \cite{Intriligator:2003jj}}, since we can always form linear combinations
of any $U(1)_R$ with other $U(1)$'s, but for the fixed point
of SUSY QCD $R_{\rm sc}$ is 
unique since we must have
\beq
R_{\rm sc}[Q] = R_{\rm sc}[{\overline Q}] ~.
\eeq
Thus we can identify the $R$ charge we have been using (as given  in table \ref{dual:table:SUN_F>N}) with  $R_{\rm sc}$.
If we denote the anomalous dimension of the squarks at the fixed point 
by $\gamma_*$ then the dimension of the meson field at the IR fixed point is
\beq
D[M]=D[\Phi {\overline \Phi}] = 
2 + \gamma_* &=& {{3}\over{2}} 2{{(F-N)}\over{F}}
\nonumber
\\
&=&3 - {{3N}\over{F}}~,
\eeq
and the anomalous dimension of the mass operator at the fixed point is
\beq
\gamma_*= 1- \frac{3N}{F}~.
\eeq
We can also check that the exact $\beta$ function (\ref{dual:exactbeta})
vanishes:
\beq
\beta \propto 3N-F(1- \gamma_*) = 0~.
\eeq
For a scalar field in a conformal theory we also have \cite{Mack}
\beq
D(\phi) \ge 1~,
\eeq
with equality holding for a free field.
Requiring that $D[M]\ge 1$ implies 
\beq
F\ge  {{3}\over{2}} N~.
\eeq
Thus the IR fixed point is an interacting conformal theory for
$\frac{3}{2} N < F < 3 N$.
Such conformal theories have no particle interpretation,
but anomalous dimensions are physical quantities.

%%%%%%%%%%%%%%%%%%%%%%%%%%%%%%%%%%%%%%%%
\subsection{An Aside on Superconformal Symmetry}
The reader may skip this section on her first time through, and return to it later
if she remains curious about how the results quoted above relating scaling
dimensions to $R$-charges were obtained. I will loosely follow the discussion
of ref. \cite{Minwalla:1997ka}.

The generators of the conformal group 
%acting on an operator in Lorentz representation $r$ and scaling dimension $d$ 
can be represented by
\beq
M_{\mu\nu} &=& -i(x_{\mu}\partial_{\nu} -x_{\nu}\partial_{\mu}) %-M^r_{\mu\nu} 
\nonumber \\
P_{\mu} &=& -i\partial_{\mu}\nonumber \\
K_{\mu} &=& -i(x^2\partial_{\mu}-2x_{\mu}x_\alpha\partial^\alpha) 
% -2 x^\alpha M_{\alpha \mu}+2 x_\mu d 
\nonumber \\
D &=& i x_\alpha\partial^\alpha %+d 
~,
\eeq
where $M_{\mu\nu}$  are the ordinary Lorentz rotations/boosts, $P_{\mu}$ are the translation generators,  $K_{\mu}$ are referred to as the ``special" conformal generators, and $D$ is 
the dilation generator. ($D$ is also know as the ``dilatation" generator by those who like extra syllables.)   In $4$ spacetime dimensions, the conformal group is isomorphic to $SO(4,2)$
We would like to find the restrictions that can be placed on conformal field theory operators
by constructing the unitary irreducible representations of the conformal group.  The simplest
method to perform
this construction is not with the generators defined above but by
a related set of generators that correspond to ``radial quantization" in Euclidean space.  These generators are defined with indices that runs from 1 to 4:
\beq
M^\prime_{ij} &=& M_{jk}\nonumber\\
M^\prime_{j4} &=& \frac{1}{2}(P_j-K_j)\nonumber\\
D^\prime &=& -\frac{i}{2}(P_0+K_0)\nonumber\\
P^\prime_j &=& \frac{1}{2}(P_j+K_j)+iM^\prime_{j0} \nonumber\\
P^\prime_4 &=& -D -\frac{i}{2}(P_0-K_0) \nonumber\\
K^\prime_j &=& \frac{1}{2}(P_j+K_j)-iM^\prime_{j0} \nonumber\\
K^\prime_4 &=& -D +\frac{i}{2}(P_0-K_0)~,
\eeq
where $j=1,2,3$ and $D^\prime$ acts as the ``Hamiltonian" of the ``radial quantization".  The eigenstates of the ``radial quantization" are in one to one correspondence with the operators of the conformal field theory.
From these definitions we see that
\beq
D^{\prime \dagger} &=& -D^\prime \nonumber\\
P^{\prime \dagger}_i &=& K^{\prime}_j ~.
\eeq

An $SO(4,2)$ representation can be specified by its decomposition into
irreducible representations of its maximal compact subgroup which is 
$SO(4) \times SO(2)$. The $SO(4) \approx SU(2) \times  SU(2)$ representations are just the usual representations of the Lorentz group which can be specified by two half-integers $(j,\tilde j)$ (Lorentz spins):
\beq
(scalar)&=&(0,0) \nonumber\\
(spinor) &=& \left(\frac{1}{2},0\right) {\rm or}  \left(0,\frac{1}{2}\right) \\
(vector) &=& \left(\frac{1}{2},\frac{1}{2}\right)~. \nonumber
\eeq
The $SO(2)$ representation is labeled by the eigenvalue of $D'$ acting on the
state (operator). I will denote this eigenvalue  by $-i d$ where $d$  is the scaling dimension of the operator. To complete the construction we
need raising and lowering operators for the $SO(2)$ group.  
We can choose a basis such that $P^\prime$ is a raising operator and therefore
$K^\prime$ is a lowering operator.  Then we can classify multiplets by the scaling dimension of
the lowest weight state (the state annihilated by $K^\prime$). The operator corresponding to the lowest weight state is also known as the primary operator. One can check (using the commutator
$[D^\prime, P^\prime_\mu]= - i P^\prime_\mu$)
that acting on the lowest weight operator with the raising operator $P^\prime$, gives a new operator (sometimes called a descendant operator) with scaling dimension $d+1$.

Unitarity requires that any linear combination of states have positive norm, so in particular
the state
\beq
a \, P^\prime_{m}| d, (j_1, \tilde j_1)\rangle +b \, P^\prime_{n}| d, (j_2, \tilde j_2)\rangle ~,
\eeq
(with $m \ne n$) must have positive norm.
Using the commutator
\beq
 [P^\prime_m,K^\prime_n] = -i (2\delta_{m n}D^\prime + 2M^\prime_{m n})~,
\eeq
we find
\beq
(|a|^2+|b|^2) d + 
2 Re( a^* b \langle d,( j_1, \tilde j_1) |  iM^\prime_{m n}  | d, ( j_2, \tilde j_2) \rangle \ge 0~.
\eeq
Given that $ iM^\prime_{m n} $ has real eigenvalues and
restricting to $|a|=|b|$ this condition can be written as
\beq
 d  
\ge \pm \langle d, (j_1, \tilde j_1) |  iM^\prime_{m n}  | d, (j_2, \tilde j_2) \rangle ~.
\label{pmbound}
\eeq
To find the eigenvalues of  $ iM^\prime_{m n} $ write (for fixed $m$, $n$)
\beq
i M^\prime_{m n} = \frac{i}{2} (\delta_{m \alpha}\delta_{n \beta}- 
\delta_{m \beta}\delta_{n \alpha}) M^\prime_{\alpha\beta} ~,
\eeq
which can be re-written as
\beq
M^\prime_{m n} = (V . M^\prime)_{m n} ~,
\eeq
where $V$ is the generator of $SO(4)$ rotations in the vector representation:
\beq
V_{\alpha\beta m n} = i (\delta_{m \alpha}\delta_{n \beta}- 
\delta_{m \beta}\delta_{n \alpha})~,
\eeq
and 
\beq
A . B \equiv \frac{1}{2} A_{\alpha\beta} B_{\alpha\beta}~.
\eeq
By a similarity transformation we can go to a Clebsch-Gordan basis
where $(V+M^\prime)$, $V$, and $M^\prime$ are ``simultaneous (commuting)  quantum numbers"
and use the relation\footnote{Readers of a certain age will recognize this technique from the calculation of the most-attractive channel in single gauge boson exchange {\protect\cite{MAC}}, while all readers should recognize the calculation of the spin-orbit term.} 
\beq
V . M = \frac{1}{2}\left[ (V+M^\prime)^2- V^2 - M^{\prime 2} \right]
\eeq
From this one can show that that the most negative eigenvalue of $V . M$ has a larger magnitude
than the most positive eigenvalue, so the most restrictive bound from the inequality
(\ref{pmbound}) is the case with the $-$ sign.
If our state $| d, j_2, \tilde j_2 \rangle$ corresponds to the representation ${\bf r}$ of $SO(4)$, and
${\bf r^\prime}$ is the representation with the smallest quadratic Casimir in the product ${\bf r} \times V$
then we have
\beq
d \ge \frac{1}{2}[ C_2({\bf r}) + C_2(V)- C_2({\bf r^\prime}) ]
\eeq
where
\beq
C_2(V) \equiv V.V
\eeq
and so on.
Using the fact that the $SO(4)$ Casimir is twice the sum of the $SU(2)$ Casimirs
(i.e. $C_2(j,\tilde j) = 2(J^2+\tilde J^2)=2(j(j+1)+\tilde j(\tilde j+1))$), one can check that 
\beq
C_2(scalar)&=&0 \nonumber\\
C_2(spinor) &=& \frac{3}{2} \\
C_2(vector) &=& 3 \nonumber~,
\eeq
and thus:
\beq
d&\ge&0~, \,\,\, (scalar)\nonumber\\
d &\ge& \frac{3}{2}~,\,\,\,(spinor)  \\
d&\ge& 3~, \,\,\,(vector)~.  \nonumber
\eeq
The first two bounds are perfectly reasonable since the identity operator is a scalar and has dimension $0$, and a free (massless) spinor has dimension $3/2$. The third bound may be a little surprising since a free massless vector (gauge) boson field has dimension 1, however such a field
is not gauge invariant, and thus unitarity cannot be applied. A conserved current is a gauge-invariant vector field and does have dimension 3.

Applying similar arguments to states with $P_{i}^\prime P_{k}^\prime$ acting on them,
one finds for scalars
\beq
d(d-1) \ge 0~,
\eeq
which means that for scalar operators with $d > 0$ (i.e. operators other than the identity)
\beq
d \ge 1~.
\eeq

Turning to superconformal symmetry, first note that in addition to the usual (Euclidian) supersymmetry generators $Q_{i \alpha}^\prime$ (where $\alpha$ is a spinor index and $i$ runs from 1 to ${\mathcal N}$) there is also a superconformal generator $S_{j \beta}^\prime$
such that
\beq
Q^{\prime \dagger}=S^\prime ~.
\eeq
$Q^\prime$ and $S^\prime$ can be chosen to be real Majorana (four component) spinors.There is also an $R$-symmetry which is $U(1)$ for ${\mathcal N}=1$, $U(2)$ for  ${\mathcal N}=2$,
and $SU(4)$ for  ${\mathcal N}=4$. 
In general we can take the matrix 
\beq
(T_{ij})_{pq}=\delta_{ip}\delta_{jq}~,
\eeq
as a generator of the full $R$-symmetry, and $R$ as the generator of
the Abelian $R$-symmetry (i.e. for  ${\mathcal N}=1$ we have  $R \equiv T_{11}$; for  ${\mathcal N}=2$,
$R \equiv \sum T_{ij}$; while for ${\mathcal N}=4$ set $R=0$).  
 
The $R$-symmetry does not commute with the supersymmetry generators, to see this explicitly
we need a few more definitions.
Define the Euclidian $\Gamma$ matrices in terms of the Lorentzian $\gamma$ matrices by
\beq
\Gamma_i &=& \gamma_i ~,\\
\Gamma_4 &=& -i \gamma_0~,
\eeq
and choose a basis where $\Gamma_a$ is real and hermitian
Then we can define left-handed and right-handed projection operators by
\beq
P_-&=& \frac{1}{2}(1-\gamma_5) ~,\\
P_+&=& \frac{1}{2}(1+\gamma_5)~.
\eeq
%By convention states that are projected to zero by $P_-$ are called chiral, while
%those that are projected to zero by $P_+$ are called anti-chiral.
The convention is that $(j,\tilde j)=(1/2,0)$ corresponds to the fermion component of chiral supermultiplet which is projected to zero by 
$P_{-}$.

Finally we can then write
\beq
[T_{ij}, Q^\prime_{m}] &=& P_{+}\, Q^\prime_i \delta_{jm} - P_{-}\,  Q^\prime_j \delta_{jm} 
\eeq
\beq
[T_{ij}, S^\prime_{m}] &=& P_{+}\, S^\prime_i \delta_{jm} - P_{-}\,  S^\prime_j \delta_{jm} 
\eeq
Most importantly for the unitarity bounds, the anticommutator
of $Q^\prime$ and $S^\prime$ is:
\beq
\{ Q_{i\alpha}^\prime, S_{j\beta}^\prime\}&=&i \frac{\delta_{ij}} { 2}
[(M^\prime_{m n}\Gamma_{m}\Gamma_{n} )_{\alpha\beta} + 2\delta_{\alpha\beta}D^\prime] 
\nonumber \\  && - 2(P_{+})_{\alpha\beta}T_{ij}
  +2 (P_{-})_{\alpha\beta}T_{ji} \  +\frac{\delta_{ij}}{2}(\gamma_5)_{\alpha\beta}R~.
\eeq
 
Now apply the requirement of unitarity to the state
\beq
a \, Q^\prime_{i\alpha}| d,R, (j_1, \tilde j_1)\rangle +
b \, Q^\prime_{j\beta}| d,R, (j_2, \tilde j_2)\rangle ~,
\eeq
(with $\alpha \ne \beta$, and for simplicity I will only consider ${\mathcal N}=1$).  The result is
 \beq
 d  
\ge \pm \langle d, R, (j_1, \tilde j_1) | \frac{i}{2} (M^\prime_{m n}\Gamma_{m}\Gamma_{n} )_{\alpha\beta} -\frac{3}{2} (\gamma_5)_{\alpha\beta}R  | d, R,(j_2, \tilde j_2) \rangle ~.
\label{scpmbound}
\eeq
The operator inside the matrix element can be split into left and right-handed parts
(i.e. proportional to $P_+$ and $P_-$):
 \beq
P_{+}(4 J . S -\frac{3}{2} R )+P_{-}(4 \tilde J . \tilde S +\frac{3}{2} R )~,
\label{LR}
\eeq
 where $S$ is the rotation generator in
 the spin-half representation of the first $SU(2)$ embedded in $SO(4)$, and $\tilde S$ is the corresponding generator of the second $SU(2)$.  The eigenvalues of $2 J . S$ are $-j-1$ and $j$ for $j >0$ and $0$ for $j=0$, so, as before, the most negative eigenvalues provide the strongest bounds.
\beq
 d  \ge P_{+}(2j+2 -2 \delta_{j0}+\frac{3}{2} R )+P_{-}(2 \tilde j+2-2 \delta_{\tilde j 0} -\frac{3}{2} R )~.
\label{nescbound}
\eeq
Although $P_{+}$ and $P_{-}$ were defined in the superconformal algebra to act on spinors,
they have been implicitly extended to act as projection operators on superpartners of spinors in the same way they act on the spinors themselves. 
 
The bound we have found above is not the whole story.  The bound (\ref{nescbound}) is a necessary condition \cite{Minwalla:1997ka} but if either $j$ or $\tilde j$ are zero there are more restrictive
conditions \cite{Flato:1983te,Dobrev:qv} that require:
\beq
d&\ge&d_{max}={\rm max}\left( 2 j +2 + \frac{3}{2}R\,,\, 
2 \tilde j + 2 -     \frac{3}{2}R\right) \ge 2+ j+\tilde j \nonumber\\
{\rm or} \,\,\,d&=&\frac{3}{2}|R|~.
\eeq
Here $d_{max}$ is the boundary of the continuous range, 
while the isolated point
(which happen only in the supersymmetric case
\cite{Flato:1983te,Dobrev:qv}) 
is achieved  for chiral and anti-chiral superfields.   We can check the result (\ref{nescbound}) by considering a chiral supermultiplet with $R$-charge $R_{sc}$.
 The primary field is the lowest component of the supermultiplet, which is just the scalar with $(j, \tilde j)=(0,0)$. Equality is achieved in the necessary condition (\ref{nescbound}) on the scaling dimension of the scalar component at:
 \beq
 d =  \frac{3}{2} R_{sc}~,
 \eeq
 in agreement with the results of \cite{Flato:1983te,Dobrev:qv}.
It is this special case which is of interest for the chiral supermultiplet. The reason is that superconformal multiplets are generally much larger than
 supermultiplets (due to the existence of the $S^\prime$ generator), but the superconformal multiplets can degenerate (``shorten" in the
 language of ref. \cite{Freedman:na}) precisely at the point where some of the states have
 zero norm, which is the point where the bound is saturated.  In other words, in a 
 superconformal theory, chiral supermultiplets are ``short" multiplets.

 %%%%%%%%%%%%%%%%%%%%%%%%%%%%%%%%%%%%%%%%%%%%
\subsection{Duality}

In a conformal theory (even if it is strongly coupled) we 
don't expect any global symmetries to break, so `t Hooft 
anomaly matching should apply to any description of 
the low-energy degrees of freedom.
The anomalies of the mesons and baryons described above do not match to
those of the quarks and gaugino. However Seiberg \cite{Seiberg:1994pq}
found a non-trivial
solution to the anomaly matching using a ``dual'' $SU(F-N)$ gauge
theory with a ``dual'' gaugino, ``dual'' quarks
and a gauge singlet ``dual mesino'' with the following quantum 
numbers\footnote{As usual only the $R$-charge of the 
scalar component is given, and $R[{\rm fermion}]=R[{\rm scalar}]-1$.}: 
\beq
\begin{tabular}{c|c|cccc}
& $SU(F-N)$ & $SU(F)$  & $SU(F)$ & $U(1)$ & $U(1)_R$ \\
\hline
$q$ & \fund & $\overline{\fund}$ & {\bf 1} &${{N}\over{F-N}}$ 
& ${{N}\over{F}}$ 
$\vphantom{\raisebox{3pt}{\asymm}}$\\
$\overline{q}$ & $\overline{\fund}$   & {\bf 1} & \fund
& $-{{N}\over{F-N}}$ & ${{N}\over{F}}$ 
$\vphantom{\raisebox{3pt}{\asymm}}$\\
${\rm mesino}$ &{\bf 1}& \fund & $\overline{\fund}$   & 0 & 
$2\,{{F-N}\over{F}}$ 
$\vphantom{\raisebox{3pt}{\asymm}}$\\
\end{tabular}
\eeq
In the language of duality the dual quarks can
be thought of as ``magnetic'' quarks, in analogy with the duality between
electrons and magnetic monopoles.

The anomalies of the two dual theories match as follows:
\beq
&&SU(F)^3 : -(F-N)+F =N \nonumber \\
&&U(1) SU(F)^2 : {{N}\over{F-N}} (F-N) {{1}\over{2}} = {{N}\over{2}} \nonumber \\
&&U(1)_R SU(F)^2 : {{N-F}\over{F}}(F-N) {{1}\over{2}} 
+{{F-2N}\over{F}} F{{1}\over{2}}= -{{N^2}\over{2 F}}\nonumber \\
&&U(1)^3 :0\nonumber \\
&&U(1) :0\nonumber \\
&&U(1) U(1)_R^2: 0 \nonumber \\
&&U(1)_R :\left({{N-F}\over{F}}\right) 2 (F-N) F +
\left({{F-2N}\over{F}}\right) F^2 +(F-N)^2-1\nonumber\\
&&\,\,\,\,\,\,\,\,\,\,\,\,\,\,\,\,\,\,\,\,\,\,= -N^2-1
\nonumber \\
&&U(1)_R^3 :\left({{N-F}\over{F}}\right)^3 2 (F-N) F +
\left({{F-2N}\over{F}}\right)^3 F^2 +(F-N)^2-1\nonumber\\
&&\,\,\,\,\,\,\,\,\,\,\,\,\,\,\,\,\,\,\,\,\,\,
= -{{2N^4}\over{F^2}}+N^2-1
\nonumber \\
&&U(1)^2 U(1)_R:\left({{N}\over{F-N}}\right)^2{{N-F}\over{F}}2 F (F-N) = -2N^2~.
\eeq

This theory admits a unique superpotential:
\beq
W=\lambda {\wtilde M}^j_i \phi_j {\overline \phi}^i~,
\eeq
where $\phi$ represents
the ``dual'' squark (that is the scalar
superpartner of the ``dual'' quark $q$) and  
${\wtilde M}$ is the dual meson (the scalar
superpartner of the ``dual'' mesino).
This ensures that the two theories have the same number of degrees of
freedom
since the ${\wtilde M}$ equation of motion removes the color singlet
$\phi {\overline \phi}$
degrees of freedom. The counting works because both theories
have the same number of massless chiral superfields at a generic
point in moduli space:
\beq
2FN-2(N^2-1) = 2 F{\wtilde N} - 2({\wtilde
N}^2-1)~,
\eeq
where ${\wtilde N}=F-N$ is the number of colors in the dual theory.

The dual theory also has baryon operators:
\beq
b^{i_1,\ldots,i_{F-N}} &=&  \phi^{n_1 i_1}\ldots\phi^{n_{F-N} i_{F-N}}
\epsilon_{n_1,\ldots,n_{F-N}}\\
{\overline b}_{\,i_1,\ldots,i_{F-N}} &=&  {\overline \phi}_{\,n_1 i_1}\ldots
{\overline\phi}_{\,n_{F-N} i_{F-N}}
\epsilon^{n_1,\ldots,n_{F-N}}~.
\eeq
 Thus the two moduli spaces have a  simple mapping
\beq
&&M \leftrightarrow {\wtilde M} \nonumber \\
&&B_{i_1, \ldots, i_N} \leftrightarrow \epsilon_{i_1,\ldots,i_N,j_1,\ldots
j_{F-N}}
b^{j_1,\ldots,j_{F_N}} \nonumber \\
&&{\overline B}^{i_1, \ldots, i_N} \leftrightarrow 
\epsilon^{i_1,\ldots,i_N,j_1,\ldots
j_{F-N}}
{\overline b}_{j_1,\ldots,j_{F_N}}~.
\label{dual:mapping}
\eeq

The one-loop $\beta$ function in the dual theory is
\beq
\beta(\wtilde g) \propto - {\wtilde g}^3 (3 {\wtilde N}-F)  
= -{\wtilde g}^3 (2F-3N)~.
\eeq
So the dual theory loses asymptotic freedom when $F\le 3N/2$.
In other words, the dual theory leaves the conformal regime to become
infrared free at exactly the point where the meson of the
original theory becomes a free field which implies very strong
coupling for the underlying quarks and squarks.

We can also apply the Banks-Zaks \cite{Banks:nn} argument to the dual
theory \cite{deGouvea}. When 
\beq
F&=&3 {\wtilde N} - \epsilon {\wtilde N}\nonumber \\
&=&\frac{3}{2}\left(1+\frac{\epsilon}{6}\right)N~,
\eeq
there is a perturbative fixed point at
\beq
{\wtilde g}_*^2 &=& {{8 \pi^2}\over{3}} {{\wtilde N}\over{{\wtilde N}^2-1}} 
\left( 1 +{{F}\over{\wtilde N}}\right) \epsilon \\
\lambda_*^2 &=& {{16 \pi^2}\over{3 {\wtilde N}}} \epsilon~.
\label{dual:dualfixedpoint}
\eeq

At this fixed point $D({\wtilde M} {\overline \phi} \phi)=3$, so the
superpotential term is marginal.
If the superpotential coupling
 $\lambda=0$, then  $\wtilde M$ has no interactions (it is a free field)
and therefore its dimension is 1.
If the dual gauge coupling is set close to the Banks-Zaks fixed point
and $\lambda \approx 0$ then 
we can calculate the dimension of $\phi {\overline \phi}$ from
the $R_{\rm sc}$ charge for $F>3N/2$:
\beq
D(\phi {\overline \phi}) = {{3(F-{\wtilde N})}\over{F}} ={{3 N}\over{F}} <2~.
\eeq
So the superpotential is a relevant operator (not marginal) and thus there
is an unstable fixed point at
\beq
{\wtilde g}^2 &=&{\wtilde g}_*^2= 
{{8 \pi^2}\over{3}} {{\wtilde N}\over{{\wtilde N}^2-1}}\, 
 \epsilon \\
\lambda^2 &=& 0 ~.
\eeq
In other words the pure Banks-Zaks fixed point in the dual theory is unstable
and the superpotential coupling flows toward the non-zero value given in
(\ref{dual:dualfixedpoint}). 

So we have found that not only does
SUSY QCD have an interacting IR fixed point for
$3N/2 < F < 3N$ there is a dual description that also has an interacting fixed
point in the same region. The original theory is weakly coupled near
$F=3N$ and moves to stronger coupling as $F$ is reduced, while the dual
theory is weakly coupled near $F=3N/2$ and moves to stronger coupling
as $F$ is increased.  

For $F \le 3N/2$ the IR fixed point of the dual theory is  trivial 
(asymptotic freedom is lost in the dual):
\beq
{\wtilde g}_*^2 &=& 0\\
\lambda_*^2 &=& 0~.
\eeq
Since $\wtilde M$ has no interactions it has scaling dimension 1,
and there is an accidental $U(1)$
symmetry in the infrared.
For this range of $F$, $R_{\rm sc}$ is a linear combination of $R$ and
this accidental  $U(1)$. This is consistent with the relation 
$D(\wtilde M) = (3/2) R_{\rm sc}(\wtilde M)$.   
Surprisingly in this range we find that the the IR is a theory
of free massless
composite gauge bosons, quarks, mesons, and their superpartners.
We can lower the number of flavors to $F=N+2$, but to go below
this requires new considerations 
since  there is no dual gauge group $SU(F-N)$ when
$F=N+1$. We will examine what happens in this case in detail later on.
%dangerous

To summarize, for $3N > F > 3N/2$ 
what we have found is two different theories that have IR fixed
points
that describe the same physics. For $3N/2 \ge F > N+1$ 
we have found that a strongly coupled theory and an IR free
theory describe the same physics.  Two theories
having the same IR physics  are referred to as
``being in the same universality class'' by condensed matter physicists. 
This phenomenon also occurs in particle theory, a well known 
example of this is QCD and the chiral Lagrangian.  Having
two different description can be very useful
if one theory is strongly coupled and the other is weakly coupled.
(Two different theories  could not both be weakly coupled and describe the
same physics.) Then we can calculate non-perturbative effects in one
theory by simple doing perturbative calculations in the other theory.
Here we see that even when the dual theory cannot be thought of
as being composites of the original degrees of freedom (since
there is no particle interpretation of conformal theories it
doesn't make sense to talk about composite particles) it still provides
a weakly coupled description in the region where the original theory
is strongly coupled. The name duality has been introduced 
because both theories are gauge theories and thus there is some
resemblance to electric-magnetic duality and the
Olive-Montonen duality \cite{Montonen:1977sn} of ${\mathcal N}=4$ SUSY gauge theories.  However.
Olive-Montonen duality is valid at all energy scales, while for
these ${\mathcal N}=1$ theories, as we go up in energy the infrared correspondence
of Seiberg duality is lost.

%%%%%%%%%%%%%%%%%%%%%%%%%%%%%%%%%%%%%%%
\subsection{Integrating out a flavor}
\label{sec:integratingoutaflavor}

If we give a mass to one flavor in the original theory
we have added a superpotential term
\beq
W_{mass} = m {\overline \Phi}^F \Phi_F~.
\eeq
In the dual theory we then have the superpotential
\beq
W_d = \lambda {\wtilde M}^j_i {\overline \phi}^i \phi_j + m {\wtilde M}^F_F~.
\eeq
Where we have made use of the mapping (\ref{dual:mapping}).
Because of this mapping it is common (though somewhat confusing) to write
\beq
\lambda {\wtilde M} = {{M}\over{\mu}}~,
\eeq
which trades the coupling
$\lambda$ for a scale $\mu$ and uses the same symbol,
$M$, for fields  in
the two different theories that have the same quantum numbers.
With this notation the dual superpotential is
\beq
W_d = \frac{1}{\mu} M^j_i {\overline \phi}^i \phi_j + m M^F_F~.
\eeq
The equation of motion for $M^F_F$ is:
\beq
{{\partial W_d}\over{\partial M^F_F}} = \frac{1}{\mu} {\overline \phi}^F
\phi_F + m=0~.
\eeq
So the dual squarks have VEVs:
\beq
 {\overline \phi}^F
\phi_F = -\mu m ~.
\eeq  
%dangerous
We saw earlier that along such a $D$-flat direction
we have a theory with one less color, one less flavor, and some singlets.
The spectrum of light degrees of freedom is:
\beq
\begin{tabular}{c|c|cc}
& $SU(F-N-1)$ & $SU(F-1)$  & $SU(F-1)$ \\
\hline
$q^\prime$ & \fund & $\overline{\fund}$ & ${\bf 1}$ \\
$\overline{q}^\prime$ & $\overline{\fund}$   & ${\bf 1}$  
& \fund\\
$M^\prime$ & ${\bf 1}$ & \fund &$\overline{\fund}$ \\
$q^{\prime\prime}$ & ${\bf 1}$ & $\overline{\fund}$ & ${\bf 1}$ \\
$\overline{q}^{\prime\prime}$ & ${\bf 1}$    & ${\bf 1}$  
& \fund\\
$S$ & ${\bf 1}$ &${\bf 1}$ & ${\bf 1}$ \\
$M^F_j$ & ${\bf 1}$ & \fund & ${\bf 1}$ \\ 
$M^j_F$ & ${\bf 1}$ & ${\bf 1}$ & $\overline{\fund}$ \\
$M^F_F$ & ${\bf 1}$ & ${\bf 1}$ & ${\bf 1}$\\
\end{tabular}
\eeq
The low-energy effective superpotential is:
\beq
W_{\rm eff} = {{1}\over{\mu}} \left(\langle{\overline \phi}^F\rangle 
M^j_F \phi^{\prime\prime}_j 
+\langle \phi_F\rangle M^F_i {\overline \phi}^{\prime\prime i} +M^F_F S\right) +\frac{1}{\mu} M^\prime 
{\overline \phi}^\prime \phi^\prime~.
\eeq
So we can integrate out $M^j_F$,  $\phi^{\prime\prime}_j$,
$M^F_i$,  ${\overline \phi}^{\prime\prime i}$,  $M^F_F$, and $S$ since
they all have mass terms in the superpotential.
This leaves just the dual of $SU(N)$ with $F-1$ flavors which has a 
superpotential
\beq
W =\frac{1}{\mu} M^\prime 
{\overline \phi}^\prime \phi^\prime ~.
\eeq
Similarly one can check that there is a consistent mapping between the two dual theories when one flavor of the original squarks and anti-squarks have $D$-flat VEVs, and the
corresponding meson VEV gives a mass to the dual quarks and dual squarks, which can
then be integrated out.  The analysis
of baryonic $D$-flat directions was done in ref. \cite{Aharony:1995qs}.

%%%%%%%%%%%%%%%%%%%%%%%%%%%%%%%%%%%%%%%
\subsection{Consistency}
We have seen that Seiberg's conjectured duality satisfies
three non-trivial consistency checks:
\begin{itemize}
\item{} The global anomalies of the original quarks and gauginos
match those of the dual quarks, dual gauginos, and ``mesons''.
\item{} The moduli spaces have the same dimensions and the gauge
invariant operators match:
\beq
&&M \leftrightarrow {\wtilde M} \nonumber\\
&&B_{i_1, \ldots, i_N} \leftrightarrow \epsilon_{i_1,\ldots,i_N,j_1,\ldots
j_{F-N}}
b^{j_1,\ldots,j_{F_N}} \nonumber \\
&&{\overline B}^{i_1, \ldots, i_N} \leftrightarrow 
\epsilon^{i_1,\ldots,i_N,j_1,\ldots
j_{F-N}}
{\overline b}_{j_1,\ldots,j_{F_N}}~.
\eeq
\item{}Integrating out a flavor in the original theory results in an
$SU(N)$ theory with $F-1$ flavors, which should have a dual with
$SU(F-N-1)$ and $F-1$ flavors.  Starting with the dual of the original
theory, the mapping of the mass term is a linear term for the ``meson''
which forces the dual squarks to have a VEV and Higgses the theory down
to $SU(F-N-1)$ with $F-1$ flavors.
\end{itemize}

The duality exchanges weak and strong coupling and also classical and
quantum effects.  For example in the original theory $M$ satisfies a
classical constraint\footnote{Discussed at the end of section
\ref{dual:classmodFgeN}.} rank$(M) \le N$.  
In the dual theory there are $F-{\rm
rank}(M)$ light dual quarks. If rank$(M) >N$ then the number of 
light dual quarks is less than $\wtilde N = F-N$, and an ADS superpotential
is generated, so there is no vacuum with 
rank$(M) > N$. Thus in the dual, rank$(M) \le N$
is enforced by quantum effects.

%%%%%%%%%%%%%%%%%%%%%%%%%%%%%%%%%
%%%%%%%%%%%%%%%%%%%%%%%%%%%%%%%%%%%
\section{Confinement in SUSY QCD}
\setcounter{equation}{0}

For certain special cases (particular values of the number of flavors)
the description in terms of a dual gauge theory breaks down since
there is no possible dual gauge group. Remarkably one finds that
the low-energy effective theory simply contains mesons and baryons
and that their properties can actually be derived from the case where
there is a dual gauge group.

%%%%%%%%%%%%%%%%%%%%%%%%%%%%%%%%%%
\subsection{$F = N$: Confinement with Chiral Symmetry Breaking}
\label{sec:ConfinementwithChiralSymmetryBreaking}
For SUSY QCD with 
$F=N$ flavors Seiberg \cite{Seiberg:1994bz} 
found that all the `t Hooft anomaly matching 
conditions can be solved with just the color singlet
meson and baryon fields, that is no dual quarks are required.
Thus this theory is confining in the sense that all of the massless
degrees of freedom are color singlet particles.

Recall that the classical moduli space of SUSY QCD is parameterized by
\beq
M^j_i &=&  \overline{\Phi}^{j n}\Phi_{n i}\\
B_{i_1,\ldots,i_N} &=&  \Phi_{n_1 i_1}\ldots\Phi_{n_N i_N}
\epsilon^{n_1,\ldots,n_N}\\
{\overline B}^{\,i_1,\ldots,i_N} &=&  {\overline \Phi}^{\,n_1 i_1}\ldots
{\overline\Phi}^{\,n_N i_N}
\epsilon_{n_1,\ldots,n_N}~.
\eeq
For $F=N$ flavors the baryons are flavor singlets:
\beq
B &=& \epsilon^{i_1,\ldots,i_F}B_{i_1,\ldots,i_F}\\
{\overline B} &=&  
\epsilon_{i_1,\ldots,i_F}{\overline B}^{\,i_1,\ldots,i_F}~.
\eeq
%dangerous
We also saw that classically these fields satisfy a constraint:
\beq
{\rm det}M =B {\overline B}~.
\eeq
%dangerous
With quark masses turned on we have:
\beq
\langle M^j_i \rangle
 =(m^{-1})^j_i \left({\rm det}\,m \Lambda^{3N-F}\right)^{{1}\over{N}}~.
\label{confine:Mvev}
\eeq
Taking a determinant of this equation (and using $F=N$) we have
\beq
{\rm det}\langle M \rangle={\rm det}\,(m^{-1}) \,{\rm det}\,m \Lambda^{2N}
=\Lambda^{2N}~,
\eeq
independent of the masses. However ${\rm det}\,m \ne 0$ sets 
$\langle B \rangle = \langle {\overline B} \rangle =0$, since
we can integrate out all the fields that have baryon number.  Thus the
classical constraint is violated!  To understand what is going on it is helpful
to use holomorphy and the symmetries of the theory.  The flavor invariants are:
\beq
\begin{tabular}{cccc}
 & $U(1)_A$ & $U(1)$ & $U(1)_R$ \\
\hline
${\rm det} M$ & $2N$ & $0$  &$0$ \\
$B$ & $N$ & $N$ & $0$ \\
$\overline{B}$ & $N$   & $-N$ &  $0$ \\
$\Lambda^{2N}$ &$2N$ & $0$   & 0 \\
\end{tabular}
\eeq
Note that the $R$-charge of the squarks, $\frac{F-N}{F}$, vanishes
since $F=N$.
Thus a generalized form of the constraint that has the correct
$\Lambda \rightarrow 0$ and $B$,${\overline B} \rightarrow 0$ limits
is
\beq
{\rm det} M - {\overline B} B = \Lambda^{2N}\left(1+\sum_{pq}\, C_{pq}\,
\frac{\left(\Lambda^{2N}\right)^p ({\overline B}B)^q}{({\rm det} M})^{p+q}\right)~,
\eeq
with $p,q >0$.
For $\langle {\overline B} B \rangle \gg \Lambda^{2N}$ the theory is
perturbative, but with $C_{pq}\ne 0$ we find solutions of the
form
\beq
{\rm det} M \approx \left({\overline B} B\right)^{{q-1}\over{p+q}}~,
\eeq
which do not reproduce the weak coupling $\Lambda\rightarrow 0$ 
limit, thus  we can
conclude $C_{pq}= 0$.
Therefore the quantum constraint is:
\beq
{\rm det} M - {\overline B} B = \Lambda^{2N} ~.
\label{confine:constraint}
\eeq
%dangerous
First note that this equation has the correct form to be an instanton
effect since
\beq
e^{-S_{\rm inst}} \propto \Lambda^b = \Lambda^{2N}~.
\eeq
  Also note that we cannot take $M=B={\overline B}=0$,
that is we cannot go to the origin of moduli space (this 
situation is
referred to as a ``deformed" moduli space).  This
means that the   global symmetries are at least partially
broken everywhere in
the quantum moduli space.  

\begin{center}
\begin{figure}
\begin{center}
\includegraphics[width=0.4\hsize]{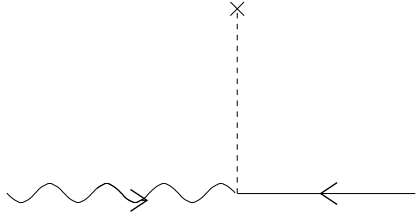}
\end{center}
\caption{Squark VEV (dashed line with cross) gives a mass
to a  quark (solid lines) and a gaugino (wavy line).}
\label{confine:fig:squarkVEV}
\end{figure}
\end{center}

For example consider the following
special points with
enhanced symmetry: $M^j_i= \Lambda^2 \delta^j_i$, $B={\overline B}=0$
and $M=0$, $B{\overline B}=-\Lambda^{2N}$.  In the first case,
with $M= \Lambda^2$,
the global  $SU(F) \times SU(F) \times U(1) \times U(1)_R$
symmetry is broken to $SU(F)_d \times U(1) \times U(1)_R$,
that is the theory undergoes chiral symmetry
breaking and, as in non-supersymmetric QCD, the chiral symmetry 
is broken to the diagonal subgroup. For $B{\overline B}=-\Lambda^{2N}$
the global symmetry is broken to $SU(F) \times SU(F) \times U(1)_R$,
that is baryon number is spontaneously broken.
For large VEVs we can understand this symmetry breaking
as a perturbative Higgs phase with squark VEVs giving masses to 
quarks and gauginos (See Figure \ref{confine:fig:squarkVEV}).  
There is no point in the moduli space where
gluons become light, so there are no singular points,
and the moduli space is smooth.  This is an example of a theory
that exhibits ``complementarity''
\cite{Banks:1979fi,Fradkin:1978dv} since we can go smoothly from a Higgs phase
(large VEVs) to a confining phase (VEVs of ${\mathcal O}(\Lambda)$) without
going through a phase transition.  Complementarity holds in any theory
where there are scalars (in our case they are squarks)
in the fundamental representation\footnote{Also known as defining or faithful
representations of the gauge
group.} of the gauge
group. In such a theory we can take any colored
field and, by multiplying by enough scalar fields, we can form a color singlet
operator\footnote{In the perturbative regime this color singlet 
operator is just
the original colored field multiplied by VEVs, while in the regime of small
VEVs it can be a complicated non-perturbative object.}. 
In other words any color charge can be screened. In a theory
where screening does not happen a Wilson loop\footnote{A Wilson loop
is a closed quark loop. If the expectation value of the loop goes like
$e^{-c_1 A}$, where $A$ is the area enclosed by the Wilson loop it is said
to obey an area law. If the expectation value of the loop goes like
$e^{-c_2 L}$, where $L$ is the length of the perimeter of the Wilson
loop it is said to obey a perimeter law.} can obey an area law (which
indicates confinement) or a perimeter law (which happens in a Higgs phase)
and it can therefore act as a gauge invariant order parameter which can 
distinguish between the area law confinement phase and the Higgs phase.
In a screening theory a Wilson
loop will always obey a perimeter law (even if confinement occurs,
since all the dynamics
takes place along the perimeter of the loop where the screening occurs)
as it does in a Higgs phase.  Thus in screening theories there is no
gauge invariant order parameter that can distinguish between a confining
phase and a Higgs phase, so there can be no phase transition between the
two regimes and the theory exhibits complementarity.

%%%%%%%%%%%%%%%%%%%%%%%%%%%%%%%
\subsection{$F = N$: Consistency Checks}
\label{confine:FeqNConsistencyChecks}
The constraint (\ref{confine:constraint}) can be put in
the form of an equation of motion arising from a
superpotential by introducing a
Lagrange multiplier field, which we can refer to as $X$:
\beq
W_{\rm constraint}
= X\left( {\rm det} M - {\overline B} B - \Lambda^{2N}\right)~.
\label{confine:Wconstraint}
\eeq
We can now check the consistency of the confined picture
by adding a mass for the $N$th flavor. It is convenient to re- write
the meson field as
\beq
M=\left( \begin{array}{cc}
{\wtilde M}^j_i & N^j \\
P_i & Y \end{array}\right)~,
\eeq
where ${\wtilde M}$ is an $(N-1)\times(N-1)$ matrix.  We can then write
the mass term for the $N$th flavor in the confined description as
$W_{\rm mass} = mY$,
so that the full superpotential is
\beq
W
= X\left( {\rm det} M - {\overline B} B - \Lambda^{2N}\right)+ mY~.
\label{confine:FeqNplusmass}
\eeq
We then have the following equations of motion:
\beq
{{\partial W}\over{\partial B}} = - X {\overline B} =0\\
{{\partial W}\over{\partial {\overline B}}} = - X  B  =0\\
{{\partial W}\over{\partial N^j}} =  X \,{\rm cof}(N^j) =0 \\
{{\partial W}\over{\partial P_i}} =  X \,{\rm cof}(P_i)  =0\\
{{\partial W}\over{\partial Y}} =  X \,{\rm det}{\wtilde M} + m =0~,
\eeq
where cof$(M^i_j)$ is the cofactor of the matrix element $M^i_j$.
The solution of these equations is:
\beq
X= -m \left({\rm det}{\wtilde M}\right)^{-1} \\
B={\overline B}=N^j=P_i=0~.
\eeq
Plugging this solution into the equation of motion for $X$ (the constraint
equation) gives
\beq
{{\partial W}\over{\partial X}} = Y {\rm det}{\wtilde M} - \Lambda^{2N} =0~.
\eeq
So, putting this result into the full superpotential 
(\ref{confine:FeqNplusmass}), we find the low-energy
effective superpotential
is
\beq
W_{\rm eff} = \frac{m\, \Lambda^{2N}}{{\rm det}{\wtilde M}}~.
\eeq
%dangerous
Using the usual matching relation for the holomorphic gauge coupling
we find
\beq
m \Lambda^{2N}=\Lambda^{2N+1}_{N,N-1}~,
\eeq
so
\beq
W_{\rm eff} = \frac{\Lambda^{2N+1}_{N,N-1}}{{\rm det}{\wtilde M}}~.
\eeq
This is just the Affleck-Dine-Seiberg  superpotential for $SU(N)$ with
$N-1$ flavors

As a further consistency check let's consider in more detail the 
points in the moduli space with
enhanced symmetry. 
When $M^j_i= \Lambda^2 \delta^j_i$, $B={\overline B}=0$
the global symmetry is broken to $SU(F)_d \times U(1) \times U(1)_R$.
In terms of the elementary fields, the
 $\Phi$ and $\overline{\Phi}$ VEVs break $SU(N) \times 
SU(F) \times SU(F)$ to $SU(F)_d$. The quarks transform as
 ${\fund}\times \overline{\fund}={\bf 1} + {\bf Ad}$ 
under this diagonal group, while
the gluino transforms as ${\bf Ad}$.
The composites have the following quantum 
numbers\footnote{As usual only the $R$-charge of the 
scalar component is given, and $R[{\rm fermion}]=R[{\rm scalar}]-1$.}:
\beq
\begin{tabular}{cccc}
 & $SU(F)_d$ & $U(1)$ & $U(1)_R$ \\
\hline
$M - {\rm Tr} M$ & ${\bf Ad}$ & $0$  &$0$ \\
${\rm Tr} M$ & {\bf 1} & 0 & 0 \\
$B$ & ${\bf 1}$ & $N$ & $0$ \\
$\overline{B}$ & ${\bf 1}$   & $-N$ &  $0$ \\
\end{tabular}
\eeq
The field ${\rm Tr} M$ gets a mass with the Lagrange multiplier
field $X$.
The non-trivial anomalies match as follows (recalling that $F=N$):
\beq
\begin{array}{lcc}
& {\rm elem.} & {\rm comp.} \\
U(1)^2 U(1)_R:& -2FN & -2N^2 \\
U(1)_R : & -2 FN +N^2 -1 & -(F^2-1) -1-1\\
U(1)_R^3 :& -2 FN +N^2 -1 & -(F^2-1) -1-1\\
U(1)_R SU(F)_d^2:& - 2N + N & - N~.
\end{array}
\eeq

At $M=0$, $B{\overline B}=-\Lambda^{2N}$
only  the $U(1)$ is broken. The composites transform as:
\beq
\begin{tabular}{cccc}
 & $SU(F)$ & $SU(F)$ & $U(1)_R$ \\
\hline
$M$ & ${\fund}$ & $\overline{\fund}$  &$0$ \\
$B$ & ${\bf 1}$ & ${\bf 1}$ & $0$ \\
$\overline{B}$ & ${\bf 1}$   & ${\bf 1}$ &  $0$ \\
\end{tabular}
\eeq
The linear combination  
$B+{\overline B}$ gets a mass with the Lagrange multiplier field $X$.
The anomalies match as follows:
\beq 
\begin{array}{lcc}
& {\rm elem.} & {\rm comp.} \\
SU(F)^3: & N & F \\
U(1)_R SU(F)^2:& -N{{1}\over{2}} & -F{{1}\over{2}} \\
U(1)_R : & -2 FN +N^2 -1 & -F^2 -1\\
U(1)_R^3 :& -2 FN +N^2 -1 & -(F^2-1) -1-1~.
\end{array}
\eeq
which, again, agree because $F=N$.

%%%%%%%%%%%%%%%%%%%%%%%%%%%%%%%%
%%%%%%%%%%%%%%%%%%%%%%%%%%%%%%%%%%
\section{S-Confinement in SUSY QCD}
\setcounter{equation}{0}

%%%%%%%%%%%%%%%%%%%%%%%%%%%%%%%%%%%%%
\subsection{$F=N+1$: Confinement without chiral symmetry breaking}
\label{sec:sconfineSQCD}
For SUSY QCD with 
$F=N+1$ flavors Seiberg \cite{Seiberg:1994bz} again found 
that all the `t Hooft anomaly matching 
conditions can be satisfied with just the color singlet
meson and baryon fields; thus this theory is also confining. This theory
also exhibits complementarity \cite{Banks:1979fi,Fradkin:1978dv} 
since screening can occur (as discussed
at the end of section \ref{sec:ConfinementwithChiralSymmetryBreaking}).
However, the theory with 
$F=N+1$ flavors does not require chiral symmetry breaking, that
is we can go to the origin of moduli space. Furthermore the theory develops
a dynamical superpotential. Confining theories that screen, do not 
spontaneously break
global symmetries, and have dynamical superpotential have been dubbed
``s-confining'' 
\cite{CsakiSchmaltzSkiba}.

To see that spontaneous chiral symmetry breaking does not occur
we begin by recalling the classical constraints on the meson and baryon
fields. For $F=N+1$ flavors the baryons are flavor anti-fundamentals
(and the anti-baryons are flavor fundamentals) since they are
anti-symmetrized in $N=F-1$ colors:
\beq
B^i &=& \epsilon^{i_1,\ldots,i_N,i}B_{i_1,\ldots,i_N} ~,\\
{\overline B}_i &=&  
\epsilon_{i_1,\ldots,i_N,i}{\overline B}^{\,i_1,\ldots,i_N}~.
\eeq
With this notation the classical constraints are:
\beq
(M^{-1})^i_j{\rm det}M =B^i {\overline B}_j~,
\label{confine:FeqNplus1constraint1}\\
M^j_i B^i = M^j_i {\overline B}_j=0 ~.
\label{confine:FeqNplus1constraint2}
\eeq
With non-zero quark masses we have:
\beq
\langle M^j_i \rangle
 &=&(m^{-1})^j_i \left({\rm det}m \Lambda^{2N-1}\right)^{{1}\over{N}}~,
\label{confine:MVEV}\\
\langle B^i \rangle &=& \langle {\overline B}_j\rangle=0~.
\label{confine:BVEV}
\eeq
So taking a determinant of equation (\ref{confine:MVEV}) gives
\beq
(M^{-1})^i_j{\rm det}M =m^i_j \Lambda^{2N-1}~.
\eeq
Thus we see that the classical constraint is satisfied as
$m^i_j \rightarrow 0$.  Taking this limit in different ways
we can cover the classical moduli space, so the classical and
quantum moduli spaces are the same. In particular chiral symmetry
remains unbroken at $M=B=\overline{B}=0$.

The most general superpotential allowed for the mesons and baryons is:
\beq
W= {{1}\over{\Lambda^{2N-1}}} \left[ \alpha B^i M^j_i {\overline B}_j
+ \beta {\rm det} M + {\rm det} M \,
f\left( {{{\rm det} M}\over{ B^i M^j_i {\overline B}_j}} \right) \right]~,
\eeq
where $f$ is an as yet unknown function.
Actually only $f=0$  reproduces the classical constraints:
\beq
{{\partial W}\over{\partial M^j_i}} &=& {{1}\over{\Lambda^{2N-1}}}
 \left[ \alpha B^i  {\overline B}_j
+ \beta (M^{-1})^i_j {\rm det} M \right] =0~,\\
{{\partial W}\over{\partial B^i}} &=& {{1}\over{\Lambda^{2N-1}}}
 \alpha M^j_i {\overline B}_j =0 ~,\\
{{\partial W}\over{\partial {\overline B}_j}} &=& {{1}\over{\Lambda^{2N-1}}}
 \alpha B^i M^j_i  =0~,
\eeq
provided that $\beta = -\alpha$.

To determine $\alpha$, consider adding a mass for one flavor 
so that we can compare with the results for $F=N$ flavors.
\beq
W= {{\alpha}\over{\Lambda^{2N-1}}} \left[ B^i M^j_i {\overline B}_j
- {\rm det} M  \right] +m X~,
\label{confine:FeqNplus1superpotentialplusmass}
\eeq
where
\beq
M&=&\left( \begin{array}{cc}
M^{\prime i}_j & Z^i \\
Y_j & X \end{array}\right) ~,\\
B&=& \left( U^i, B^\prime \right)~, \\
{\overline B}& =& \left(
\begin{array}{c}
{\overline U_j}~, \\
{\overline B^\prime}
\end{array}\right)~.
\eeq
For this theory with one massive flavor and $F=N$ light flavors we
have the following equations of motion:
\beq
{{\partial W}\over{\partial Y}} &=&  {{\alpha}\over{\Lambda^{2N-1}}}
\left( B^\prime {\overline U}-{\rm cof}(Y)\right) =0\\
{{\partial W}\over{\partial Z}} &=&  {{\alpha}\over{\Lambda^{2N-1}}}
\left( U {\overline B^\prime}-{\rm cof}(Z)\right)=0 \\
{{\partial W}\over{\partial U}} &=&  {{\alpha}\over{\Lambda^{2N-1}}}
 Z {\overline B^\prime}=0 \\
{{\partial W}\over{\partial {\overline U}}} &=&  {{\alpha}\over{\Lambda^{2N-1}}}
 B^\prime  {\overline Y }=0 \\
{{\partial W}\over{\partial X}} &=&  {{\alpha}\over{\Lambda^{2N-1}}}
\left( B^\prime {\overline B^\prime}-{\rm det}M^\prime \right) +m =0~.\\
\eeq
The solution of these equations is:
\beq
Y=Z=U={\overline U}=0~,\\
{\rm det}M^\prime -B^\prime {\overline B^\prime}= 
{{m\Lambda^{2N-1}}\over{\alpha}} = {{1}\over{\alpha}}\Lambda^{2N}_{N,N}~.
\label{confine:alphaconstraint} 
\eeq
Equation (\ref{confine:alphaconstraint})
gives the correct quantum constraint (\ref{confine:constraint}) 
for $F=N$ flavors if and only if $\alpha=1$.

Plugging the solutions of the equations of motion into
the superpotential (\ref{confine:FeqNplus1superpotentialplusmass}) 
we find the effective superpotential
is
\beq
W_{\rm eff} = {{X}\over{\Lambda^{2N-1}}}
\left( B^\prime {\overline B^\prime}-{\rm det}M^\prime
+m\Lambda^{2N-1}\right)~.
\eeq
%dangerous
With the usual matching relation for the holomorphic gauge coupling
we find
\beq
m \Lambda^{2N-1}=\Lambda^{2N}_{N,N}~,
\eeq
so
\beq
W_{\rm eff} = {{X}\over{\Lambda^{2N-1}}}
\left( B^\prime {\overline B^\prime}-{\rm det}M^\prime
+\Lambda^{2N}_{N,N}\right)~.
\eeq
Holding $\Lambda_{N,N}$ fixed as $m\rightarrow \infty$ implies that
$\Lambda \rightarrow 0$, so $X$ becomes a Lagrange multiplier field 
in this limit. Thus we can completely reproduce the
superpotential (\ref{confine:Wconstraint}) used to 
discuss $F=N$ flavors in section \ref{confine:FeqNConsistencyChecks}.

Thus to summarize we have determined that the correct superpotential
for the confined description of SUSY QCD with $F=N+1$ flavors
is:
\beq
W= {{1}\over{\Lambda^{2N-1}}} \left[ B^i M^j_i {\overline B}_j
- {\rm det} M  \right]~,
\label{confine:FeqNplus1superpotential}
\eeq

Since the point  $M=B={\overline B}=0$ in on the quantum moduli space,
we should worry about what singular behavior occurs there.
Naively gluons and gluinos should become massless.
What actually happens is that only the components of  
$M$, $B$, ${\overline B}$ become massless.  That is
we simply have confinement without chiral symmetry breaking.
This is the type of theory that `t Hooft was searching for
when he proposed his anomaly matching conditions \cite{tHooft}.

 The color singlet composites transform 
as\footnote{As usual only the $R$-charge of the 
scalar component is given, and $R[{\rm fermion}]=R[{\rm scalar}]-1$.}:
\beq\label{confine:FeqNcompositespectrum}
\begin{tabular}{c|cccc}
 & $SU(F)$ &  $SU(F)$  & $U(1)$ & $U(1)_R$ \\
\hline
$M$ & $\fund$ & $\overline{\fund}$ &$0$  &${{2}\over{F}}$ \\
$B$ &  $\overline{\fund}$ & {\bf 1} & $N$ &${{N}\over{F}}$ \\
$\overline{B}$ &{\bf 1}& \fund   & $-N$ &  ${{N}\over{F}}$ \\
\end{tabular}
\eeq
Some of the anomaly matchings go as follows:
\beq 
\begin{array}{lcc}
& {\rm elem.} & {\rm comp.} \\
SU(F)^3: & N & F-1 \\
U(1) SU(F)^2:& N{{1}\over{2}} & N{{1}\over{2}} \\
U(1)_R SU(F)^2:& -{{N}\over{F}}{{N}\over{2}} & {{2-F}\over{F}}{{F}\over{2}}
+{{N-F}\over{F}} \\
U(1)_R : & -N^2 -1 & {{2-F}\over{F}}F^2
+2(N-F) \\
U(1)_R^3 :& -\left({{N}\over{F}}\right)^3 2 NF +N^2 -1 & 
\left({{2-F}\over{F}}\right)^3 F^2
+\left({{N-F}\over{F}}\right)^3 2 F ~,
\end{array}
\eeq
which agree because $F=N+1$.

%%%%%%%%%%%%%%%%%%%%%%%%%%%%%%%%%%%
\subsection{Connection to theories with $F >N+1$}

We can also check that the confined 
descriptions of the theories with $F=N$ and $F=N+1$ flavors are 
consistent with dual descriptions of the theories with more flavors.
Consider the dual theory for $F=N+2$:
\beq
\begin{tabular}{c|c|cccc}
& $SU(2)$ & $SU(N+2)$  & $SU(N+2)$ & $U(1)$ & $U(1)_R$ \\
\hline
$q$ & \fund & $\overline{\fund}$ & {\bf 1} &${{N}\over{2}}$ 
& ${{N}\over{N+2}}$ 
$\vphantom{\raisebox{3pt}{\asymm}}$\\
$\overline{q}$ & $\overline{\fund}$   & {\bf 1} & \fund
& $-{{N}\over{2}}$ & ${{N}\over{N+2}}$ 
$\vphantom{\raisebox{3pt}{\asymm}}$\\
$M$ &{\bf 1}& \fund & $\overline{\fund}$   & 0 & ${{4}\over{N+2}}$ 
$\vphantom{\raisebox{3pt}{\asymm}}$\\
\end{tabular}
\eeq
The dual theory has a superpotential
\beq
W =\frac{1}{\mu} M 
{\overline \phi} \phi ~.
\eeq
As we have seen in section \ref{sec:integratingoutaflavor},
%
%dangerous unlabeled reference
%
giving a mass to one flavor in the corresponding 
SUSY QCD theory produces a dual squark vev 
\beq
\langle {\overline \phi}^F \phi_F \rangle = -\mu m~,
\eeq
which completely breaks the $SU(2)$ gauge group. 

The spectrum of the low-energy effective theory\footnote{That is
after integrating out the
gauge singlet fields which are massive.} is:
\beq
\begin{tabular}{ccccc}
&  $SU(N+1)$  & $SU(N+1)$ & $U(1)$ & $U(1)_R$ \\
\hline
$q^\prime $ &  $\overline{\fund}$ & {\bf 1} &$N$ 
& ${{N}\over{N+1}}$ 
$\vphantom{\raisebox{3pt}{\asymm}}$\\
$\overline{q^\prime}$    & {\bf 1} & \fund
& $-N $ & ${{N}\over{N+1}}$ 
$\vphantom{\raisebox{3pt}{\asymm}}$\\
$M^\prime$ & \fund & $\overline{\fund}$   & 0 & ${{2}\over{N+1}}$ 
$\vphantom{\raisebox{3pt}{\asymm}}$\\
\end{tabular}
\eeq
Comparing with the confined spectrum (\ref{confine:FeqNcompositespectrum})
we see that we should identify
\beq
q^{\prime i} = c B^i ~,\\
\overline{q^\prime}_j = \overline{c}\overline{B}_j~,
\eeq
where $c$ and $\overline{c}$ are some constant re-scalings.

After integrating out the massive gauge singlet fields
the remnant of the tree-level dual superpotential is
\beq
W_{\rm tree}
 = {{c  \overline{c}}\over{\mu}} B^i M^{\prime j}_i {\overline B}_j~.
\eeq

\begin{center}
\begin{figure}
\begin{center}
\includegraphics[width=0.8\hsize]{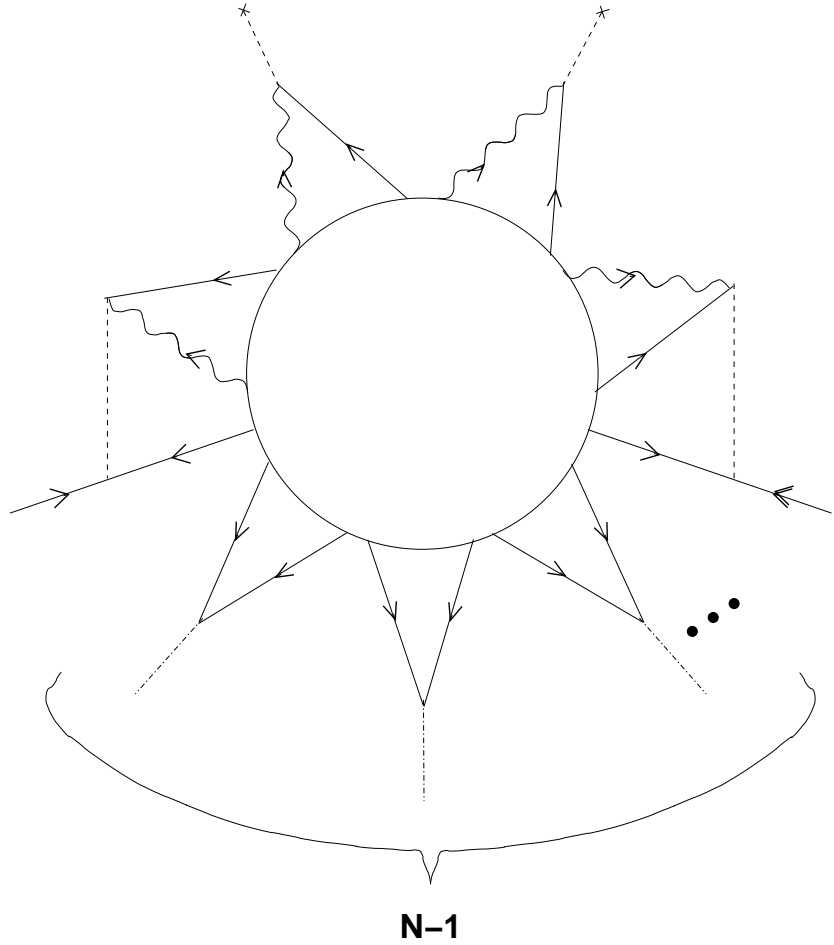}
\end{center}
\caption{Instanton contribution to the interaction between two mesinos
(external straight lines) and $N-1$ mesons (dash-dot lines).
The instanton has 4 gaugino legs (internal wavy lines) and $N+2$ quark
and anti-quark legs (internal straight lines).
A squark VEV and an anti-squark VEV are shown
(dashed line with cross) gives a mass
to a  quark and a gaugino or a anti-quark and gaugino, however they can be
arbitrarily many insertions of these VEVs which must be re-summed.}
\label{confine:fig:instantonsuperpot}
\end{figure}
\end{center}

%dangerous
Since we have completely broken the dual gauge group we expect that
instantons will generate extra terms in the superpotential.
Indeed one finds:
\beq
W_{\rm inst.} &=& 
\frac{\wtilde{\Lambda}^{\wtilde b}_{N,N+2}}
{ \langle {\overline \phi}^F \phi_F \rangle}
{\rm det}\left( {{M^\prime}\over{\mu}}\right) ~,\\
 &=& -
{{\wtilde{\Lambda}^{4-N}_{N,N+2}}\over
{ m}}
{{{\rm det} M^\prime}\over{\mu^{N+2}}}~.
\eeq
where $\wtilde{\Lambda}$ is the intrinsic holomorphic scale of the
dual gauge theory.
A simple way to verify that this superpotential is generated by
instantons is
to check that the corresponding interaction between two fermion
components of $M$ (the mesinos) and $N-1$ mesons is actually generated.
The instanton contribution to this interaction can be seen in
Figure \ref{confine:fig:instantonsuperpot}.

So the effective superpotential agrees with the result for
$F-N+1$ flavors (\ref{confine:FeqNplus1superpotential}):
\beq
W_{\rm eff}= 
{{1}\over{\Lambda^{2N-1}}} \left[ B^i M^{\prime j}_i {\overline B}_j
- {\rm det} M^\prime \right]~,
\eeq
if and only if
\beq
c  \overline{c}= {{\mu}\over{\Lambda^{2N-1}}}~,\\
{{\wtilde{\Lambda}^{4-N}_{N,N+2}}\over
{\mu^{N+2} m}}= {{1}\over{\Lambda^{2N-1}}}~.
\label{confine:dualscalerelation}
\eeq
The second relation (\ref{confine:dualscalerelation})
actually follows from a more general relation
\beq
\wtilde{\Lambda}^{3 \wtilde{N}-F}\Lambda^{3 N-F} = (-1)^{F-N} \mu^F~.
\label{confine:special}
\eeq
To see why this relation is true
consider generic values of $\langle M \rangle$
in the dual of SUSY QCD.
All the dual quarks are massive, so we have a pure $SU(F-N)$ gauge theory.
The intrinsic scale of the dual low-energy effective theory is given 
by matching:
\beq
\wtilde{\Lambda}_L^{3 \wtilde{N}} = \wtilde{\Lambda}^{3 \wtilde{N}-F}
 {\rm det}\left({{ M}\over{\mu}}\right)~.
\eeq
%dangerous
This theory undergoes gaugino condensation, and as we have seen the
effective superpotential is:
\beq
W_L &=& \wtilde{N} \wtilde{\Lambda}_L^{3}\\
&=& (F-N) \left( {{\wtilde{\Lambda}^{3 \wtilde{N}-F}{\rm det} M}\over{\mu}}
\right)^{{1}\over{F-N}}\\
&=&(N-F) \left( {{\Lambda^{3 N -F}}\over{ {\rm det} M}}
\right)^{{1}\over{N-F}}~,
\eeq
where we have used equation (\ref{confine:special}).
Adding a mass term $m^i_j M^j_i$ gives:
\beq
M^j_i 
 =(m^{-1})^j_i \left({\rm det}m \Lambda^{3N-F}\right)^{{1}\over{N}}~,
\eeq
which we have already seen is the correct result.

If we consider the dual of the 
dual\footnote{Which has $F-\wtilde{N}=N$ colors.} 
of SUSY QCD then (with the assumption
that ($\wtilde{\wtilde{\Lambda}}=\Lambda$) equation 
(\ref{confine:special})
implies
\beq
\Lambda^{3 N-F} \wtilde{\Lambda}^{3 \wtilde{N}-F}= (-1)^{F-\wtilde{N}} 
\wtilde{\mu}^F~.
\eeq
So, since $F-\wtilde{N}=N$, we must have for consistency
\beq
\wtilde{\mu}= -\mu~.
\eeq
If we write the composite meson of the dual quarks as:
\beq
N^i_j \equiv \overline{\phi}^i \phi_j~,
\eeq
and the dual-dual squarks as $d$, 
then the superpotential of the dual of the dual is
\beq
W_{dd} = {{N^j_i}\over{\wtilde{\mu}}}  \overline{d}^i d_j + 
{{M^i_j}\over{\mu}} N^j_i~.
\label{confine:dualdualsuperpotential}
\eeq
The equations of motion give
\beq
{{\partial W}\over{\partial M^i_j}} &=& {{1}\over{\mu}}N^j_i=0
 \label{confine:dualdual}\\
{{\partial W}\over{\partial N^j_i}} &=& {{1}\over{\wtilde{\mu}}} \overline{d}^i
d_j +{{1}\over{\mu}}M^i_j =0 
\eeq
So, since $\wtilde{\mu}=-\mu$,
we can identify the original squarks with the dual-dual squarks:
\beq
\Phi_j = d_j~.
\eeq
Plugging the solution (\ref{confine:dualdual}) back into the dual-dual
superpotential (\ref{confine:dualdualsuperpotential}) we find that
it vanishes
and we conclude that the dual of the dual of SUSY QCD is just SUSY QCD.

%%%%%%%%%%%%%%%%%%%%%%%%%%%%%%%%%%%%%%
%%%%%%%%%%%%%%%%%%%%%%%%%%%%%%%%%%%%%%%
\section{Duality for $SO(N)$}
\setcounter{equation}{0}

Dualities for $SO(N)$ theories have also been found \cite{Intriligator:1995},
and they exhibit some interesting new features since they can be non-screening
if there are no spinor representations. Also there are massless composites made of
squarks and gluinos.

%%%%%%%%%%%%%%%%%%%%%%%%%%%%%%%%%%%%%%%%%
\subsection{The $SO(N)$ Theories and Their Classical Moduli Spaces}
 
An $SO(N)$ gauge theory with $F$ quarks in the vector representation
has a global $SU(F) \times U(1)_R$ symmetry as follows:
\beq
\begin{tabular}{c|c|cc}
& $SO(N)$ & $SU(F)$ & $U(1)_R$ \\
\hline
$Q$ & \fund & \fund &  ${{F+2-N}\over{F}}$ 
$\vphantom{\raisebox{3pt}{\asymm}}$\\
\end{tabular}
\eeq
Since there are no dynamical spinors in this theory, static spinor
sources cannot be screened, so there is a distinction between 
area-law confining
and Higgs phases.

Recall that the adjoint of $SO(N)$ is the two-index
antisymmetric tensor.
For odd $N$, there is one spinor 
representation, while for even $N$ there are two in-equivalent spinors.
For $N=4k$ the spinors are self-conjugate, while for $N=4k+2$ the
two spinors are complex conjugates.  The simplest  irreducible
representations
are summarized in the following two tables.

\beq \begin{array}{|c|c|c|} \hline 
\multicolumn{3}{|c|}{SO(2N+1)} \\ \hline
{\rm Irrep} & d({\bf r}) & 2T({\bf r}) \\ \hline
\Yfund & 2N+1 & 2 \\
S & 2^N & 2^{N-2} \\
\Yasymm & N(2N+1) & 4N-2 \\
\Ysymm & (N+1)(2N+1)-1& 4N+6 \\ \hline \end{array}
\eeq
 
\[ \begin{array}{|c|c|c|} \hline 
\multicolumn{3}{|c|}{SO(2N)} \\ \hline
{\rm Irrep} & d({\bf r}) & 2T({\bf r}) \\ \hline
\Yfund & 2N & 2 \\
S & 2^{N-1} & 2^{N-3} \\
\bar{S}, (S') &  2^{N-1} & 2^{N-3} \\
\Yasymm & N(2N-1) & 4N-4 \\
\Ysymm & N(2N+1)-1& 4N+4 \\ \hline \end{array}
\]
 
The one-loop $\beta$ function coefficient for $N >4$ is
\beq
b= 3(N-2) -F~.
\eeq

Solving the D-flatness conditions one finds that
up to flavor transformations, the classical vacua for $F<N$
are given by
\beq
\langle \Phi\rangle =
\left( \begin{array}{ccc}
{\overline v}_1 & & \\
& \ddots &  \\
& & {\overline v}_F \\
0 & \ldots & 0 \\
\vdots  & & \vdots \\
0 & \ldots & 0 \\
\end{array} \right)~.
\eeq
At a generic point in the classical moduli space the $SO(N)$
gauge symmetry is broken broken to $SO(N-F)$ and there are
$NF -N(N-1)+ (N-F)(N-F-1)$ massless chiral supermultiplets.
For $F \ge N$ the vacua are:
\beq
\langle \Phi \rangle = \left( \begin{array}{cccccc}
v_1 & & & 0 & \ldots & 0\\
& \ddots & &\vdots & & \vdots  \\
& & v_N & 0   & \ldots & 0\\
\end{array} \right)~.
\eeq
At a generic point in the moduli space the $SO(N)$
gauge symmetry is broken completely and there are
$NF -N(N-1)$ massless chiral supermultiplets.
We can describe these light degrees of freedom in 
a gauge invariant way by scalar ``meson'' and 
(for $F\ge N$) ``baryon''
fields and their superpartners:
\beq
M_{j i} &=&  \Phi_j \Phi_i~,\\
B_{[i_1,\ldots,i_N]} &=&  \Phi_{[i_1}\ldots\Phi_{i_N]}~,
\eeq
where $[\,]$ denotes antisymmetrization.

Up to flavor
transformations
the moduli space is described by:
\beq
\langle M\rangle &=&
\left( \begin{array}{cccccc}
 v_1^2 & & & & &\\
& \ddots &  & & &   \\
& & v_N^2  &  & & \\
& & & 0 & &\\
& & & & \ddots & \\
& & & & & 0
\end{array} \right) ~,\\
\langle B_{1,\ldots,N} \rangle &=&  v_1\ldots v_N~,\\
\eeq
with all other components set to zero.
The rank of $M$ is at most $N$. 
If the rank of $M$ is $N$, then $B = \pm \sqrt{{\rm det}^\prime M}$,
where ${\rm det}^\prime$ is the product of non-zero eigenvalues.

%%%%%%%%%%%%%%%%%%%%%%%%%%%%%%%%%%%%%%
\subsection{The Dynamical  Superpotential for $F < N-2$}
To construct the effective superpotential we should
look at how the  chiral superfields transform under the 
anomalous axial $U(1)_A$.
\beq
\begin{array}{ccc}
 &U(1)_A & U(1)_R \\
W^a & 0 & 1 \\
\Lambda^b & 2F & 0 \\
{\rm det}M & 2F & 2(F+2-N) \\
\end{array}
\eeq
So we see it is possible to generate a dynamical superpotential (the analogue
of the ADS superpotential for $SU(N)$):
\beq
W_{\rm dyn} = c_{N,F} \left( {{\Lambda^b}\over{ {\rm det}M
}}\right)^{{1}\over{N-2-F}} ~,
\eeq
for $F<N-2$.

%%%%%%%%%%%%%%%%%%%%%%%%%%%%%%%%%%%%%%%%
\subsection{Duality}

For $F \ge 3(N-2)$ we lose asymptotic freedom, so the theory can
be understood as a weakly coupled low-energy effective theory.
For $F$ just below $3N$ we have an infrared fixed point.

A solution to the anomaly matching for $F >  N-2$ is given by:
\beq\nonumber
\begin{tabular}{c|c|cc}
& $SO(F-N+4)$ & $SU(F)$ & $U(1)_R$ \\
\hline
$q$ & \fund & $\overline{\fund}$ &  ${{N-2}\over{F}}$ 
$\vphantom{\raisebox{3pt}{\asymm}}$\\
$M$ & {\bf 1} & \Ysymm &  ${{2(F+2-N)}\over{F}}$ 
$\vphantom{\raisebox{3pt}{\asymm}}$\\
\end{tabular}
\eeq

For $N > N-3$ this theory admits a unique superpotential:
\beq
W={{M_{j i}}\over{2 \mu}} \phi^j \phi^i~.
\eeq

The dual theory also has baryon operators:
\beq
{\wtilde B}^{[i_1,\ldots,i_{\wtilde N}]} &=&  \phi^{[i_1}\ldots\phi^{i_{\wtilde N}]}~.
\eeq

There are additional hybrid ``baryon'' operators in both theories
since the adjoint is an antisymmetric tensor. In the original $SO(N)$
theory
we have:
\beq
h_{[i_1,\ldots,i_{N-4}]} &=& W_\alpha^2 \Phi_{[i_1}\ldots\Phi_{i_{N-4}]}
\nonumber\\
H_{[i_1,\ldots,i_{N-2}] \alpha} &=& W_\alpha \Phi_{[i_1}\ldots\Phi_{i_{N-4}]}~.
\eeq
While in the dual theory we have:
\beq
{\wtilde h}^{[i_1,\ldots,i_{{\wtilde N}-4}]} &=& 
{\wtilde W}_\alpha^2 \phi^{[i_1}\ldots\phi^{i_{{\wtilde N}-4}]}
\nonumber\\
{\wtilde H}^{[i_1,\ldots,i_{{\wtilde N}-2}]}_\alpha &=& 
{\wtilde W}_\alpha \phi^{[i_1}\ldots\phi^{i_{{\wtilde N}-4}]}~.
\eeq

 The two theories  thus have a  mapping of mesons, baryons, and hybrids:
\beq
&&M \leftrightarrow M \nonumber \\
&&B_{i_1, \ldots, i_N} \leftrightarrow \epsilon_{i_1,\ldots,i_F}
{\wtilde h}^{i_1,\ldots,i_{{\wtilde N}-4}} \nonumber \\
&&h_{i_1, \ldots, i_{N-4}} \leftrightarrow \epsilon_{i_1,\ldots,i_F}
{\wtilde B}^{i_1,\ldots,i_{{\wtilde N}}} \nonumber \\
&&H^{[i_1,\ldots,i_{N-2}]}_\alpha \leftrightarrow \epsilon_{i_1,\ldots,i_F}
{\wtilde H}^{[i_1,\ldots,i_{{\wtilde N}-2}]}_\alpha~.
\eeq

The dual $\beta$ function is
\beq
\beta(\wtilde g) \propto - {\wtilde g}^3 (3 ({\wtilde N}-2)-F)  
= -{\wtilde g}^3 (2F-3(N-2))~.
\eeq
So the dual theory loses asymptotic freedom when $F\le 3(N-2)/2$.
When 
\beq
F=3 {\wtilde N} - \epsilon {\wtilde N}~,
\eeq
there is a perturbative IR fixed point in the dual theory.  
One can check that the exact
$\beta$ function vanishes in this range using the relation between
dimensions and $R$ charges in a superconformal theory.
So we have found that $SO(N)$ with $F$ vectors
has an interacting IR fixed point for
$3(N-2)/2 < F < 3(N-2)$.  

For $N-2\le F \le 3(N-2)/2$ the IR fixed point of the dual theory is  trivial 
and we find in the IR free massless
composite gauge bosons, quarks, mesons, and their superpartners.

One can check that adding a mass term in the original theory
and the corresponding linear meson term in the dual theory leads to the correct
reduction of flavors and dual colors.

%%%%%%%%%%%%%%%%%%%%%%%%%%%%%%%%%%%%%%
\subsection{Some Special Cases}
For $F \le N-5$, $SO(N)$ breaks to $SO(N-F) \supset SO(5)$, which 
undergoes gaugino condensation and produces the dynamical superpotential:
\beq
W_{\rm dyn} \propto \langle \lambda \lambda \rangle \propto
\left( {{16 \Lambda^{3(N-2)-F}}\over{ {\rm det}M
}}\right)^{{1}\over{N-2-F}} ~.
\eeq

For $F = N-4$, $SO(N)$ breaks to $SO(4) \sim SU(2)_L \times SU(2)_R$, so
there are two gaugino condensates
\beq
W_{\rm condensate} =2 \langle \lambda \lambda \rangle_L +2 \langle \lambda
\lambda \rangle_R ={{1}\over{2}}(\epsilon_L +\epsilon_R)
\left( {{16 \Lambda^{2N-1}}\over{ {\rm det}M
}}\right)^{{1}\over{2}} ~,
\eeq
where
\beq
\epsilon_s = \pm 1~.
\eeq
So there are four vacua corresponding to two physically distinct
branches: one with $(\epsilon_L +\epsilon_R)= \pm 2$ and the other
with $(\epsilon_L +\epsilon_R)=0$.  The first branch has runaway
vacua, while the second has a quantum moduli space.
At $M=0$, the composite $M$ satisfies the `t Hooft 
anomaly matching conditions.  One can check that this only happens
for $F=N-4$.  So we have another example of confinement without chiral
symmetry breaking, this time without any baryons.
Integrating out a flavor on the first branch gives the correct 
runaway vacua of the previous case ($F+N-5$), while on the second branch we
find no supersymmetric vacua after integrating out a flavor, 
which is a consistency check.

For $F = N-3$, $SO(N)$ breaks to $SO(4) \sim SU(2)_L \times SU(2)_R$, which
then breaks to $SU(2)_d \sim SO(3)$ so
there are instanton effects (since $\Pi_3(G/H) = \Pi_3(SU(2))=Z$) and
gaugino condensation
\beq
W_{\rm condensate}= 4(1+\epsilon) 
 {{ \Lambda^{2N-3}}\over{ {\rm det}M}} ~,
\eeq
where
\beq
\epsilon = \pm 1~,
\eeq
corresponding to the two phases of the gaugino condensate.
So there are  two physically distinct
branches: one with $\epsilon = 1$ and the other
with $\epsilon=-1$. The first has runaway vacua, while
the second has a quantum moduli space.  Integrating out
a flavor, we would need to find two branches again, so
$W \ne 0$ even on the second branch.  For this to be true we
must have some other fields that interact with $M$.  We also
know that $M$ does not match the anomalies by itself.
The solution of the anomaly matching is given by:
\beq\nonumber
\begin{tabular}{c|cc}
& $SU(F)$ & $U(1)_R$ \\
\hline
$q$ & $\overline{\fund}$ &  ${{N-2}\over{F}}$ 
$\vphantom{\raisebox{3pt}{\asymm}}$\\
$M$ & \Ysymm &  ${{2(F+2-N)}\over{F}}$ 
$\vphantom{\raisebox{3pt}{\asymm}}$\\
\end{tabular}
\eeq
The most general superpotential is
\beq
W= {{1}\over{2 \mu}} M qq f\left( {{ {\rm det} M \,
Mqq}\over{{\Lambda^{2N-2}}}}\right)~,
\eeq
where $f(t)$ is an unspecified function.
Adding a mass term  gives 
\beq
q_F=\pm i v~,
\eeq
which gives us the correct number of ground states.
Note that the operator mapping must be:
\beq
q \leftrightarrow h = Q^{N-4} W_\alpha W^\alpha~,
\eeq
which is a hybrid operator.  For $N=4$ this is a gluino-ball.
This is an example of confinement without chiral symmetry breaking
with hybrids.

Starting with the $F=N$ dual which has an $SO(4)$ gauge group,
and integrating out a flavor there will be instanton effects
when we break to $SO(3)$ so the dual superpotential is modified in the
case $F=N-1$:
\beq
W={{M_{j i}}\over{2 \mu}} \phi^j \phi^i - {{1}\over{64 \Lambda^{2N-5}}} 
{\rm  det} M~.
\eeq

For $F=N-2$ we see in the original theory that we can generically
break to $SO(2)$,
while in the dual we also break so $SO(2) \sim U(1)$.  This case
needs a separate treatment \cite{Intriligator:1995}.

%%%%%%%%%%%%%%%%%%%%%%%%%%%%%%%%%%%%%%
%%%%%%%%%%%%%%%%%%%%%%%%%%%%%%%%%%%%%%
\section{$Sp(2N)$ and Chiral Theories}
\setcounter{equation}{0}

%%%%%%%%%%%%%%%%%%%%%%%%%%%%%%%%%%%%%%%
\subsection{Duality for $Sp(2N)$}
An $Sp(2N)$ gauge theory\footnote{We'll use the notation that
$Sp(2) \sim SU(2)$.} with $2F$ quarks ($F$ flavors)
in the fundamental representation
has a global $SU(2F) \times U(1)_R$ symmetry as follows:
\beq
\begin{tabular}{c|c|cc}
& $Sp(2N)$ & $SU(2F)$ & $U(1)_R$ \\
\hline
$Q$ & \fund & \fund &  ${{F-1-N}\over{F}}$ 
$\vphantom{\raisebox{3pt}{\asymm}}$\\
\end{tabular}
\eeq

Recall that the adjoint of $Sp(2N)$ is the two-index
symmetric tensor. The simplest irreducible representations
are summarized in the following table.

\beq \begin{array}{|c|c|c|} \hline
{\rm Irrep} & d(r) & T(r) \\ \hline
\Yfund & 2N & 1 \\
\Yasymm & N(2N-1) -1& 2N-2 \\
\Ysymm & N(2N+1)& 2N+2 \\
\Ythreea & \frac{N(2N-1)(2N-2)}{3}-2N & \frac{(2N-3)(2N-2)}{2}-1 \\
\Ythrees & \frac{N(2N+1)(2N+2)}{3} & \frac{(2N+2)(2N+3)}{2} \\
\Yadjoint & \frac{2N(2N-1)(2N+1)}{3}-2N & (2 N)^2-4 \\ \hline
\end{array}
\eeq

The one-loop $\beta$ function coefficient for $N >4$ is
\beq
b= 3(2N+2) -2F~.
\eeq

The moduli space is parameterized by a ``meson''
\beq
M_{j i} &=&  \Phi_j \Phi_i~,
\eeq
which is antisymmetric in the flavor indices $i,j$.

The holomorphic intrinsic scale (spurion) and the flavor singlet chiral superfield transform 
under the global symmetry as:
\beq
\begin{array}{c|cc}
 &U(1)_A & U(1)_R \\ \hline
\Lambda^{b/2} & 2F & 0 \\
{\rm Pf}M & 2F & 2(F-1-N) \\
\end{array}
\eeq
Where the Pfaffian
of a $2F \times 2F$ matrix $M$ is given by
\beq
{\rm Pf}M = \epsilon^{i_1\ldots i_{2F}} M_{i_1i_2}\ldots M_{i_{2F-1}i_{2F}}~.
\eeq
So we see it is possible to generate a dynamical superpotential
\beq
W_{\rm dyn}\propto  \left( {{\Lambda^{{b}\over{2}}}\over{ {\rm Pf}M
}}\right)^{{1}\over{N+1-F}} ~,
\eeq
for $F<N+1$.
For $F=N+1$ one finds confinement with chiral symmetry breaking
\beq
{\rm Pf}M = \Lambda^{2(N+1)}~.
\eeq
For $F=N+2$ one finds confinement without chiral symmetry breaking
with a superpotential:
\beq
W = {\rm Pf}M~.
\eeq

A solution to the anomaly matching for $F >  N-2$ is given by:
\beq\nonumber
\begin{tabular}{c|c|cc}
& $Sp(2(F-N-2))$ & $SU(2F)$ & $U(1)_R$ \\
\hline
$q$ & \fund & $\overline{\fund}$ &  ${{N+1}\over{F}}$ 
$\vphantom{\raisebox{3pt}{\asymm}}$\\
$M$ & {\bf 1} & \asymm &  ${{2(F+2-N)}\over{F}}$ 
$\vphantom{\raisebox{3pt}{\asymm}}$\\
\end{tabular}
\eeq

This dual theory admits a unique superpotential:
\beq
W={{M_{j i}}\over{\mu}} \phi^j \phi^i~.
\eeq
For $3(N+1)/2 < F<3(N+1)$ we have an IR point.
For $N+3 \le F \le 3(N+1)/2$ the dual is IR free and the low
energy theory is made up of quarks, gluons, mesons, and their
superpartners. A complete discussion of $Sp(N)$ theories with fundamentals is
given in \cite{Intriligator:1995ne}.

%%%%%%%%%%%%%%%%%%%%%%%%%%%%%%%%%%%%
\subsection{Why Chiral Gauge Theories are Interesting} 
We would eventually
like to use non-perturbative methods like duality for understanding
SUSY gauge theories that dynamical break SUSY.
Usually vector-like gauge theories don't break SUSY, while chiral
gauge theories can.  If a theory is vector-like we can give
masses to all the matter fields. If these masses are large
we have a pure gauge theory that has gaugino condensation
but no SUSY breaking.  By Witten's index argument \cite{Witten:df} we can vary
the mass but the number of bosonic minus fermionic vacua 
doesn't change. If taking the mass to zero does not move some
vacua in from or out to infinity, then the massless theory has
the same number of vacua, and SUSY is not broken.

As our first example of a chiral gauge theory 
consider
\beq\nonumber
\begin{tabular}{c|c|cc}
& $SU(N)$ & $SU(N+4)$  \\
\hline
$\overline{Q}$ & $\overline{\fund}$ & \fund  
$\vphantom{\raisebox{3pt}{\asymm}}$\\
$T$  & \Ysymm & {\bf 1}  
$\vphantom{\raisebox{3pt}{\asymm}}$\\
\end{tabular}
\eeq
I will not write down the charges under the two global $U(1)$'s.
This theory is dual to \cite{Pouliot:1996sk}
\beq\nonumber
\begin{tabular}{c|c|cc}
& $SO(8)$ & $SU(N+4)$  \\
\hline
$q$ & \fund & $\overline{\fund}$  
$\vphantom{\raisebox{3pt}{\asymm}}$\\
$p$ & {\bf S} & {\bf 1}  
$\vphantom{\raisebox{3pt}{\asymm}}$\\
$U \sim {\rm det}T$  & {\bf 1} & {\bf 1}  
$\vphantom{\raisebox{3pt}{\asymm}}$\\
$M\sim \overline{Q}T\overline{Q}$ & {\bf 1} & \Ysymm   
$\vphantom{\raisebox{3pt}{\asymm}}$\\
\end{tabular}
\eeq
with a superpotential
\beq
W=Mqq +Upp~.
\eeq
This theory is vector-like!
The dual $\beta$ function coefficient is:
\beq
b=3(8-2)-(N+4)-1=13-N~.
\eeq
So the dual is IR free for $N>13$.

%%%%%%%%%%%%%%%%%%%%%%%%%%%%%%%%%%%%%%%%%
%%%%%%%%%%%%%%%%%%%%%%%%%%%%%%%%%%%%%%%%
\section{S-Confinement}
\label{sec:sconfine}
We need some semi-systematic way to survey chiral gauge
theories.  One way to do this is to generalize well understood
dual descriptions.  The simplest of these is confinement without
chiral symmetry breaking in $SU(N)$ with $N+1$ flavors.
%dangerous
Recall (from section \ref{sec:sconfineSQCD}) the confined description had a superpotential
\beq
W= {{1}\over{\Lambda^{2N-1}}}\left( {\rm det} M - B M \overline{B} \right)~.
\eeq
The crucial features of this description were
that since there was no chiral symmetry breaking and that
the meson-baryon description was valid over the whole moduli space.
That is, there was a smooth description with no phase-transitions.
This is because the theory obeyed complementarity \cite{Banks:1979fi,Fradkin:1978dv},
 since every static source
could be screened by squarks.  To generalize this we will need to 
have fields that are fundamentals of $SU$ or $Sp$ and spinors of $SO$.
We will only consider theories that have a superpotential in the 
confined description.  This requirement gives us an index constraint.
Theories that satisfy these conditions are called s-confining
\cite{CsakiSchmaltzSkiba}.

Consider a gauge theory with one gauge group and arbitrary matter fields.
Choose an anomaly free $U(1)_R$ such that $\phi_i$ that transforms in
the representation ${\bf r_i}$ of the gauge group and has $R$-charge $q$ while all 
other fields
have zero charge.  The charge $q$ is determined by anomaly cancellation:
\beq
0 &=& (q-1)T({\bf r_i}) +T(Ad) - \sum_{j\ne i}T({\bf r_j})\\
&=&q T({\bf r_i}) +T(Ad) - \sum_j T({\bf r_j})~.
\eeq
Since we can do this for any field, and for each choice the superpotential
has $R$-charge 2, we have
\beq
W \propto \Lambda^3 \left[ \Pi_i \left( {{\phi_i}\over{\Lambda}}\right)^{T({\bf r_i})}
\right]^{2/(\sum_jT({\bf r_j})-T(Ad))}~.
\eeq
There may in general be a sum of terms corresponding to different
contractions of gauge indices.
Requiring that this superpotential be holomorphic at the origin
means there should be integer powers of the composite fields, which
implies integer powers of the fundamental fields.  Unless
all the $T(r_i)$ have a common divisor we must have
\beq
\sum_jT(r_j)-T(Ad) &=& 
1 \,\,{\rm or}\,\,2  \,\,{\rm for}\,\,SO\,\, {\rm or}\,\, Sp \nonumber\\
2(\sum_jT(r_j)-T(Ad)) &=&1 \,\,{\rm or}\,\,2 \,\,{\rm for}\,\,SU~.
\eeq
The differing cases come from the different conventions for 
normalizing generators, for $SO$ and $Sp$ we have $T(\fund)=1$,
while for $SU$ we have $T(\fund)=1/2$.
Anomaly cancellation for $SU$ and $Sp$ require that the left hand side be
even.  This condition is necessary for  s-confinement, but not sufficient.
One has to check explicitly (by exploring the moduli space, as discussed below) 
that for $SO$ none of the candidate theories  where the sum is 2 turn out to be s-confining.
Thus we have
\beq
\sum_jT(r_j)-T(Ad) &=& 
\left\{ \begin{array}{l}
1   \,\,{\rm for}\,\,SU \,\,{\rm or}\,\, SO \\
2 \,\,{\rm for}\,\,Sp
\end{array}\right.~.
\eeq
This condition gives a finite list of candidate s-confining theories.

We can check whether candidate theories that satisfy the index constraint
really are s-confining by going out in moduli space.  Generically we break
to theories with smaller gauge groups and singlet fields that decouple in the
infrared.  If the smaller gauge theory is not s-confining the the original
theory was not s-confining.  Alternatively if we have
an s-confining theory and we go out in moduli space we must end up with
another s-confining theory.  Using these checks one can go through the
list of candidates.
For $SU$ one finds that the following theories are the only ones
that satisfy the conditions for s-confinement.
\beq 
\begin{array}{l|l} 
SU(N) & (N+1) (\Yfund + \overline{\Yfund}) \\
SU(N) & \Yasymm + N\, \overline{\Yfund} + 4\, \Yfund  \\
SU(N) & \Yasymm + \overline{\Yasymm} + 3 (\Yfund + \overline{\Yfund})  \\
SU(5) &  3 \left(\Yasymm + \overline{\Yfund}\right)  \\
SU(5) &  2\, \Yasymm + 2\, \Yfund + 4\, \overline{\Yfund}  \\
SU(6) & 2\, \Yasymm + 5\, \overline{\Yfund} + \Yfund  \\
SU(6) & \Ythreea + 4 (\Yfund + \overline{\Yfund})  \\
SU(7) & 2 \left(\Yasymm + 3\, \overline{\Yfund}\right)
\end{array}~.
\eeq

Let's consider the special case:
\beq 
\begin{array}{c|c|ccccc}
& SU(2N+1)  & SU(4) & SU(2N+1) & U(1)_1 & U(1)_2 & U(1)_R \\ \hline
A & \Yasymm & 1 & 1 & 0 & 2N+5 & 0 \\
\overline{Q} & \overline{\Yfund} & 1 & \Yfund & 4 & -2N+1 & 0 \\
Q & \Yfund & \Yfund & 1 & -2N-1 & -2N+1 & \frac{1}{2} 
\end{array}
\eeq
This theory has a confined description in terms of the following composite fields:
\beq
\begin{array}{c|ccccc}
 & SU(4) & SU(2N+1) & U(1)_1 & U(1)_2 & U(1)_R \\ \hline
Q\overline{Q} &  \Yfund & \Yfund & 3-2N & -4N+2 & \frac{1}{2} \\
A\overline{Q}^2 & 1 &   \Yasymm & 8 & -2N+7 & 0 \\
A^{N}Q  & \Yfund &  1 & -2N-1 & 2N^2+3N+1 & \frac{1}{2} \\
A^{N-1}Q^3 & \overline{\Yfund} &  1 & -6N-3 & 2N^2-3N-2 & 
\frac{3}{2} \\
\overline{Q}^{2N+1} &  1 & 1 & 4(2N+1) & -4N^2+1 & 0 \end{array}
\eeq
with a superpotential
\beq
 W &=& \frac{1}{\Lambda^{2N}} \Big[ (A^NQ)(Q\overline{Q})^3 
(A\overline{Q}^2)^{N-1}+(A^{N-1}Q^3)(Q\overline{Q})(A\overline{Q}^2)^{N}
\nonumber \\ && \,\,\,\,\,\,\,\,\,\,\,\,\,+
 (\overline{Q}^{2N+1})(A^N Q)(A^{N-1} Q^3)
\Big]~.
\eeq
One can check that the equations of motion reproduce the classical constraints, and that
integrating out
a flavor gives confinement with chiral symmetry breaking.

%%%%%%%%%%%%%%%%%%%%%%%%%%%%%%%%%%%%%
%%%%%%%%%%%%%%%%%%%%%%%%%%%%%%%%%%%%%
\section{Deconfinement}

Given that we can find confined descriptions of certain SUSY gauge theories, we
can also reverse engineer a dual description of a given theory where some of
the fields correspond to mesons or baryons of the dual description.  Such
a procedure is referred to as ``deconfinement" \cite{Berkooz:1995km}.

Consider the $SU(N)$ gauge theory with an antisymmetric tensor \cite{Pouliot:1996me,Terning:1997jj} 
for odd $N$ theory with $F\ge 5$ flavors:
\[
\begin{array}{c|c|ccccc}
 & SU(N)  & SU(F)  & SU(N+F-4) & U(1)_1 & U(1)_2 & U(1)_R
\\
\hline
A & \Yasymm & {\bf 1} & {\bf 1} & 0  & -2F\vbr &
{{-12}\over{N}}\vbr \\
Q & \Yfund & \Yfund & {\bf 1}  & 1  & N-F & 2-{{6}\over{N}}\vbr \\
\overline{ Q} &  \overline{\Yfund}  & {\bf 1}  & \Yfund & {{-F}\over{N+F-4}}  &
F & {{6}\over{N}}\vbr \\
\end{array}
\]
We can find a dual description of this theory by taking $A$ 
to be  a composite meson of a s-confining $Sp$ theory
\beq
\begin{array}{c|cc|ccccc}
 & SU(N)  & Sp(N-3)  & SU(F)  & SU(N+F-4) & U(1)_1 & U(1)_2 & U(1)_R
\\
\hline
Y & \Yfund & \Yfund & {\bf 1} & {\bf 1} & 0  & -F\vbr &
{{-6}\over{N}}\vbr \\
Z & {\bf 1} & \Yfund & {\bf 1} & {\bf 1} & 0  & FN\vbr &
8 \vbr \\
\overline{P} & \overline{\Yfund} & {\bf 1} & {\bf 1} & {\bf 1} & 0  & F(1-N)\vbr &
6-{{6}\over{N}}\vbr \\
Q & \Yfund & {\bf 1} & \Yfund & {\bf 1}  & 1  & N-F & 2-{{6}\over{N}}\vbr \\
\overline{ Q}  & \overline{\Yfund} & {\bf 1} & {\bf 1}  & \Yfund & {{-F}\over{N+F-4}}  &
F & {{6}\over{N}}\vbr \\
\end{array}
\eeq
with a superpotential
\beq
W=Y Z \overline{P}~.
\eeq
The $SU(N)$ group has $N+F-3$ flavors so we can using our standard duality:
to find another dual:
\beq
\begin{array}{c|cc|cc}
 & SU(F-3)  & Sp(N-3)  & SU(F)  & SU(N+F-4) 
\\
\hline
y & \Yfund & \Yfund & {\bf 1} & {\bf 1}  \\
\overline{p} & \overline{\Yfund} & {\bf 1} & {\bf 1} & {\bf 1}  \\
q & \Yfund & {\bf 1} & \overline{\Yfund} & {\bf 1}   \\
\overline{ q} & \overline{\Yfund} & {\bf 1}  & {\bf 1}  & \overline{\Yfund}  \\
M & {\bf 1} & {\bf 1} & \Yfund & \Yfund \\
L & {\bf 1} & \Yfund & {\bf 1} & \Yfund \\
B & {\bf 1} & {\bf 1} & \Yfund & {\bf 1} \\
\end{array}
\eeq
with 
\beq
W= M q \overline{q} + B q \overline{p} +L y \overline{q}~.
\eeq
But $Sp(N-3)$ with $N+2F-7$ fundamentals has an  $Sp(2F-8)$ dual:
\beq
\begin{array}{c|cc|cc}
 & SU(F-3)  & Sp(2F-8)  & SU(F)  & SU(N+F-4) 
\\
\hline
\wtilde y & \overline{\Yfund} & \Yfund & {\bf 1} & {\bf 1}  \\
\overline{p} & \overline{\Yfund} & {\bf 1} & {\bf 1} & {\bf 1}  \\
q & \Yfund & {\bf 1} & \overline{\Yfund} & {\bf 1}   \\
\overline{ q}  & \overline{\Yfund}& {\bf 1}  & {\bf 1}  & \overline{\Yfund}  \\
M & {\bf 1} & {\bf 1} & \Yfund & \Yfund \\
l & {\bf 1} & \Yfund & {\bf 1} & \overline{\Yfund}\\
B & {\bf 1} & {\bf 1} & \Yfund & {\bf 1} \\
a & \Yasymm & {\bf 1} & {\bf 1} & {\bf 1} \\
H & {\bf 1} & {\bf 1} & {\bf 1} & \Yasymm  \\
(L y) & \Yfund & {\bf 1} & {\bf 1} & \Yfund \\
\end{array}
\eeq
with 
\beq
W= a \wtilde{y} \wtilde{y} + H ll + (L y) l \wtilde y +
M q \overline{q} + B q \overline{p} + (L y) \overline{q}~,
\eeq
which, after integrating out $(L y)$ and  $\overline{q}$
becomes
\beq
W= a \wtilde{y} \wtilde{y} + H ll  +
M q l \wtilde y + B q \overline{p} ~.
\eeq
With $F=5$ we have a gauge group $SU(2) \times SU(2)$ and
one can show (using the fact that gauge invariant scalar operators
have dimensions larger than or equal to one) that for $N>11$ 
this theory has an IR fixed point \cite{Terning:1997jj}. One can also show that
some of the fields are IR free.  Integrating out
one flavor completely breaks the gauge group and the light
degrees of freedom are just the composites of the s-confining
%dangerous
description discussed at the end of section \ref{sec:sconfine}. 
With the other dual descriptions we would have
to discuss strong interaction effects to see that we get the
correct confined description.

%%%%%%%%%%%%%%%%%%%%%%%%%%%%%%%%%%%%%%%%%%%%%
%%%%%%%%%%%%%%%%%%%%%%%%%%%%%%%%%%%%%%%%%%%%%
\section{Gauge Mediation}
\setcounter{equation}{0}

The basic idea of gauge mediation\footnote{For an excellent review see
\cite{Giudice:1998bp}.}
 as proposed\footnote{For early work along these lines see ref. \cite{OldGaugeMed}.}  
 by Dine, Nelson, Nir, and Shirman\cite{GaugeMediation} 
is that there are three sectors
in the theory, a dynamical SUSY breaking sector, a messenger
sector, and the Minimal Supersymmetric Standard Model (MSSM).  SUSY breaking is communicated to the
messenger sector so that the messengers have a SUSY breaking spectrum.
They also have standard model gauge interactions, 
which then communicate SUSY breaking
to the ordinary superpartners. This mechanism has the great advantage that
since the gauge interactions are flavor blind it does not introduce
flavor changing neutral currents which are an enormous problem for
supergravity mediation models.  

%%%%%%%%%%%%%%%%%%%%%%%%%%%%%%%%%%%%%%%%%%
\subsection{Messengers of SUSY breaking}

We will first consider a model with $N_f$ messengers
$\phi_i$, $\overline{\phi}_i$ and a Goldstino multiplet $X$ 
with an expectation value:
\beq
\langle X \rangle = M +\theta^2 F~,
\eeq
so the scale of SUSY breaking is set by $\sqrt{F}$.  
The Goldstino  is coupled to the messenger fields via a superpotential 
\beq
W= X \overline{\phi}_i \phi_i~.
\eeq
In order to preserve gauge unification, $\phi_i$ and $\overline{\phi}_i$
should form complete Grand Unified Theory (GUT) multiplets.  
The existence of the messengers shifts the coupling at the GUT scale, $\mu_{\rm GUT}$, 
relative to that in the MSSM by
\beq
\delta \alpha_{\rm GUT}^{-1}= -{{N_m}\over{2 \pi}}
\ln\left( {{\mu_{\rm GUT}}\over{M}}\right)~,
\eeq
where
\beq
N_m= \sum_{i=1}^{N_f} 2 T(r_i) ~.
\eeq
For the unification to remain perturbative we need
\beq
N_m < {{150}\over {\ln\left( {{\mu_{\rm GUT}}\over{M}}\right)}}~.
\eeq
The VEV of $X$ gives each messenger fermion a mass $M$, and
the scalars squared masses equal to $M^2\pm F$.  We will be interested in the 
case that $F \ll M^2$.  

We can construct an effective theory by
integrating out the messengers. At one-loop this gives a mass to 
the gauginos\footnote{The super-partners of the the standard model
gauge bosons.} of order 
\beq
M_{\lambda i} \sim \frac{\alpha_i}{4 \pi} \frac{F}{M}~.
\eeq
At two-loop order one finds squared masses for squarks and sleptons of order
\beq
M_s^2 \sim \sum_i \left(\frac{\alpha_i}{4 \pi} \frac{F}{M}\right)^2~,
\eeq
the sum indicates that there are contributions from all the gauge interactions
that the scalar couples to.
This is very nice since the squark and slepton masses turn out to be
of the same order as the gaugino masses, however doing two-loop calculations
is quite tedious. Fortunately since we are only interested in the finite 
parts of two-loop graphs, there is a simple method for doing these calculations
using the renormalization group (RG) and holomorphy 
\cite{Giudice:1998bp,Giudice:1997ni}.

%%%%%%%%%%%%%%%%%%%%%%%%%%%%%%%%%%%%
\subsection{RG Calculation of Soft Masses}
Below the mass scale of the messengers
we can integrate  them out and write the pure gauge part of the Lagrangian
as:
\beq
{\mathcal L}_G =-{{i}\over{16 \pi}} \int d^2 \theta\, \tau(X,\mu) W^\alpha W_\alpha~,
\eeq
where the effects of the messengers are included through their effects
on the one-loop running of the holomorphic gauge coupling. 
Taylor expanding in the $F$ component of $X$ we find a gaugino mass
given by
\beq
M_\lambda & =& {{i}\over{2\tau}} {{\partial \tau}\over{\partial X}} |_{X=M} F \nonumber \\
&=& {{i}\over{2}} {{\partial \ln \tau}\over{\partial \ln X}} |_{X=M} {{F}\over{M}}~.
\eeq
The holomorphic coupling is given by 
\beq
\tau(X,\mu) = \tau(\mu_0) +i {{b^\prime }\over{2 \pi}} \ln \left({{X}\over{\mu_0}}
\right) +i {{b }\over{2 \pi}} \ln \left({{\mu}\over{X}}
\right)~,
\eeq
where, as usual, $b^\prime$ is the $\beta$ function coefficient of the theory including
the messenger fields
and $b$ is the $\beta$ function coefficient in the effective
theory (i.e. the MSSM) below the mass scale of the messengers.  These two coefficients
are related by:
\beq
b^\prime = b -N_m~.
\eeq
So the gaugino mass is simply given by
\beq
M_\lambda = \frac{\alpha(\mu)}{4 \pi} N_m \frac{F}{M}~.
\eeq
Note that in the MSSM, where we have three gauginos, the ratio
of the gaugino mass to the gauge coupling is universal:
\beq
\frac{M_{\lambda 1}}{\alpha_1}=\frac{M_{\lambda 2}}{\alpha_2}=
\frac{M_{\lambda 3}}{\alpha_3}= N_m \frac{F}{M} ~.
\eeq
This type of relation was originally found in supergravity mediation
models and, at the time, was thought to be a signature of such models.

Next consider the wavefunction renormalization for the 
matter fields of the MSSM:
\beq
{\mathcal L} = \int d^4 \theta \, Z(X,X^\dagger) Q^{\prime \dagger} Q^\prime~,
\eeq
where $Z$ must be real and the superscript $^\prime$ is meant to indicate
that we have not yet canonically normalized the field.
Taylor expanding in the superspace coordinate $\theta$ we have
\beq
{\mathcal L} &= &\int d^4 \theta \left(Z+{{\partial Z}\over{\partial X}} F \theta^2
+{{\partial Z}\over{\partial X^\dagger}} F^\dagger \theta^{\dagger 2}
+{{\partial^2 Z}\over{\partial X \partial X^\dagger}} F \theta^2 
F^\dagger \theta^{\dagger 2}\right)|_{X=M}Q^{\prime \dagger} Q^\prime~.\nonumber \\
&&
\eeq
Canonically normalizing we have:
\beq
Q = Z^{1/2} \left(1+{{\partial Z}\over{\partial X}} F \theta^2\right)|_{X=M} 
Q^\prime~,
\eeq
so
\beq
{\mathcal L} = \int d^4 \theta \left[1-\left({{\partial \ln Z}\over{\partial X}} 
{{\partial \ln Z}\over{\partial X^\dagger}} 
-{{1}\over{Z}}{{\partial^2 Z}\over{\partial X \partial X^\dagger}} \right)F \theta^2 
F^\dagger \theta^{\dagger 2}\right]|_{X=M}
Q^{\dagger} Q.
\eeq
Thus we have a sfermion mass term:
\beq
m_Q^2 = - {{\partial^2 \ln Z}\over{\partial \ln X \partial \ln X^\dagger}}|_{X=M}
{{FF^\dagger}\over{M M^\dagger}}~.
\eeq
Rescaling the matter fields also introduces an $A$ term in the effective potential
from Taylor expanding the superpotential:
\beq
Z^{-1/2} {{\partial \ln Z}\over{\partial X}} |_{X=M} F Q 
{{\partial W}\over{\partial Z^{-1/2} Q }}~,
\eeq
which is suppressed by a Yukawa coupling.
To calculate $Z$, we do a 
supersymmetric calculation and replace $M$ by $\sqrt{X X^\dagger}$.
At $l$ loops an RG analysis gives
\beq
\ln Z = \alpha(\mu_0)^{l-1}f(\alpha(\mu_0) L_0,  \alpha(\mu_0) L_X)~,
\eeq
where
\beq
L_0 = \ln \left( {{\mu^2}\over{\mu_0^2}}\right)~, \\
L_X = \ln \left( {{\mu^2}\over{X X^\dagger}}\right) ~,
\eeq
so
\beq
{{\partial^2 \ln Z}\over{\partial \ln X \partial \ln X^\dagger}}
= \alpha(\mu)^{l+1}h(\alpha(\mu) L_X)~.
\eeq
Thus the two-loop scalar masses are determined by a one-loop
RG equation.

At one-loop we have
\beq
{{d \ln Z}\over{d \ln \mu}} = {{C_2(r)}\over{\pi}} \alpha(\mu)~,
\eeq
so
\beq
Z(\mu) = Z_0 \left( {{\alpha(\mu_0)}\over{\alpha(X)}}\right)^{{2 C_2(r)}\over{b^\prime}}
\left( {{\alpha(X)}\over{\alpha(\mu)}}\right)^{{2 C_2(r)}\over{b}}~,
\eeq
where
\beq
\alpha^{-1}(X)= \alpha^{-1}(\mu_0) + {{b^\prime}\over{4 \pi}} \ln
\left( {{X X^\dagger}\over{\mu_0^2}}\right)~, \\
\alpha^{-1}(\mu)= \alpha^{-1}(X) + {{b}\over{4 \pi}} \ln
\left( {{\mu^2}\over{X X^\dagger}}\right) ~.
\eeq
So we finally obtain
\beq
m_Q^2 = 2 C_2(r) \frac{\alpha(\mu)^2}{16 \pi^2} N_m 
\left( \xi^2 +\frac{N_m}{b}
(1-\xi^2)\right)\left(\frac{F}{M}\right)^2~,
\eeq
where
\beq
\xi = {{1}\over{1+{{b}\over{2\pi}} \alpha(\mu) \ln {{M}\over{\mu}} }}~.
\eeq
What is particularly impressive about this calculation \cite{Giudice:1998bp,Giudice:1997ni}
is that a two-loop result is obtained from a one-loop calculation.

%%%%%%%%%%%%%%%%%%%%%%%%%%%%%%%%%%%%%%%%%
\subsection{Gauge Mediation and the $\mu$ Problem}
Recall that in order to obtain a viable mass spectrum 
for the electroweak sector in the MSSM,
the two Higgs doublets, $H_u$ and
$H_d$, needed two types of masses terms: a supersymmetric $\mu$ term:
\beq
W= \mu H_u H_d ~,
\eeq
and a soft SUSY-breaking $b$ term:
\beq
V= b H_u H_d~,
\eeq
with a peculiar relation between these two seemingly unrelated parameters
\beq
b \sim \mu^2~.
\eeq
In gauge mediated models we need to be of the same order as the squark and slepton masses:
\beq
\mu \sim {{1}\over{16 \pi^2}} {{F}\over{M}}~.
\eeq
If we introduce a coupling of the Higgses to the SUSY breaking field $X$,
\beq
W= \lambda X H_u H_d ~,\\
\eeq
we get
\beq
\mu &=& \lambda M ~,\\
b&=& \lambda F \sim 16 \pi^2 \mu^2~,
\eeq
so $b$ is much too large for this type of model to be viable.

A more indirect coupling
\beq
W= X (\lambda_1 \phi_1 \overline{\phi}_1 + \lambda_2 \phi_2 \overline{\phi}_2)
 \lambda H_u \phi_1 \phi_2 
 +\overline{\lambda} H_d \overline{\phi_1} \overline{\phi_2}~,
\eeq
yields a one-loop correction to the effective Lagrangian:
\beq
\Delta {\mathcal L}= \int d^4 \theta {{\lambda \overline{\lambda}}\over{16 \pi^2}}
f(\lambda_1/\lambda_2) H_u H_d {{X}\over{X^\dagger}}~.
\eeq
This unfortunately still
gives the same, non-viable, ratio for $b/\mu^2$.  

The correct ratio can
be arranged with the introduction of two additional singlet fields $S$ and $N$:
\beq
W=S(\lambda_1 H_u H_d +\lambda_2 N^2 + \lambda \phi \overline{\phi} -M_N^2)
+X\phi \overline{\phi}~,
\eeq
then
\beq
\mu = \lambda_1 \langle S \rangle ~,\\
b = \lambda_1 F_S~.
\eeq
A VEV for $S$ is generated at one loop
\beq
\langle S \rangle \sim {{1}\over{16 \pi^2}} {{F_X^2}\over{M M_N^2}}~,
\eeq
but $F_S$ is only generated at two loops:
\beq
F_S \sim {{1}\over{(16 \pi^2)^2}} {{F_X^2}\over{M^2}} \sim 
{{1}\over{16 \pi^2}} \mu{{M_N^2}\over{M }}~.
\eeq
Thus we can obtain
 $b \sim \mu^2$ provided that we arrange for $M_N^2 \sim F_X$.  So 
 it seems that somewhat elaborate models
are needed to solve the $\mu$ problem in the gauge mediated scenario.

%%%%%%%%%%%%%%%%%%%%%%%%%%%%%%%%%%%%%%
%%%%%%%%%%%%%%%%%%%%%%%%%%%%%%%%%%%%%%
\section{Dynamical SUSY Breaking}
\setcounter{equation}{0}

Non-perturbative SUSY techniques can be used to understand how
to dynamically break gauge symmetries (like the the GUT symmetry of the electroweak
gauge symmetry of the standard model), however the most important
application is to break SUSY itself.  We would like to have a theory
of dynamical breaking so that the SUSY breaking scale can be naturally
much smaller than the Planck or string scale without fine-tuning the
ratio of these scales by hand\footnote{For an excellent review see \cite{Shadmi:1999jy}.}.  
This is what naturally happens in
asymptotically free gauge theories since the coupling at high energies 
can be perturbative and slowly running and then get strong at some much
lower scale.  This is just what happens in QCD: the coupling is perturbative
near the (putative) unification scale, but gets strong in the IR, producing a large
but natural ratio between  the GUT scale and the proton mass.

%%%%%%%%%%%%%%%%%%%%%%%%%%%%%%%%%%%%%
\subsection{A Rule of Thumb for SUSY Breaking}

A theory that has no flat directions
and spontaneously breaks a continuous
global symmetry generally breaks SUSY
\cite{Affleck:1983vc,Affleck:1985xz}.  This is because
there must be a Goldstone boson (which has no interactions
in the potential), and by SUSY it must have a scalar partner
(a modulus),
but if there are no flat directions this is impossible.
(Unless the modulus is also a Goldstone boson.)  In the
early days people looked for theories that had no
classical flat directions (assuming that quantum
corrections would not cancel the classical potential)
and tried to make them break global symmetries in the perturbative
regime.  This method produced
a handful of dynamical SUSY breaking theories.  With duality we can find
many examples of dynamical SUSY breaking. An important
twist is that we will find that non-perturbative
quantum effects can lift flat directions  both at the origin of moduli space as well as
for large VEVs.

%%%%%%%%%%%%%%%%%%%%%%%%%%%%%%%%%%%%%%%%
\subsection{The 3-2 Model}
\label{3-2Model}
In the mid 1980s, Affleck, Dine, and Seiberg \cite{Affleck:1985xz}
found the simplest known model
of dynamical SUSY breaking. Their model has a gauge group
$SU(3)\times SU(2)$ and two global $U(1)$ symmetries
with the following chiral supermultiplets:
\beq
\begin{array}{c|cc|cc}
 & SU(3) & SU(2) & U(1) & U(1)_R\\ \hline
Q & \fund & \fund & 1/3 & 1\\
L &  {\bf 1} &  \fund & -1 & -3 \\
\overline{U}  &  \overline{\fund} &  {\bf 1} & -4/3 & -8 \ \\
\overline{D} & \overline{\fund}  &  {\bf 1} &2/3 & 4 
\end{array}
\eeq
I will write $\overline{Q}=(\overline{U},\overline{D})$,
and denote the intrinsic scales of the two gauge groups by
$\Lambda_3$ and $\Lambda_2$ respectively.
For $\Lambda_3 \gg \Lambda_2$ (that is when the $SU(3)$ interactions
are much stronger than the $SU(2)$ interactions), instantons  give
the standard Affleck-Dine-Seiberg superpotential:
\beq
W_{\rm dyn} = {{\Lambda_3^{7}}\over{{\rm det} (\overline{Q} Q) }} ~,
\label{DSB:dynADS32}
\eeq
which has a runaway vacuum.
Adding a tree-level trilinear term to the superpotential
\beq
 W =  {{\Lambda_3^{7}}\over{{\rm det} (\overline{Q} Q) }} +
\lambda \, Q \bar{D} L~,
\eeq
removes the classical flat directions and produces a stable
minimum for the potential. Since the vacuum is driven away from
the point where the VEVs vanish by the dynamical ADS potential
(\ref{DSB:dynADS32}), the global $U(1)$ symmetries are broken
and we expect (by the rule of thumb described above) that SUSY is
broken.

  The $L$ equation of motion
\beq
\frac{\partial W}{\partial L_\alpha} = \lambda  \epsilon^{\alpha \beta} 
Q_{m \alpha} \overline{D}^m=0~,
\eeq
tries to set 
${\rm det} \overline{Q} Q$ to zero
since
\beq
{\rm det} \overline{Q} Q &=& {\rm det}\left(\begin{array}{cc}
\overline{U} Q_1 & \overline{U} Q_2 \\
\overline{D} Q_1 & \overline{D} Q_2 
\end{array}\right)\nonumber \\
&=& \overline{U}^m Q_{m \alpha}\overline{D}^n
Q_{n \beta}\epsilon^{\alpha \beta}~. 
\eeq
Thus the
potential can't have a zero-energy  minimum since the
dynamical term blows up at  ${\rm det} \overline{Q} Q$=0.  
Therefore SUSY is indeed broken.

We can crudely estimate the vacuum energy for by taking all the VEVs to
be of order $\phi$.  For $\phi \gg \Lambda_3$ and $\lambda \ll 1$ 
we are in a perturbative regime. The potential is
then given by
\beq
V&=& |\frac{\partial W}{\partial Q}|^2
+|\frac{\partial W}{\partial \overline{U}}|^2
+|\frac{\partial W}{\partial \overline{D}}|^2
+|\frac{\partial W}{\partial L}|^2 \\
&\approx& \frac{\Lambda_3^{14}}{\phi^{10}} 
+ \lambda\frac{\Lambda_3^{7}}{\phi^{3}} + \lambda^2 \phi^4~,
\label{DSB:32potential}
\eeq
where in the last line we have dropped the numerical factors
since we are only interested in the scaling behavior of the solution.
This potential has a minimum near
\beq
\langle \phi \rangle \approx \frac{\Lambda_3} {\lambda^\frac{1}{7}}~,
\eeq
so we see that this solution is self-consistent:
the vacuum is weakly coupled for small $\lambda$ since this ensures
that $\phi \gg \Lambda_3$.  Plugging the solution back into the potential
(\ref{DSB:32potential}) we find the vacuum energy is of order
\beq
V \approx \lambda^\frac{10}{7} \Lambda_3^4~,
\eeq
which vanishes as $\lambda$ or $\Lambda$ go to zero, as it must.

Using duality Intriligator and Thomas \cite{Intriligator:1996fk}
showed that we can also understand
the case where $\Lambda_2 \gg \Lambda_3$ and supersymmetry is broken
non-perturbatively.
The $SU(2)$ gauge group has
4 doublets which is equivalent to
 2 flavors, so we have  confinement with chiral symmetry breaking.
 The $SU(3)$ gauge group has two flavors and is completely broken
for generic VEVs.
It is simpler to consider $SU(2)$ as  an $SU$ group rather than
an $Sp$ group, so we write the gauge invariant composites as
mesons and baryons:
\beq
\begin{array}{ccl}
M & \sim & \left( \begin{array}{cc}
L Q_1 & L Q_2 \\
Q_3 Q_1 & Q_3 Q_2
\end{array} \right)~,\\
 B & \sim & Q_1 Q_2 ~,\\
 \bar{B} & \sim & Q_3 L~. 
\end{array}
\eeq
In this notation the effective superpotential is
\beq
W = X \, \left( {\rm det} M -  B \overline{B} - \Lambda_2^{4} \right)  +
\lambda\left(\sum_{i=1}^2 M_{1{\rm i}} \overline{D}^{\rm i} +
\overline{B} \,\overline{D}^3 \right) ~,
\eeq
where $X$ is a Lagrange multiplier field that imposes the constraint
for confinement with chiral symmetry breaking.
The $\overline{D}$ equations of motion try to force $M_{1i}$ 
and $\overline{B}$ to zero while the 
constraint means that at least one of $M_{11}$, $M_{12}$, or $\overline{B}$
is non-zero, so we see that SUSY is broken at tree-level in the dual (confined)
description.  We can estimate the vacuum energy as
\beq
V\approx \lambda^2 \Lambda_2^4~.
\eeq
Comparing the vacuum energies in the two cases we see that the $SU(3)$
interactions dominate when $\Lambda_3 \gg \lambda^\frac{1}{7} \Lambda_2$.

Without making the approximation that one gauge group is much stronger than
the other we should consider the full superpotential
\beq
W = A \, \left( {\rm det} M -  B \bar{B} - \Lambda_2^{4} \right)  +
{{\Lambda_3^{7}}\over{{\rm det} (\overline{Q} Q) }} 
+ \lambda \, Q \bar{D} L~,
\eeq
which still breaks SUSY, although the analysis is more complicated.

%%%%%%%%%%%%%%%%%%%%%%%%%%%%%%%%%%%%%%
\subsection{The $\mbox{SU(5)}$ Model}
\label{SU5Model}
Another simple model analyzed by Affleck, Dine, and Seiberg
\cite{Affleck:1983vc} as well as Meurice and Veneziano \cite{Meurice:1984ai}
is  $SU(5)$  
with matter content $\overline{\fund}  + \Yasymm$, that is
an antifundamental and an antisymmetric tensor representation. 
This chiral gauge theory has no classical flat directions,
since there are no gauge invariant operators that we can write down.
Affleck, Dine, and Seiberg tried to match the anomalies in order to 
find a confined
description but found only ``bizarre", ``implausible''
solutions.  This lead to the
belief that the at least one of the global $U(1)$ symmetries was broken
and that therefore SUSY was broken (using the rule of thumb
described earlier).  Adding extra flavors
$(\fund +\overline{\fund})$
with masses Murayama \cite{Murayama:1995ng} showed
 that SUSY is broken, but taking the masses to $\infty$
takes the theory to a strongly coupled regime, so the possibility remained that
non-perturbative effects induced some phase transition as the mass
was varied.   With duality  arguments Pouliot
\cite{Pouliot:1995me} showed that SUSY is indeed broken
at strong coupling.

The simplest way to see SUSY breaking is to consider the case that where
there are enough flavors that the theory s-confines. 
Csaki, Schmaltz, and Skiba\cite{CsakiSchmaltzSkiba} showed 
that with four flavors 
 \beq
\begin{array}{c|c|ccccc}
& SU(5)  & SU(4) & SU(5) & U(1)_1 & U(1)_2 & U(1)_R \\ \hline
A & \Yasymm & 1 & 1 & 0 & 9 & 0 \\
\overline{Q} & \overline{\Yfund} & 1 & \Yfund & 4 & -3 & 0 \\
Q & \Yfund & \Yfund & 1 & -5 & -3 & \frac{1}{2} 
\end{array}
\eeq
the theory  s-confines.  To simplify the notation, rather than introducing
new symbols for each color-singlet composite I will simply label it by
it's constituents between parenthesis.  Thus the meson is denoted by
$(Q\overline{Q})$. The spectrum of massless composites is:
\beq
\begin{array}{c|ccccc}
 & SU(4) & SU(5) & U(1)_1 & U(1)_2 & U(1)_R \\ \hline
(Q\overline{Q}) \vphantom{  \Yasymm } &  \Yfund & \Yfund & -1 & -6 & \frac{1}{2} \\
(A\overline{Q}^2) & 1 &   \Yasymm & 8 & 3 & 0 \\
(A^{2}Q)  & \Yfund &  1 & -5 & 15 & \frac{1}{2} \\
(AQ^3) & \overline{\Yfund} &  1 & -15 & 0 & 
\frac{3}{2} \\
(\overline{Q}^{5}) &  1 & 1 & 20 & -15 & 0 \end{array}
\eeq
with a superpotential
\beq
 W_{dyn} &=& \frac{1}{\Lambda^{9}} \Big[ (A^2Q)(Q\overline{Q})^3 
(A\overline{Q}^2)+(AQ^3)(Q\overline{Q})(A\overline{Q}^2)^{2}
\\ && \,\,\,\,\,\,\,\,\,\,\,\,+ (\overline{Q}^{5})(A^2 Q)(A Q^3)
\Big]~.
\eeq
Note that by examining the global symmetry index structure one can
see that the first term in $W_{dyn}$ 
is  anti-symmetrized in both the $SU(5)$ and $SU(4)$
indices, while the second term is anti-symmetrized in just the $SU(5)$
indices.

 We can add mass terms and Yukawa couplings for the extra flavors:
 \beq
 \Delta W = \sum_{i=1}^4 m Q_i \overline{Q}_i + \sum_{i,j\le 4} 
 \lambda_{ij} A \overline{Q}_i\overline{Q}_j~,
 \eeq
 which lift all the flat directions.
 
 The equations of motion  give
 \beq
 {{\partial W}\over{\partial(\overline{Q}^{5})}} &=& (A^2 Q)(A Q^3) =0 ~,
\label{DSB:SU5eq1} \\
  {{\partial W}\over{\partial(Q\overline{Q})}}&=&3(A^2Q)(Q\overline{Q})^2 
(A\overline{Q}^2)+(AQ^3)(A\overline{Q}^2)^{2}+ m=0~.
\label{DSB:SU5eq2}
\eeq
Assuming that $(A^2Q) \ne 0$ then the first equation of motion 
(\ref{DSB:SU5eq1})
requires $(A Q^3)=0$ and multiplying the second equation 
of motion (\ref{DSB:SU5eq2}) by $(A^2Q)$ we see that because of the
antisymmetrizations the first term vanishes and therefore
\beq
 (AQ^3)(A\overline{Q}^2)^{2}=-m~,
\label{DSB:m5}
\eeq
but this contradicts $(A Q^3)=0$, so there is no solution with our assumption
 $(A^2Q) \ne 0$.

Assuming that $(A Q^3) \ne 0$ then the first equation of motion
requires $(A^2Q)=0$, and plugging this into the second equation of motion
 (\ref{DSB:SU5eq2}) we find equation (\ref{DSB:m5}) directly.
Multiplying equation (\ref{DSB:m5}) by $(AQ^3)$ we find that the 
left hand side vanishes again due to 
antisymmetrizations, so $(AQ^3)=0$ but this contradicts equation 
(\ref{DSB:m5}) and our assumption. Therefore
the equations of motion cannot be satisfied, and SUSY is broken at tree
level in the dual description.

%%%%%%%%%%%%%%%%%%%%%%%%%%%%%%%%%%
\subsection{SUSY Breaking and Deformed Moduli Spaces}

\label{sec:deformed}

The Intriligator-Thomas-Izawa-Yanagida 
\cite{Intriligator:1996pu,Izawa:1996pk} model is a vector-like theory
which consists of an $SU(2)$ SUSY gauge theory with two flavors\footnote{Since 
doublets and anti-doublets of $SU(2)$ are equivalent, an $SU(2)$ theory
with $F$ flavors has a global $SU(2F)$ symmetry rather than 
an $SU(F) \times SU(F)$ as one finds for a larger number of colors.}
and a gauge singlet:
\beq 
\begin{array}{cc|c}
& SU(2)  & SU(4) \\
 \hline
Q  &\Yfund & \Yfund  \\
S  & {\bf 1} &\Yasymm  \\
\end{array}
\eeq
with  a superpotential
\beq
W=\lambda S^{ij} Q_i Q_j~.
\eeq
The strong $SU(2)$ dynamics enforces a constraint
\beq
{\rm Pf}(QQ) = \Lambda^4~.
\label{DSB:PFconstraint}
\eeq
Where the Pfaffian\footnote{If we artificially divided $Q$ up into 
$q$ and $\overline{q}$ then we could follow our previous notation
$M=q \overline{q}$, $B = \epsilon^{ij}q_iq_j$,  
$\overline{B} = \epsilon^{ij}\overline{q}_i\overline{q}_j$
and write the constraint as ${\rm Pf}(QQ) = {\rm det}M - B \overline{B}=
\Lambda^4$.}
of a $2F \times 2F$ matrix $M$ is given by
\beq
{\rm Pf}(M) = \epsilon^{i_1\ldots i_{2F}} M_{i_1i_2}\ldots M_{i_{2F-1}i_{2F}}~.
\eeq
The equation of motion for for the gauge singlet $S$ is
\beq
{{\partial W}\over{\partial S^{ij}}}  = \lambda Q_i Q_j =0~.
\eeq
Since this equation is incompatible with the constraint 
(\ref{DSB:PFconstraint})
we see that SUSY is broken.

Another way to see this is that at least for large values of $\lambda S$ we
can integrate out the quarks, leaving
an $SU(2)$ gauge theory with no flavors which has gaugino condensation:
\beq
\Lambda_{\rm eff}^{3N}=\Lambda^{3N-2}\left(\lambda  S\right)^{2} ~,
\\
W_{\rm eff}=2 \Lambda_{\rm eff}^{3}= 2 \Lambda^{2}  \lambda S~, \\
\frac{\partial W_{\rm eff}}{\partial S^{ij}}  = 2 \lambda \Lambda^{2}~,
\eeq
so again we see that the vacuum energy is non-zero.  

%For general values of $\lambda S$ we can write:
%\beq
%W_{\rm eff}=\lambda S^{ij} Q_i Q_j+X({\rm Pf}M-\Lambda^4)~,
%\eeq
%where $X$ is a Lagrange multiplier field.
%For $\lambda \ll 1$ the vacuum is close to the SUSY QCD vacuum given
%by the $X$ equation of motion, and we can treat the first term
%in the superpotential as a small mass perturbation. 
%
%The potential energy is given by:
%\beq
%V=\sum_i|\frac{\partial W_{\rm eff}}{\partial Q_{i}}|^2
% +\sum_{ij}|\frac{\partial W_{\rm eff}}{\partial S^{ij}}|^2 ~.
%\eeq
%A supersymmetric vacuum exists if all the terms vanish.
%Treating $\lambda S$ as a mass perturbation
%we can set the derivatives with respect to $Q$ to zero simply
%by
%solving for the squark VEVs in the standard way.  This gives
%\beq
%Q_{i}Q_{j} = \left( {\rm Pf}(\lambda S) \Lambda^{3N-F}\right)^{{1}\over{N}} 
%\left(
%{{1}\over{\lambda S}}\right)_{ij}~.
%\eeq
%Plugging this back in to the potential gives
%\beq
%V&=&\sum_{ij}|{{\partial W_{\rm eff}}\over{\partial S^{ij}}}|^2 =
%|\lambda|^2\sum_{ij}|M_{ij}|^2 \nonumber ~,\\
%&=&|\lambda|^2 |{\rm Pf}S \Lambda^4| \sum_{ij}|\left({{1}\over{S}}\right)_{ij}%|^2
%~,
%\eeq
%which is minimized at 
%\beq
%S^{ij}=({\rm Pf}S)^{{1}\over{2}} \epsilon^{ij}
%\eeq
%so
%\beq
%V = 4 |\lambda|^2 \Lambda^4
%\eeq
%which agrees with our gaugino condensation calculation.

Since this theory is vector-like (it admits mass terms for all the quarks
and for $S$)
one would naively expect that this model could not break SUSY.
This is because the Witten index Tr$(-1)^F$ is non-zero with mass terms turned on so
there is at least one supersymmetric vacuum. Since the index is topological,
it does not change under variations of the mass.
However Witten noted that there is a loop-hole
in the index argument since the potential for large field
values are very different with $\Delta W=m_s S^2$ 
from the theory with $m_s\rightarrow 0$,
since in this limit  vacua can come in from or go out to $\infty$
and thus change the index.

At the level of analysis described above
$S$ appears to be a flat direction
but  
%As we saw in the O'Raifeartaigh model 
a general feature of SUSY breaking theories is
that flat directions become pseudo-flat.
With a non-zero vacuum energy, 
flat directions can be modified by corrections from the \Kahler  potential.
For large values of 
$\lambda S$ in this model there is a wavefunction
renormalization for $S$ \cite{Arkani-Hamed:1997ut,Dimopoulos:1997fv}
\beq
Z_S= 1+c \lambda \lambda^\dagger \ln\left( {{\mu_0^2}\over{\lambda^2 S^2}} \right)~.
\eeq
So the vacuum energy is corrected to be
\beq
V &=& \frac{4 |\lambda|^2}{|Z_S|} \Lambda^4 \nonumber\\
&\approx&|\lambda|^2 \Lambda^4\left[1+c \lambda \lambda^\dagger 
\ln\left( {{\lambda^2 S^2}\over{\mu_0^2}} \right) \right]~.
\eeq
So we see that the potential slopes towards the origin.  This can be stabilized
by gauging a subgroup of $SU(4)$.  Otherwise there is a calculable 
low-energy effective theory \cite{Chacko:1998si} near $\lambda S \approx 0$
(essentially an O'Raifeartaigh model \cite{O'Raifeartaigh:pr})
which has a local minimum at $S=0$.
The effective theory becomes non-calculable 
near $\lambda S \approx \Lambda$.  The behavior near this region is
unknown.

%%%%%%%%%%%%%%%%%%%%%%%%%%%%%%%%%%%%
\subsection{SUSY Breaking from Baryon Runaways}
\label{DSB:baryonrunaway}

Consider a generalization \cite{Luty:1998nq} of the 3-2 model:

\beq \begin{array}{c|cc|ccc}
& Sp(2N) & SU(2N - 1) & SU(2N - 1) & U(1) & U(1)_R \\
 \hline
Q &\Yfund & \Yfund  & {\bf 1}& 1 & 1 \\
L &\Yfund & {\bf 1}& \Yfund & -1 &-{{3}\over{2N - 1}} \\
\overline{U} & {\bf 1}& \overline{\Yfund} & \overline{\Yfund}& 0&
 {{2N + 2}\over{2N - 1}}\\
\overline{D} & {\bf 1}& \overline{\Yfund} & {\bf 1}& -6& -4N \\
\end{array}
\eeq
with a tree-level superpotential
\beq
W = \lambda Q L \overline{U}.
\eeq
If we turn off the $SU(2N - 1)$ gauge coupling and the superpotential,
$Sp(2N)$ is in a non-Abelian Coulomb phase for $N \ge 6$,
it has a weakly-coupled dual description for $N = 4,5$,
it s-confines for $N = 3$, and confines with a quantum-deformed moduli
space for $N = 2$. If we turn off the $Sp(2N)$ gauge coupling and the superpotential,
$SU(2N - 1)$ s-confines for any $N \ge 2$. Here we will consider the case that
$\Lambda_{SU} \gg\Lambda_{Sp}$.

Including the effects of the tree-level superpotential,
this theory has a classical moduli space that can be parameterized by
the gauge-invariants:
 
\beq
 \begin{array}{c|ccc}
 & SU(2N - 1) & U(1) & U(1)_R \\
 \hline
M = (L L)
& \Yasymm & -2& -{{6}\over{2N - 1}}
\\
B = (\overline{U}^{2N - 2} \overline{D})
& \Yfund & -6 & -{{4(N^2 - N + 1)}\over{2N - 1}}
\\
b = (\overline{U}^{2N - 1})
& {\bf 1}& 0& 2N + 2 \\
\end{array}
\eeq
subject to the constraints
\beq
M_{jk} B_l \epsilon^{k l m_1 \cdots m_{2N - 3}} = 0,
\qquad
M_{jk} b = 0.
\eeq
These constraints split the moduli space into two branches:
on one of them $M = 0$ and $B, b \ne 0$,
and on the other $M \ne 0$ and $B, b = 0$.

% -----------------------------------------------------------------------
First consider the branch where $M=0$.  We will see that later that
the true vacuum actually lies on this branch.
In terms of the elementary fields, this corresponds to
the VEVs (up to gauge and flavor transformations)
\beq
\langle \overline{U} \rangle = \pmatrix{ v \cos\theta & \cr
& v {\bf 1}_{2N - 2} \cr},
\qquad
\langle \overline{D}\rangle
 = \pmatrix{ v \sin\theta \cr 0 \cr \vdots \cr 0 \cr},
\eeq

For large VEVs, $v>\Lambda_{SU}$, $SU(2N-1)$ is generically broken and the superpotential gives masses to $Q$ and $L$
or order $\lambda v$.  The low-energy effective theory is pure $Sp(2N)$,
which has gaugino condensation. The intrinsic scale of the low-energy
theory is
\beq
\Lambda_{\rm eff}^{3(2N+2)}=\Lambda_{Sp}^{3(2N+2)-2(2N-1)}\left(\lambda 
\overline{U}\right)^{2(2N-1)}~.
\eeq
The effective superpotential is given by
\beq
W_{\rm eff}\propto \Lambda_{\rm eff}^{3}\sim \Lambda_{Sp}^{3}\left({{\lambda 
\overline{U}}\over{\Lambda_{Sp}}}\right)^{{2N-1}\over{N+1}}~.
\eeq
For $N>2$ this forces $\langle \overline{U} \rangle$ to zero.

Now consider the case of small VEVs, $v < \Lambda_{SU}$, 
then $SU(2N-1)$ s-confines so we have
the following effective theory
\beq
\begin{array}{c|cc}
& Sp(2N)  & SU(2N - 1) \\
 \hline
L &\Yfund & \Yfund  \\
(Q \overline{U}) &\Yfund   & \overline{\Yfund} \\
(Q \overline{D}) &\Yfund   & {\bf 1} \\
(Q^{2N-1}) &\Yfund   & {\bf 1} \\
B & {\bf 1}& \Yfund \\
b & {\bf 1}&{\bf 1}\\
\end{array}
\eeq
with a superpotential
\beq
W&=&\lambda (Q\overline{U})L \nonumber \\
&&+{{1}\over{\Lambda_{SU}^{4N-3}}}
\left[(Q^{2N-1})(Q \overline{U})B+(Q^{2N-1})(Q \overline{D})b - {\rm det} 
\overline{Q}Q \right].
\eeq
The first term in the superpotential is a mass term, 
so $(Q\overline{U})$ and $L$ can be integrated out with $(Q\overline{U})=0$,
and we have a low-energy superpotential
\beq
W={{1}\over{\Lambda_{SU}^{4N-3}}}
(Q^{2N-1})(Q \overline{D})b ~.
\eeq
On this branch of the moduli space $\langle b \rangle = \langle \overline{U}^{2N-1}
\rangle \ne 0$, and this VEV gives a mass to $(Q^{2N-1})$ and $(Q \overline{D})$
which leaves an pure $Sp(2N)$ as the low energy effective theory.
So we again find gaugino condensation with a scale given by
\beq
\Lambda_{\rm eff}^{3(2N+2)}=\Lambda_{Sp}^{3(2N+2)-2(2N-1)}\left(\lambda 
\Lambda_{SU}\right)^{2(2N-1)} \left({{b}\over{\Lambda_{SU}}}\right)^2~,
\eeq
and a effective superpotential
\beq
W_{\rm eff}\propto \Lambda_{\rm eff}^{3}\sim b^{{1}\over{N+1}}\left(
\Lambda_{Sp}^{N+4} \lambda^{2N-1}
\Lambda_{SU}^{(2N-2)}\right)^{{1}\over{N+1}}~.
\eeq
Which forces $b \rightarrow \infty$, (this is a baryon runaway vacuum)
but the effective theory is only valid
for scales below $\Lambda_{SU}$, since we assumed   $v < \Lambda_{SU}$.
We have already seen that beyond this point the potential
starts to rise again, so the vacuum is around 
\beq
\langle b \rangle =\langle \overline{U}^{2N-1}\rangle  \sim \Lambda_{SU}^{2N-1}~.
\eeq
With some more work \cite{Luty:1998nq}  one can also see that SUSY is also
broken when $\Lambda_{Sp}
\gg \Lambda_{SU}$.  

An especially interesting case is when $N=3$, where the
$Sp(2N)$ s-confines and we have
the following effective theory
\beq 
\begin{array}{cc|c}
& SU(5)  & SU(5) \\
 \hline
(Q Q) &\Yasymm  & {\bf 1}  \\
(L L)  & {\bf 1} &\Yasymm  \\
(Q L) &\Yfund   &\Yfund \\
(Q^{2N-1}) &\Yfund   & {\bf 1} \\
\overline{U} & \overline{\Yfund}&  \overline{\Yfund} \\
\overline{D} &  \overline{\Yfund} &{\bf 1}\\
\end{array}
\eeq
with
\beq
W=\lambda (Q L)\overline{U} +Q^{2N-1}L^{2N-1}~.
\eeq
The non-Abelian global symmetry group
in this case is $SU(5)$ which is just large enough
to embed the standard model gauge groups, and hence this is a candidate
model for gauge mediation \cite{GaugeMediation}.
After integrating out  $(Q L)$ and $\overline{U}$ we find $SU(5)$ with
an antisymmetric tensor, an antifundamental, and some gauge singlets,
which we have already seen breaks SUSY in section \ref{SU5Model}.

Returning to general $N$ and considering 
the other branch of the moduli space 
where $M= (LL) \ne 0$, one can see that the
D-flat directions for $L$ break $Sp(2N)$  to $SU(2)$, and
the spectrum of
the effective theory is 
\beq
\begin{array}{c|c|c}
& SU(2)  & SU(2N - 1) \\
 \hline
Q^\prime  &\Yfund   & \Yfund \\
L^\prime  &\Yfund & {\bf 1}  \\
\overline{U}^\prime  & {\bf 1}&\overline{\Yfund} \\
\overline{D} & {\bf 1}&\overline{\Yfund}\\
\end{array}
\eeq
and some singlets with a superpotential
\beq
W=\lambda Q^\prime\overline{U}^\prime L^\prime~.
\eeq
This is a generalized 3-2 model (see section \ref{3-2Model})
with a dynamical superpotential. For
$\langle L \rangle \gg \Lambda_{SU}$  the vacuum energy is independent
of the $SU(2)$ scale and proportional
to $\Lambda_{SU(2N-1)}^4$ which itself is proportional to
a positive power of 
$\langle L \rangle$, thus the effective potential
in this region drives $\langle L \rangle$ smaller.
For $\langle L \rangle \ll \Lambda_{SU}$ we can use the s-confined
description above, and find again that the baryon $b$ runs away.  For
 $\langle L \rangle \approx \Lambda_{SU}$, the vacuum energy is
 \beq
 V \sim \Lambda_{SU}^4~.
 \eeq
 which is larger than the vacuum energy on the other branch, so we
 see that the global minimum is on the baryon branch with 
$b= (\overline{U}^{2N - 1})\ne 0$.

Further examples of dynamical SUSY breaking can be found in references 
\cite{Shadmi:1999jy,Affleck:1984uz,Csaki:1996gr,Poppitz:1996wp}.

%%%%%%%%%%%%%%%%%%%%%%%%%%%%%%%%%%%%%%%%%%
\subsection{Direct Gauge Mediation}

A more elegant (and more difficult from the model building view point)
approach to gauge mediation is direct gauge mediation where the fields
that break SUSY have SM gauge couplings.  These types of models only have
two sectors (a SUSY breaking sector and an MSSM sector) rather than three. Consider the model \cite{Dimopoulos:1997ww}
\[ \begin{array}{c|cc|c}
& SU(5)_1 & SU(5)_2 & SU(5) \\
 \hline
Y & {\bf 1}& \Yfund & \overline{\Yfund}\\
\phi & \overline{\Yfund} & {\bf 1} & \Yfund   \\
\overline{\phi} &\Yfund &\overline{\Yfund}& {\bf 1} \\
\end{array}
\]
with a superpotential
\beq
W= \lambda Y^i_j \overline{\phi}^j \phi_i~.
\eeq
We can weakly gauge the global $SU(5)$ with the SM gauge groups.
For large $Y \gg \Lambda_1, \Lambda_2$, $\phi$ and $\overline{\phi}$ get a mass and, by the usual
matching argument, the scale of the effective gauge theory is
\beq
\Lambda_{\rm eff}^{3 \cdot 5}=\Lambda_{1}^{3\cdot 5-5}\left(\lambda 
X \right)^{5}~,
\eeq
where $X=({\rm  det} Y)^{1/5}$.
The effective gauge theory undergoes gaugino condensation, so the superpotential
is given by
\beq
W_{\rm eff}=\Lambda_{\rm eff}^{3}\sim \lambda X \Lambda_{1}^{2}~,
\eeq
so SUSY is broken a la the Intriligator-Thomas-Izawa-Yanagida model discussed in 
section \ref{sec:deformed}.  
The vacuum energy is given by
\beq
V \approx\frac{ | \lambda  \Lambda_{1}^{2}|^2}{Z_X}~,
\eeq
where $Z_X$ is the wavefunction renormalization for $X$.
For large $X$ the vacuum energy grows monotonically.
A local minimum occurs at the point where the anomalous dimension
$\gamma=0$.  For $\langle X \rangle > 10^{14}$ GeV, the Landau pole for
$\lambda$ is above the Planck scale. 

The theoretical problem with this model is that for small values of $X$ there is
a supersymmetric minimum along a baryonic direction \cite{yuri}.  For this regime it is appropriate to look at the constrained mesons and baryons of $SU(5)_1$.  The superpotential is
\beq
W = A( {\rm det} M - B \overline{B} - \Lambda_1^{10})  + \lambda Y M~.
\eeq
There is a supersymmetric minimum at $B \overline{B}= - \Lambda_1^{10}$, $Y=0$, $M=0$.
This supersymmetric minimum would have to be removed, or the non-supersymmetric minimum
made sufficiently metastable (with a lifetime of the order of the age of the Universe) by adding
appropriate terms to the superpotential that force  $B \overline{B}=0$

 The phenomenological problem with this model is that
heavy gauge boson messengers can give negative contributions to 
squark and slepton squared masses.  Consider the general case where
a Goldstino VEV
\beq
\langle X \rangle &=& M +\theta^2 F~,
\eeq
breaks SUSY and breaks
two gauge groups down to the SM gauge groups
\beq
G \times H \rightarrow SU(3)_c \times SU(2)_L \times U(1)_Y~,
\eeq
with
\beq
{{1}\over{\alpha(M)}} &=& {{1}\over{\alpha_G(M)}}+{{1}\over{\alpha_H(M)}}~.
\eeq
Analytic continuation in superspace gives
\beq
M_\lambda = {{\alpha(\mu)}\over{4 \pi}} (b-b_H -b_G) {{F}\over{M}}~,
\eeq
and
\beq
m_Q^2 &=& 2 C_2(r) {{\alpha(\mu)^2}\over{(16 \pi^2)^2}} \left({{F}\over{M}}\right)^2 
\nonumber \\&& \left[ (b+(R^2-2)b_H -2 b_G) \xi^2
+{{b-b_H-b_G}\over{b}}(1-\xi^2) \right]~,
\eeq
where
\beq
\xi &=& {{\alpha(M) }\over{\alpha(\mu)} }~, \\
R &=& {{\alpha_H(M) }\over{\alpha(M)} }~.
\eeq
This typically gives a negative mass squared for right-handed sleptons.

Another danger for direct mediation models arises if not all the messengers are
heavy.  Then two-loop RG evolution gives:
\beq
\mu {{d}\over{d \mu}} m_Q^2 \propto -g^2 M_\lambda^2 +c g^4 
{\rm Tr}\left((-1)^{2F} m_{i}^2 \right)~,
\eeq
the one-loop term proportional to the gaugino mass squared drives the
scalar mass positive as the renormalization scale is run down, while the
two-loop term
can drive the mass squared negative.  This effect is maximized when the
gaugino is light, so in the standard case where the gluino is
heavy, then again it is the sleptons that receive dangerous negative contributions.
This effect is also dangerous \cite{Arkani-Hamed:1997ab} in models where the squarks and sleptons of the first two
generations are much heavier that 1 TeV.

%%%%%%%%%%%%%%%%%%%%%%%%%%%%%%%%%%%%%%%%
\subsection{Single Sector Models}
Another appealing approach to gauge mediation
is to have the strong
dynamics that break SUSY also produce composite MSSM particles
\cite{Luty:1998vr}. Thus rather that having three sectors (one for the
standard model, one hidden sector that dynamically breaks SUSY,
and one ``messenger'' sector that couples to the SUSY breaking
and has standard model gauge interactions)
there is really just one sector.

Consider a SUSY model with gauge group $SU(k) \times SO(10)$ with
the following matter content\footnote{Note that the spinor representation
of $SO(10)$ has dimension 16.}:
\beq
\begin{array}{c|cc|cc}
& SU(k) & SO(10) & SU(10) & SU(2) \\
 \hline
Q &\Yfund & \Yfund  & {\bf 1}& {\bf 1} \\
L &\overline{\Yfund} & {\bf 1}& \Yfund & {\bf 1} \\
\overline{U} & {\bf 1}& \Yfund & \overline{\Yfund}&{\bf 1}\\
S & {\bf 1}& {\bf 16}& {\bf 1}&\Yfund  \\
\end{array}
\eeq
and a superpotential
\beq
W= \lambda Q L \overline{U}~.
\eeq
The $SU(10)$ global symmetry is large enough to embed
the standard model gauge group, or a complete Grand Unified Theory
gauge group ($SO(10)$ or $SU(5)$) can be embedded.

This is a baryon runaway model (similar to the models described in
section \ref{DSB:baryonrunaway}).
For large ${\rm det}\overline{U}\gg \Lambda_{10}$
\beq
W_{\rm eff} \sim \overline{U}^{10/k}~,
\eeq
while for small ${\rm det}\overline{U}\ll \Lambda_{10}$:
\beq
W_{\rm eff} \sim \overline{U}^{10(1-\gamma)/k}~,
\eeq
where $\gamma$ is the anomalous dimension of $\overline{U}$
which depends on the as yet not fully understood details of  $SO(10)$ gauge
theories with spinors \cite{Berkooz:1997bb}, but which seem to have
infrared fixed points \cite{Terning:1997jj,Berkooz:1997bb}.
For $10 \ge k > 10 (1-\gamma)$ SUSY is broken.  There are two composite 
generations corresponding to the spinor $S$.  The composite
squarks and sleptons have masses of order
\beq
m_{\rm comp}\approx {{F}\over{\overline{U}}}~.
\eeq
where $F$ is the value of the $F$-component of $\overline{U}$ which
breaks SUSY (i.e. the vacuum energy is $|F|^2$).
This can be thought of as gauge mediation via the strong $SO(10)$ interactions.
The global $SU(2)$ symmetry enforces a degeneracy that suppresses flavor
changing neutral currents.
The composite fermions can only get couplings to the Higgs fields
 from higher dimension operators,
so they are light.
The fundamental gauginos and third generation scalars get masses from
gauge mediation.  Since the superpartners of the the first two (composite) generations are 
much heavier than the superpartners of the third generation, the spectrum is very similar
to that of the ``More Minimal" Supersymmetric Standard Model \cite{Cohen:1996vb}.

%%%%%%%%%%%%%%%%%%%%%%%%%%
\section*{Acknowledgments}
I would like to  thank Howard Georgi and the Harvard and MIT students and post-docs who attended Physics 284 in the fall 0f 1999:  they were the first guinea pigs for these lectures.
I would also like to thank the organizers, lecturers, and students from TASI 2002 who provided
stimulating and fun environment for these lectures. Finally I must  thank Csaba Cs\'aki, Vladimir Dobrev, Josh Erlich,
Shiraz Minwalla, Stavros Mouslopoulos, Ann Nelson, Ricardo Rattazzi,  Yuri Shirman, and Witold
Skiba
for useful discussions and comments.
J.T. is supported by in part by the US Department of Energy under contract W-7405-ENG-36, 
and during the original stages of preparation was supported by in part by the NSF under grant PHY-98-02709.

\end{document}